\documentclass[twocolumn,aps,prd,eqsecnum,nofootinbib]{revtex4}
\usepackage{epsfig}
\usepackage{amssymb}
\usepackage{amsmath}
\usepackage{amsfonts}
\usepackage{bm}

\newcommand{\nn}{\nonumber}
\newcommand{\cmz}{\,^{\scriptscriptstyle \text{cm}}\!z}
\newcommand{\cmbmz}{\,^{\scriptscriptstyle \text{cm}}\!{\bm{z}}}
\newcommand{\cmddotz}{\,^{\scriptscriptstyle \text{cm}}\!{\ddot z}}
\newcommand{\cmdotz}{\,^{\scriptscriptstyle \text{cm}}\!{\dot z}}
\newcommand{\nM}{\,^{\scriptscriptstyle \text{n}}\!M}
\newcommand{\ndotM}{\,^{\scriptscriptstyle \text{n}}\!{\dot M}}
\newcommand{\nddotM}{\,^{\scriptscriptstyle \text{n}}\!{\ddot M}}
\newcommand{\ndddotM}{\,^{\scriptscriptstyle \text{n}}\!{\dddot M}}
\newcommand{\pndotM}{\,^{\scriptscriptstyle \text{pn}}\!{\dot M}}
\newcommand{\pnddotM}{\,^{\scriptscriptstyle \text{pn}}\!{\ddot M}}
\newcommand{\nG}{\,^{\scriptscriptstyle \text{n}}\!G}
\newcommand{\ndotG}{\,^{\scriptscriptstyle \text{n}}\!{\dot G}}
\newcommand{\nddotG}{\,^{\scriptscriptstyle \text{n}}\!{\ddot G}}
\newcommand{\ndddotG}{\,^{\scriptscriptstyle \text{n}}\!{\dddot G}}

\newcommand{\nP}{\,^{\scriptscriptstyle \text{n}}\!P}
\newcommand{\pnP}{\,^{\scriptscriptstyle \text{pn}}\!P}
\newcommand{\ncalT}{\,^{\scriptscriptstyle \text{n}}\!{\cal T}}
\newcommand{\pncalT}{\,^{\scriptscriptstyle \text{pn}}\!{\cal T}}
\newcommand{\pnM}{\,^{\scriptscriptstyle \text{pn}}\!M}
\newcommand{\pnG}{\,^{\scriptscriptstyle \text{pn}}\!G}
\newcommand{\barnM}{\,^{\scriptscriptstyle \text{n}}\!\bar{M}}
\newcommand{\barnG}{\,^{\scriptscriptstyle \text{n}}\!\bar{G}}
\newcommand{\barpnM}{\,^{\scriptscriptstyle \text{pn}}\!\bar{M}}
\newcommand{\barpnG}{\,^{\scriptscriptstyle \text{pn}}\!\bar{G}}

\newcommand{\bodyM}{\mathcal M}
\newcommand{\bodyS}{\mathcal S}
\def\ve{\varepsilon}
\def\be{\begin{equation}}
\def\ee{\end{equation}}
\def\bfzeta{\mbox{\boldmath $\zeta$}}
\def\bfnabla{\mbox{\boldmath $\nabla$}}
\DeclareMathSymbol{\R}{\mathbin}{AMSb}{"52}

\begin{document}

\title{Post-1-Newtonian equations of motion for systems of arbitrarily structured bodies}
\author{\'{E}tienne Racine}
\author{ \'{E}anna \'{E}.\ Flanagan}
\affiliation{Center for Radiophysics and Space Research, Cornell University, Ithaca, New York, 14853}
\date{\today}
%%%%%%%%%%%%%%%%%%%%%%%%%%%%%%%%%%%%%%%%%%%%%%%%%%%%%%%%%%%%%%%%%%%%%%%%%%%%%%%
\begin{abstract}

We give a surface integral derivation of post-1-Newtonian
translational equations of motion for a system of arbitrarily
structured bodies, including the coupling to all the bodies' mass and
current multipole moments.   The derivation requires only that the
post-1-Newtonian vacuum field equations are satisfied in weak-field
regions between the bodies; the bodies' internal gravity can be
arbitrarily strong. In particular black holes are not excluded.
The derivation extends previous results due to Damour, Soffel
and Xu (DSX) for weakly self-gravitating bodies in which the
post-1-Newtonian field equations are satisfied everywhere.
The derivation consists of a number of steps:  (i) The
definition of each body's current and mass multipole moments and
center-of-mass worldline in terms of the behavior of the metric in a
weak-field region surrounding the body.  (ii) The definition for each
body of a set of
gravitoelectric and gravitomagnetic tidal moments that act on that body, again
in terms of the behavior of the metric in a weak-field region
surrounding the body.  For the special case of weakly self-gravitating
bodies, our definitions of these multipole and tidal
moments agree with definitions given previously by DSX.  (iii) The
derivation of a formula, for any given body, of the second time
derivative of its mass dipole moment in terms of its other multipole
and tidal moments and their time derivatives.  This formula was
obtained previously by DSX for weakly self-gravitating bodies.  (iv)
A derivation of the relation between the tidal moments acting on each
body and the multipole moments and center-of-mass worldlines of all
the other bodies.  A formalism to compute this relation was developed
by DSX; we simplify their formalism and compute the relation
explicitly.  (v) The deduction from the previous steps of the explicit
translational equations of motion, whose form has not been previously
derived.

\end{abstract}
%%%%%%%%%%%%%%%%%%%%%%%%%%%%%%%%%%%%%%%%%%%%%%%%%%%%%%%%%%%%%%%%%%%%%%%%%%%%%%%
\maketitle
%%%%%%%%%%%%%%%%%%%%%%%%%%%%%%%%%%%%%%%%%%%%%%%%%%%%%%%%%%%%%%%%%%%%%%%%%%%%%%%%%%%%%%%%%%%%%%%%%%%%%%%%%%%%%%%%%%%%%%%%%%%%%%%%%%%%%%%%%%%%%%%%%%%%%%%%%%%%%%

\section{Introduction and summary}

\subsection{Background and motivation}

For slow motion sources in weak gravitational fields, general
relativity can be accurately described in terms of a post-Newtonian
approximation scheme.  This approximation scheme is extremely useful
in applications and is very well developed.  Reviews of post-Newtonian
theory can be found in Refs.\ \cite{damour,dsxI,blanchet} and in the
textbook by Will \cite{Will}.

There are several different types of equations that arise in
post-Newtonian theory.  First, one has continuum field equations,
which are usually specialized to gravity coupled to perfect or
imperfect fluids.  These have been derived up to post-2.5-Newtonian order \cite{chandrasekhar}.  At post-1-Newtonian
order, they have been extended beyond general relativity to encompass
the class of theories of gravity described by the parameterized
post-Newtonian framework \cite{Will}.

A second type of equation of motion applies to systems consisting of
$N$ interacting, extended bodies moving under their mutual gravitational
interactions, in the limit where the bodies' sizes are small compared
to their mutual separations.  For such systems one has ``point
particle'' equations of motion.  Such equations were first derived
\cite{damour} at post-1-Newtonian order by Lorentz and Droste
\cite{ld}, and later independently by Einstein, Infeld and Hoffmann
(EIH) \cite{eih}.  They were also independently derived by Petrova \cite{petrova}
using a method devised by Fock \cite{fock}.  These equations are
usually called the EIH equations.
In recent years the advent of gravitational wave astronomy
\cite{detectors,CutlerThorne} has spurred renewed interest
in such equations of motion.  For coalescing binary systems,
the waveforms of the emitted gravitational waves are expected to carry
a great deal of information, and full exploitation of the expected observations
will require accurate theoretical models of the waveforms
\cite{CutlerThorne}.  This requirement has prompted the computation of
point-particle equations of motion (as well as radiation reaction
effects) to higher and higher post-Newtonian orders.
Most recently the coalescence waveform's phase has been computed up to
post-3.5-Newtonian order \cite{frenchgw}; see also Refs.\
\cite{americangw,japanesegw}.
At post-5-Newtonian order and higher, the concept of point-particle
equations of motion will break down due to effects related to the
finite sizes of the bodies \cite{Damour83}.  However
an argument due to Damour \cite{Damour83} indicates that
the point-particle equations should be well defined
at lower orders, up to and including post-4.5-Newtonian.

A third type of equation of motion applies to systems of $N$
interacting bodies whose sizes cannot be neglected.  These equations
consist of the point particle equations of motion supplemented by
tidal interaction terms.  In principle, if one included tidal
interactions to all multipole orders, and in addition coupled the
equations of motion to a dynamical description of the internal degrees
of freedom in each body, one would obtain a complete description of
the system, equivalent to that provided by the continuum
equations of motion (up to radiative effects).

For a system of bodies of typical size $\sim R$, of typical mass $\sim
M$, and with typical separations $\sim D$, the force $F$ that acts on one of
the bodies can be written schematically as\footnote{These scalings
apply to generic bodies; if the bodies are spherically symmetric the
scalings are of course altered.}
\begin{eqnarray}
F &\sim& \frac{M^2}{D^2} \bigg\{ 1 + O \left( \frac{M}{D} \right) +
O\left( \frac{M^2}{D^2} \right) + \ldots \nn \\
\mbox{} && + O\left[ \left( \frac{R}{D} \right)^l \right] +
O\left[ \frac{M}{D} \left( \frac{R}{D} \right)^l \right] + \ldots
\bigg\}.
\label{eq:eom_schematic}
\end{eqnarray}
Here we use geometric units with $G = c=1$.  The terms inside the
curly brackets are as follows.  On the first
line, the 1 is the usual Newtonian force between two point particles and the second and third
terms are the post-1-Newtonian and post-2-Newtonian point-particle
corrections.  On the second line, the first term is the correction due
to Newtonian tidal couplings; the minimum value of $l$ allowed is
$l=2$ corresponding to quadrupolar coupling.  The second term
describes the post-1-Newtonian tidal couplings.  Here the minimum
allowed value $l$ is lower than in the Newtonian case due to
gravitomagnetic interactions
which have no Newtonian analogs.  This minimum value is $l=1/2$,
corresponding to spin-orbit couplings (assuming that the bodies
internal velocities are maximal, $v \sim \sqrt{M/R}$).

The purpose of this paper is to compute in detail the post-1-Newtonian
tidal interaction terms in Eq.\ (\ref{eq:eom_schematic}), for all
values of $l$, for a system of $N$ bodies.  The explicit form of
these terms has not been derived before, although there is a
substantial literature on this topic
\cite{dsxI,dsxII,dsxIII,dsxIV,dsxV,skpw,Futamasenew,Kopeikin0,Brumberg1989,Brumberg1991,Klioner,KlionerII}.

There are a number of motivations for this computation.
First, as described in Refs.\ \cite{dsxI,skpw}, in the area of
celestial mechanics future experiments and
observations in the solar system will provide very high precision
data.  For example, there are current plans to increase the accuracy
of lunar laser ranging from the current centimeter level to the millimeter
level \cite{lunar}.  The future astrometric missions SIM (Space Interferometry Mission) and GAIA (Global Astrometric Interferometer for Astrophysics) are
expected to measure angles to an accuracy of a few microarcseconds, as
compared to the current accuracy of milliarcseconds.
In the radio, VLBI (Very Long Baseline Interferometry) observations currently can yield precisions of
order 10 microarcseconds \cite{vlbi}.
Also, the proposed future laser astrometric test
of relativity (LATOR) mission \cite{LATOR} would be sensitive to
post-2-Newtonian effects, and therefore would likely require detailed
modeling of post-1-Newtonian tidal effects.

Second, gravitational wave measurements of coalescing binary compact
stars will likely have some ability to detect finite size effects for
sufficiently strong signals \cite{Rasio}.  Although post-Newtonian
tidal effects will in many cases be small compared to Newtonian tidal
effects, there are some situations where the post-1-Newtonian effects
dominate.  An example is the gravitomagnetic resonant excitation of
Rossby modes in neutron stars that are spinning at $\sim 100 \, {\rm
Hz}$, which could be detectable with LIGO for moderately strong
detected inspirals \cite{Racine1}.

\subsection{Tidal coupling in post-Newtonian theory}

The textbook treatment of post-1-Newtonian gravity \cite{Will} is
inadequate for the treatment of tidal interactions for several
reasons, as explained by Damour et. al. \cite{dsxI}.  First, the
standard treatment uses a single global coordinate system.
Although one can write down the continuum equations of motion for a
given body in that coordinate system, it is very difficult to separate
out the gravitational influences of the other bodies from the
self-field of the body, since the fractional distortions of the
coordinate system produced by the other bodies can be large even when
the tidal distortion of the star is negligible.  The development of
approximation schemes such as linear perturbations about an
equilibrium state is hindered by the fact that the equilibrium state
is not described in the usual way in the global coordinates.

This difficulty has been comprehensively addressed in a series of
papers by Brumberg and Kopeikin (BK) \cite{Kopeikin0,Brumberg1989,Brumberg1991}
and by Damour, Soffel and Xu (DSX) \cite{dsxI,dsxII,dsxIII,dsxIV}.
These authors developed a detailed theory of post-1-Newtonian reference
frames, in which each body has associated with it a coordinate system
naturally adapted to that body.
DSX also developed a formalism to compute translational equations of
motion including the coupling to all the mass and current multipole
moments of each body\footnote{The BK and DSX formalisms have recently been
generalized to the parameterized post-Newtonian framework for
scalar-tensor theories of gravity by Kopeikin and
Vlasov\protect{\cite{Kopeikin}}.}.
They applied their formalism to compute equations of motion including
spin and quadrupole couplings.  In this paper, we extend the DSX
results in two ways.  First, by simplifying their formalism we are
able to compute the explicit form of the translational equations of
motion, including all the multipole couplings.
Second, we give a
derivation that is valid for strongly self-gravitating objects as well
as weakly self-gravitating objects\footnote{That is, we show that the
dominant fractional errors scale as $O(M^2/D^2)$; global
post-Newtonian methods \protect{\cite{dsxI,dsxII,dsxIII}} show only
that these errors are $O(M^2/R^2)$ or smaller.}.  We need only assume that the
post-1-Newtonian field equations are satisfied in a weak field
region surrounding each body.  The bodies' internal gravitational
fields can be arbitrarily strong; in particular our assumptions do not
exclude black holes.  By contrast, DSX assumed the global validity of
the post-1-Newtonian continuum field equations, and so their
derivation applies only to weakly self-gravitating objects.
Our result also generalizes existing derivations
of the Newtonian \cite{Futamasenew} and
post-1-Newtonian \cite{ifa,thornehartle}
equations of motion
for strongly self-gravitating objects that incorporate only a few low
order multipoles. Similar derivations to higher post-Newtonian orders
including monopole terms only can be found in Refs.\ \cite{damourprl,ifa1}.

\begin{figure}
\begin{center}
\epsfig{file=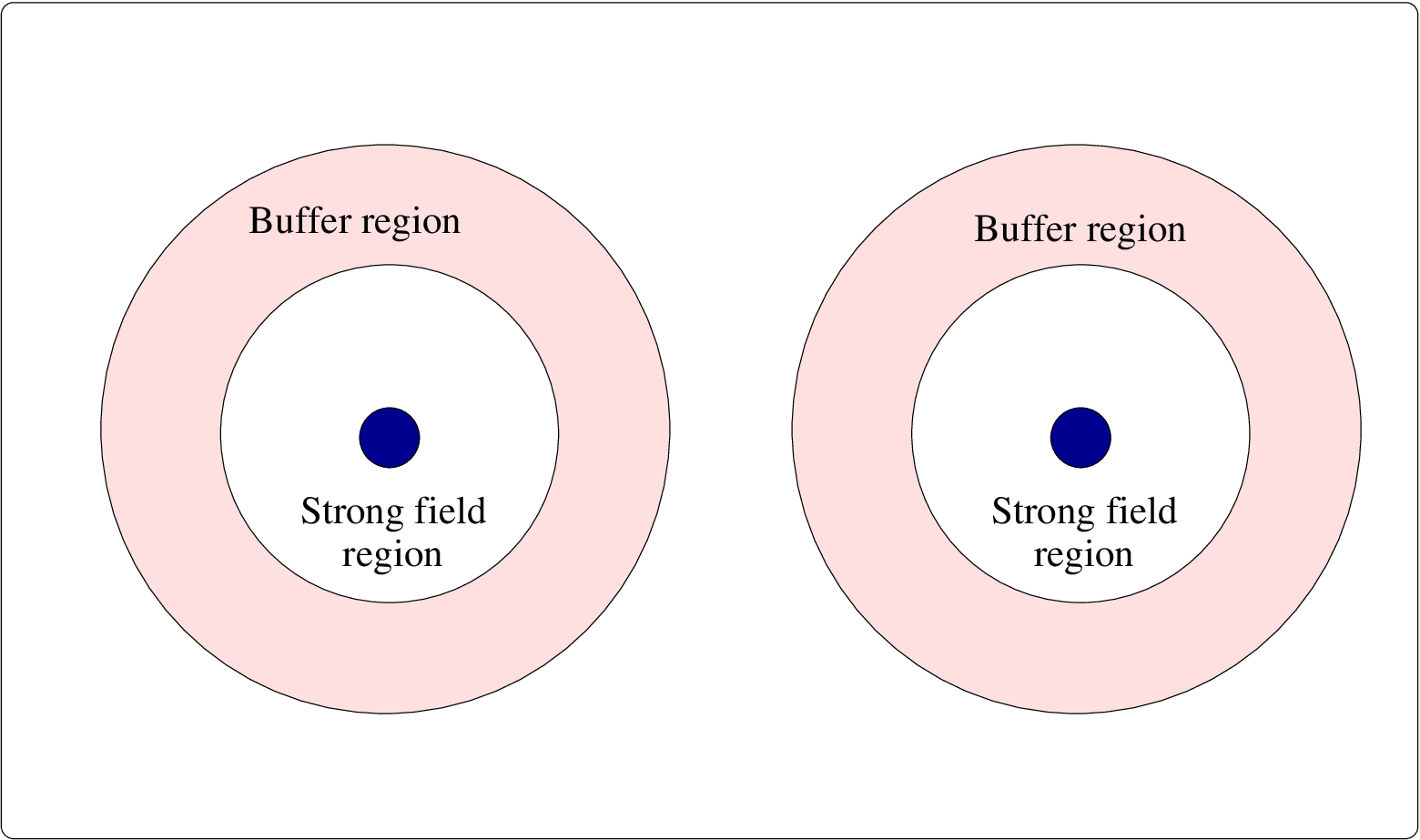,width=8.5cm}
\caption{An illustration of our assumptions for a system of $N$ bodies.
Each body is surrounded by a strong field region which is excluded from
our analysis.  Surrounding these strong field regions are weak field
buffer regions.  Each body's center of mass worldline and mass and
current multipole moments are defined in terms of the behavior of the
metric in that body's buffer region.  We assume that the vacuum
post-1-Newtonian field equations are satisfied in all the buffer
regions and in the regions of space between the buffer regions.}
\label{fig1}
\end{center}
\end{figure}

One of the key ideas underlying our derivation is that the equations
of motion are determined entirely by the local field equations in weak
field regions between the bodies.  This was originally pointed out by
Weyl and by Einstein and Grommer; see Thorne and Hartle
\cite{thornehartle} and references therein.
Each body is surrounded by a
vacuum, weak field
region called a ``buffer region''
\cite{thornehartle}, and the quantities entering into the equations of
motion are defined in terms of the behavior of the metric in those
buffer regions (see Fig.\ \ref{fig1}).
In particular, our multipole moments are defined in terms of the
behavior of the metric in the buffer regions.  Our definition of
multipole moments is thus more general than the
definition in terms of integrals over sources used by DSX.
However, our multipole moments do coincide
with those of DSX in the case of weakly self-gravitating bodies.

Our derivation consists of a number of steps:  (i) The
definition of each body's current and mass multipole moments and
center-of-mass worldline in terms of the behavior of the metric in
that bodies' buffer region.  (ii) The definition for each
body of a set of gravitoelectric and gravitomagnetic tidal moments
that act on that body, again
in terms of the behavior of the metric in that bodies' buffer region.
For the special case of weakly self-gravitating
bodies, our definitions of these multipole and tidal
moments agree with definitions given previously by DSX.  (iii) The
derivation of a formula, for any given body, of the second time
derivative of its mass dipole moment in terms of its other multipole
and tidal moments and their time derivatives.  This formula was
obtained previously by DSX for weakly self-gravitating bodies.  (iv)
A derivation of the relation between the tidal moments acting on each
body and the multipole moments and center-of-mass worldlines of all
the other bodies.  A formalism to compute this relation was developed
by DSX; we simplify their formalism and compute the relation
explicitly.  (v) The deduction from the previous steps of the explicit
translational equations of motion, whose form has not been previously
derived.

\subsection{Results for equations of motion}\label{sec:resultseom}

We next describe our results for the equations of motion.  We label
each body by an integer $A$, with $1 \le A \le N$.  We use a harmonic
coordinate system $(t,x^i)$ that covers all of spacetime except for
the strong field regions near each body.
The position of body $A$ in this coordinate system is parameterized by
a function $x^i = \cmz^A_i(t)$ called the ``center of mass
worldline''.  This function is defined precisely in Sec.\ \ref{sec:configvars}
below.  It does not correspond to an actual worldline in spacetime;
rather it parameterizes the location of the local asymptotic rest
frame (see below) attached to the $A$th body.  That is, it is encoded
in the behavior of the metric in a weak field region surrounding body
$A$ in the same way that the actual center-of-mass worldline of a
weakly self-gravitating body would be encoded.

Associated with each body $A$ is a coordinate system $(s_A, y_A^i)$
which is defined only in that body's buffer region, and which is
adapted to the body in the sense that it
minimizes the coordinate effects of the external gravitational field
due to the other bodies as much as possible.  This
coordinate system is discussed in detail in Secs.\
\ref{sec:bodyadapted} and \ref{sec:assumptions} below.
We will call the corresponding
reference frame the ``body frame'' or, following Thorne and Hartle
\cite{thornehartle}, the body's ``local asymptotic rest
frame''.
The details of the transformation between the body-adapted coordinates
$(s_A,y_A^i)$ and the global
coordinates $(t,x^i)$ are important for the purpose of deriving the
translational equations of motion.  However, for the purpose of using
the equations of motion, one only needs to know the following.  First,
the time coordinate $s_A$ is a ``proper time'' associated with body
$A$.  It corresponds to the proper time that would be measured by an
observer in the local asymptotic rest frame of body $A$.  In
that local asymptotic rest frame it is related to the global frame
time coordinate $t$ by
\begin{equation}
s_A = s_A(t),
\end{equation}
where the function $s_A(t)$ is determined by a differential equation
[Eqs.\ (\ref{eq:eomtime}) and (\ref{eq:eomtimeexplicit}) below].
Second, the leading order relation between the spatial
coordinates $y_A^i$ and $x^i$ is just a translation together with a
time-dependent rotation [cf.\ Eqs.\
(\ref{coordinatetransformationII}), (\ref{eq:com3}) and (\ref{eq:rotUdef}) below]:
\begin{equation}\label{eq:dragging}
x^i = \cmz^A_i(t) + U^{A\,j}_i(t) y_A^j.
\end{equation}
The rotation matrix $U^{A\,j}_i(t)$ describes dragging of inertial
frames\footnote{The time derivative of this rotation matrix
is actually of post-1-Newtonian order, so to Newtonian order this is a
constant matrix.  In the body of the paper we assumed that this
constant, Newtonian order rotation matrix is the unit matrix.  The
description of our results given here allows this constant
matrix to be arbitrary; this slight generalization would be useful to
describe systems that evolve for a time long enough that the
accumulated rotation due to frame dragging is of order
unity.}; a differential equation for its evolution is given below.
The body-adapted coordinates $(s_A,y_A^i)$ rotate with respect to
distant stars, while the global coordinates $(t,x^i)$ do not.

Each body $A$ has associated with it a unique set of mass multipole moments
\begin{equation}
M^A_L(s_A) = M^A_{a_1 \ldots a_l}(s_A),
\end{equation}
for $l=0,1,2 \ldots$ which are symmetric, tracefree, spatial tensors
with $l$ indices, of which mass dipole $M^A_i(s_A)$ vanishes identically.
It also has a unique set of current multipole moments
\begin{equation}
S^A_L(s_A) = S^A_{a_1 \ldots a_l}(s_A),
\end{equation}
for $l=1,2, \ldots$.  These quantities are functions of the body's
proper time $s_A$.  In the absence of interactions with other bodies
the mass monopole $M^A$
%, the time derivative of the mass dipole ${\dot M}^A_i$,
and the spin $S^A_i$ are conserved.

We will obtain below coupled equations of motion for the center of
mass worldlines of all the bodies.  Appearing in these equations as
unknowns will be the mass multipole moments
$M^A_L(s_A)$ for $l\ge2$, and the current multipole moments
$S^A_L(s_A)$ for $l\ge2$.  In order to obtain a closed system of
equations, one would need to supplement the equations of this paper
with equations determining the evolution of these multipole moments.
We discuss further below various circumstances and approximations in
which the evolution of the multipoles can be computed.

Next, we define the moments $\bodyM^A_L(t)$ and $\bodyS^A_L(t)$ to be
the body's mass and current multipole moments, transformed to the
non-rotating frame, and expressed as functions of the global time $t$.
These moments are given by the equations [cf.\ Eqs.\
(\ref{eq:calMdef}) and (\ref{eq:calSdef}) below]
\begin{subequations}
\begin{eqnarray}\nn
\bodyM^A_{a_1 \ldots a_l}(t) &=& U_{a_1}^{A\,a'_1}(t) \ldots U_{a_l}^{A\,a'_l}(t)
M^A_{a'_1 \ldots a'_l}[s_A(t)], \,\,\,\,\,\,\,
\\ \label{eq:calMdef0} \\ \mbox{}
\bodyS^A_{a_1 \ldots a_l}(t) &=& U_{a_1}^{A\,a'_1}(t) \ldots U_{a_l}^{A\,a'_l}(t)
S^A_{a'_1 \ldots a'_l}[s_A(t)].\,\,\,\,\,\,\,
\label{eq:calSdef0}
\end{eqnarray}
\end{subequations}

We can now write down the schematic form of the equations of motion.
They can be written as
\begin{subequations}
\begin{eqnarray}
\cmddotz^A_i(t) &=& {\cal F}^A_i[ \cmz^B_i, \cmdotz^B_i, \bodyM^B_L, {\dot
\bodyM}^B_L, {\ddot \bodyM}^B_L, \bodyS^B_L, {\dot \bodyS}^B_L ] \nn
\\
\label{eq:eomschematic}
\\
{\dot \bodyM}^A(t) &=& {\cal F}^A[ \cmz^B_i, \cmdotz^B_i, \bodyM^B_L, {\dot
\bodyM}^B_L]
\label{eq:eomenergy}
\\
{\dot \bodyS}^A_i(t) &=& {\bar {\cal F}}^A_i[ \cmz^B_i, \bodyM^B_L ]
\label{eq:eomspin}
\\
\frac{ds_A}{dt} &=& {\bar {\cal F}}^A[ \cmz^B_i,
  \cmdotz^B_i,\bodyM^B_L],\label{eq:eomtime}
\end{eqnarray}
and
\begin{eqnarray}
\left[{\dot {\bf U}}^A \cdot \left({\bf U}^A\right)^{-1}\right]_{ij} &=&
{\cal F}^A_{ij}[\cmz^B_i, \cmdotz^B_i, \bodyM^B_L, {\dot
\bodyM}^B_L, \bodyS^B_L ]. \nn \\
\label{eq:Uevol}
\end{eqnarray}
\end{subequations}
Here ${\cal F}^A_i$, ${\cal F}^A$, ${\bar {\cal F}}^A_i$,
${\bar {\cal F}}^A$ and ${\cal F}_{ij}^A$ are
functions of their argument whose specific forms are discussed below.
In these equations the dependencies on the time derivatives ${\dot
\bodyM}^B_L$, ${\ddot \bodyM}^B_L$ and ${\dot \bodyS}^B_L$ only occur
for $l\ge 2$.  Also the mass dipoles $\bodyM^B_i$ vanish identically.
Therefore, if we assume
that the moments $M^A_L(s_A)$ and $S^A_L(s_A)$ are known for $l \ge
2$, Eqs.\ (\ref{eq:calMdef0}) -- (\ref{eq:Uevol}) form a closed set of
evolution equations which can be solved to obtain the center of mass
worldlines as well as the rotation matrices ${\bf U}^A$ and time functions
$s_A(t)$.

We remark that the three equations (\ref{eq:eomschematic}) --
(\ref{eq:eomspin}) by themselves form a closed set of evolution
equations for the variables $\cmz^A_i(t), \bodyM^A(t)$, and
$\bodyS^A_i(t)$, if we assume that the moments $\bodyM^A_L(t)$ and
$\bodyS^A_L(t)$ are known for $l \ge 2$.  However,
approximation schemes for computing the multipole moments for $l\ge2$ usually
yield the variables $M^A_L(s_A)$, $S^A_L(s_A)$ rather than the
variables $\bodyM^A_L(t)$,
$\bodyS^A_L(t)$.  This is because the moments $M^A_L(s_A)$ and
$S^A_L(s_A)$ are the physical moments that would be measured by an
observer in the the local asymptotic rest frame of body $A$.
In such cases we must enlarge the set of variables
$\cmz^A_i(t), \bodyM^A(t), \bodyS^A_i(t)$
to include the rotation matrices ${\bf U}^A(t)$ and time functions
$s_A(t)$ in order to obtain a closed set of equations.

We also note that it is formally consistent to post-1-Newtonian
accuracy to replace Eq.\ (\ref{eq:calSdef0}) with the simpler relation
$\bodyS^A_L(t) = S^A_L(t)$.
Nevertheless it can be useful in some circumstances to use
the more accurate relation (\ref{eq:calSdef0}), for example for
systems which evolve for sufficiently long times that the rotation
matrices $U_a^{A\,a'}$ become significantly different from unity.

We now discuss the functions ${\cal F}^A_i$, ${\cal F}^A$, ${\bar
{\cal F}}^A_i$, ${\bar {\cal F}}^A$ and ${\cal F}_{ij}^A$ that
appear in Eqs.\ (\ref{eq:eomschematic}) --
(\ref{eq:Uevol}).  The functional form of ${\cal F}^A_i$ is one
of our key results.  It is given by Eq.\ (\ref{fulleom}) below,
with the coefficients modified according to the substitutions given in
Eq.\ (\ref{eq:monopolesub}) and in Appendix
\ref{sec:newcoeffs}.  These modified coefficients are obtained by
combining the results of this paper with those of
the second paper in this series \cite{Racine}, which we will call
paper II.
The functions ${\cal F}^A$ and ${\bar {\cal F}}^A_i$ are standard
functions that can be derived from Newtonian stress-energy
conservation for weakly
self-gravitating bodies, and their explicit functional forms are
respectively given in section IV of paper II \cite{Racine} and in
Eq.\ (\ref{spinev}) below.  The validity of Eqs.\
(\ref{eq:eomenergy}) and (\ref{eq:eomspin}) for strongly
self-gravitating bodies is derived in paper II \cite{Racine}.
The function ${\bar {\cal F}}^A$ is given by
[cf.\ Eqs.\ (\ref{coordinatetransformationII}),
(\ref{eq:alphacans}), (\ref{globaltidalnG}), and (\ref{eq:calTKdef0}) below]
\begin{eqnarray}
{\bar {\cal F}}^A &=& 1 - \frac{1}{2} \cmdotz^A_i \cmdotz^A_i \nn \\
\mbox{} && - \sum_{B\ne A}
\sum_{k=0}^\infty \frac{(2k-1)!!}{k!} \frac{ \bodyM^B_K}{r_{BA}^{k+1}}
n^{BA}_K. \label{eq:eomtimeexplicit}
\end{eqnarray}
Here $K$ is the multi-index $b_1 \ldots b_k$, $r_{BA} = | \cmbmz^B -
\cmbmz^A|$, ${\bm n}^{BA} = (\cmbmz^B - \cmbmz^A)/ r_{BA}$, and
$n^{BA}_K = n^{BA}_{b_1} \ldots n^{BA}_{b_k}$.
Finally, the function ${\cal F}^A_{ij}$ is given by Eq.\ (\ref{eq:Uevol1})
below.

We next discuss various approximation schemes in which the equations of
motion (\ref{eq:eomschematic}) -- (\ref{eq:Uevol})
can be supplemented by methods for obtaining the evolution of
the mass and current multipole moments $\bodyM^A_L(t)$ and
$\bodyS^A_L(t)$ for $l\ge2$ in order to obtain a complete, closed set
of equations.  Some examples of such approximations are as follows:

\begin{itemize}

\item The simplest case is when the effect of all the $l\ge2$
multipoles is negligible, and one can set $\bodyM^A_L = \bodyS^A_L
=0$ for all $l \ge 2$.  This yields the monopole-spin truncated
equations of motion discussed in Refs.\ \cite{thornehartle,dsxII}.

\item Another simple case is when the evolution of the multipoles of
each body is dominated by dynamics internal to that body, and is
negligibly influenced by the tidal fields of the other bodies.
In this case, one can solve for the evolution of the multipoles
$M^A_L(s_A)$, $S^A_L(s_A)$ of each body separately, and then insert
those multipoles into the equations of motion
(\ref{eq:calMdef0}) -- (\ref{eq:Uevol}).  This application will be
valid only if the timescale over which the bodies' multipoles evolve
is sufficiently long \cite{thorne}; see Sec.\
\ref{sec:domain_of_validity} below for further discussion of this point.

\item Another useful case to consider is that of rigid bodies.
As noted by Thorne and G\"ursel \cite{thornegursel}, in general
relativity a body's rotation can be rigid only if its angular velocity
(with respect to its local asymptotic rest frame) is constant.  If
the angular velocity is changing, for example due to precession, then
the body cannot be rigid due to Lorentz contraction effects.
However, to linear order in the body's angular velocity the motion is
rigid \cite{thornegursel}. The analysis of Thorne and
G\"ursel can be adapted to the present context, if the bodies'
rotations are slow enough that they can be idealized as rigid.  In
this case, the time dependence of
the mass multipole moments $M^A_L(s_A)$
for $l\ge2$ can be parameterized in terms of a time-dependent rotation matrix
${\cal U}_a^{A\, {\bar a}}(s_A)$:
\begin{eqnarray}
\;\;\;\;\;\;\;\;\; M^A_{a_1 \ldots a_l}(s_A) &=& {\cal U}_{a_1}^{A\,{\bar a}_1}(s_A) \ldots
{\cal U}_{a_l}^{A\,{\bar a}_l}(s_A)
M^A_{{\bar a}_1 \ldots {\bar a}_l}.
\nn
%\\ \mbox{}
%S^A_{a_1 \ldots a_l}(s_A) &=& {\cal U}_{a_1}^{A\,{\bar a}_1}(s_A) \ldots
%{\cal U}_{a_l}^{A\,{\bar a}_l}(s_A)
%S^A_{{\bar a}_1 \ldots {\bar a}_l}.\nn
\end{eqnarray}
Here the moments $M^A_{{\bar a}_1 \ldots {\bar a}_l}$
are constant; these are the
moments in the co-rotating frame which rotates with the body.
We define the angular velocity $\Omega^A_a(s_A)$ in the usual way
as ${\dot {\cal U}}_a^{A\,{\bar b}} {\cal U}_b^{A\,{\bar b}} =
\epsilon_{acb} \Omega^A_c$.
Then, the co-rotating frame spin
$S^A_{\bar a} = {\cal U}_a^{A\,{\bar a}} S^A_a$
is related to the co-rotating frame angular velocity
$\Omega^A_{\bar a} = {\cal U}_a^{A\,{\bar a}} \Omega^A_a$
via \cite{thornegursel,rigidbody}
$$
S^A_{\bar a}(s_A) = I^A_{{\bar a}{\bar b}} \Omega^A_{\bar b}(s_A),
$$
where $I^A_{{\bar a}{\bar b}}$ is the (constant) moment of inertia
tensor\footnote{Thorne and G\"ursel \protect{\cite{thornegursel}} have
shown that for fully relativistic stars, as for Newtonian stars, the
moment of inertia tensor is constant, independent of the angular
velocity, up to linear order in the angular velocity.}.
Similarly the higher order current multipole moments are given
by
$$
S^A_{{\bar a}_1 \ldots {\bar a}_l}(s_A) = I^A_{{\bar a}_1 \ldots {\bar
    a}_l{\bar b}} \Omega^A_{\bar b}(s_A),
$$
where $I^A_{{\bar a}_1 \ldots {\bar a}_l{\bar b}}$ is a higher order
generalization of the moment of inertia tensor \cite{thornegursel}.
Combining these relations with the equations of motion
(\ref{eq:calMdef0}) -- (\ref{eq:Uevol}) yields a closed system of
equations which can be solved for the center of mass worldlines, the
rotation ${\cal U}_a^{A\,{\bar a}}(s_A)$ of each body with respect to
its local asymptotic rest frame $(s_A,y_A^i)$, and the rotation ${\bf
U}^A(t)$ of that local asymptotic rest frame with respect to distant
stars.  These equations describe torqued precession of relativistic
objects, generalizing the free precession equations of
Thorne and G\"ursel \cite{thornegursel}.

\item For weakly self-gravitating bodies one can use the formalism
developed by DSX \cite{dsxI,dsxII,dsxIII} to obtain a post-1-Newtonian
description of the internal dynamics of each body, for example by
using post-1-Newtonian stellar perturbation theory.  Coupling such a
description to the equations of motion yields a closed system of
equations.

\item Lastly, for fully relativistic, spherical stars, one can
compute the leading order effects of tidal interactions by combining
the results of this paper with linear relativistic stellar
perturbation theory using matched asymptotic expansions; see, for
example, Refs.\ \cite{death,damour2,thorne1,flanagan}.  For example, if one
is interested only
in the mass quadrupoles, and one restricts attention to the dominant,
fundamental $l=2$ modes with no radial nodes, then one has a relation
of the form
$$
M^A_{ij}(s_A) = \int ds_A^\prime K(s_A - s_A^\prime) G^A_{ij}(s_A^\prime).
$$
Here $K(s_A - s_A^\prime)$ is a Green's function which can be computed
from stellar
perturbation theory, and $G^A_{ij}$ is the body-frame gravitoelectric
tidal moment that acts on body $A$, which is defined in Sec.\
\ref{sec:body_frame_moments}
below and which can be computed in terms of the worldlines and
multipole moments of the other bodies.
Combining this relation with the equations of motion
(\ref{eq:eomschematic}) -- (\ref{eq:Uevol}) again yields a closed
system of equations, if one neglects the mass multipoles for $l \ge 3$
and the current multipoles.

\end{itemize}

\subsection{Domain of validity of our results}
\label{sec:domain_of_validity}

As mentioned above, the key assumption which we make in
deriving our results is that the post-1-Newtonian vacuum field equations are
satisfied in a weak field region between the bodies; see Sec.\
\ref{sec:assumptions} below for more details.
We are unable to give a derivation of this assumption from first principles.
However, in this subsection we discuss various physical effects which can
cause our assumption to break down, and we make
estimates of the sizes of these effects.  We believe that the
assumption should be generally valid aside from the effects discussed
in this subsection.

The first type of correction are post-2-Newtonian corrections to the
metric in the weak field, vacuum region between the bodies.  These
will give rise to fractional corrections of order $M^2/D^2$, where $M$
is a typical mass and $D$ a typical separation of the bodies, cf. the
third term on the first line of Eq.\ (\ref{eq:eom_schematic}).
We can estimate as follows when these corrections will be larger than
the tidal coupling terms which we retain.
The estimate given in the last term of Eq.\ (\ref{eq:eom_schematic})
can be refined by multiplying it by
the dimensionless measure $\varepsilon_l = \bodyM_L / (M R^l)$ of the
$l$th mass multipole.  Demanding that this quantity be larger than the
post-2-Newtonian, point particle term in Eq.\ (\ref{eq:eom_schematic})
yields the criterion
\begin{equation}
D \alt \varepsilon_l^{\frac{1}{l-1}} \left( \frac{R}{M} \right)^{\frac{1}{l-1}} \,
R.
\end{equation}
Thus, the post-2-Newtonian terms will always dominate at sufficiently
large separations $D$, but for any given $l\ge2$ there will be a range of
values of $D$ for which the post-1-Newtonian tidal terms dominate, as
long as $\varepsilon_l \agt M/R$.  In particular this will be true for
generic (non-symmetric) bodies for which $\varepsilon_l \sim 1$.
Similar estimates apply to current multipole couplings for $l\ge2$.
Thus, there is a nonempty regime in which the
post-1-Newtonian tidal couplings computed here dominate over post-2-Newtonian,
point-particle effects.

Note, however, that this range of values of $D$ gets smaller as the
strength of internal gravity $\sim M/R$ increases.  In the limit of $M
\sim R$ of a black hole, the post-2-Newtonian terms are always
comparable to or larger than the post-1-Newtonian tidal terms.
Therefore, our results cannot be applied consistently to black
holes without including post-2-Newtonian and higher terms in the equations of
motion.

Another type of correction, which is also formally of post-2-Newtonian
order, is that due to the time dependence of the
mass and current multipole moments of the individual bodies
\cite{thorne}.  The post-1-Newtonian solutions [Eqs.\ (\ref{basicmodelA}) --
(\ref{reducedzeta}) below] do not exhibit the correct retarded
dependence on these moments, that is, they are functions of $M_L(t)$
and $S_L(t)$ rather than $M_L(t-r)$ and $S_L(t-r)$.
If these moments vary on a timescale $\tau$, then the
corresponding fractional corrections to the mass moments $M^A_L$
scale as $D^4/\tau^4$, and the fractional corrections to the current
moments $S^A_L$ scale as $D^2/\tau^2$.  Demanding that these
corrections
be smaller than the post-1-Newtonian accuracies of these quantities
($\sim M/D$ and $\sim 1$ respectively) yields the criterion
$\tau \gg D$ for the current moments, and the more stringent criterion
\begin{equation}
\tau \gg D \left( \frac{D}{M} \right)^{1/4}
\label{eq:criterion}
\end{equation}
for the mass moments.
Fractional corrections to the post-1-Newtonian tidal interactions will
be of order unity for $\tau \sim D (D/M)^{1/4}$.  The criterion
$\tau \gg D$ essentially says that all of the bodies lie in
the near zone of the gravitational radiation produced by any one body,
and not in the wave zone \cite{thorne}, and the criterion
(\ref{eq:criterion}) is a somewhat stronger requirement than this.

To illustrate the criterion (\ref{eq:criterion}) it is useful to
consider some examples.  First,
if the time evolution of a body's moments is driven by
tidal interactions with other bodies, then the timescale is of order
$\tau \sim D (D/M)^{1/2}$, and the criterion is
satisfied.  Second, suppose that we have a 3-body system consisting of
two black holes in a tight binary, together with a third body in orbit
around the binary.  We can model such a system as a 2 body system
using the formalism of this paper, treating the black hole binary as a
single body\footnote{Although our formalism cannot be consistently applied to
individual black holes unless supplemented with post-2-Newtonian and
higher order point particle terms, our formalism can
be applied to black hole binaries treated as a single body.  This is
because binaries are less compact than black holes.}
whose mass and current multipole moments are
evolving with time due to internal dynamics.  Then, early in the gravitational-wave
driven inspiral of the black hole binary, the criterion
(\ref{eq:criterion}) will be satisfied and our results for the
equation of motion will be valid.  As the black hole binary gets
tighter however, eventually the orbital period will become shorter
than (\ref{eq:criterion}) and our results will no longer be
applicable.

This second example illustrates that the post-1-Newtonian
approximation can sometimes completely break down, even in the
supposed weak-field region between the bodies.  During the final
coalescence of the black holes the gravitational radiation metric
perturbation will become temporarily as large as the Newtonian
potential in the region between the binary and the loosely bound
companion.   Our results are not applicable to such systems, in which
one of the bodies emits a strong burst of gravitational radiation.
Further work is required to deduce the form of the translational
equations of motion in this type of situation.

There are two other assumptions made in our derivation
which slightly restrict the domain of validity of our results.
First, we assume that a coordinate system which covers the weak field
region between the bodies can be smoothly extended to cover the
bodies' interiors (see Sec.\ \ref{sec:pnsolns} below).  This
assumption essentially restricts the spatial topology of the bodies'
interiors, and excludes objects like eternal black holes, wormhole
mouths and naked singularities.  It does not exclude realistic,
astrophysical black holes for the reason explained in Sec.\
\ref{sec:pnsolns} \cite{thornegw}.
Second, in order for a body's multipole moments to be definable, it is
necessary that there exist two concentric coordinate spheres
surrounding the object, such that the region between the two spheres is
vacuum, in a particular coordinate system centered on the body (see
Sec.\ \ref{sec:pnsolns} below).  This assumption might break down when
two bodies get within one or two radii of one another, slightly before
they actually touch.

\subsection{Organization of this paper}

The structure of the paper is as follows.  In Sec. \ref{sec:cfe} we
introduce our notations for the post-1-Newtonian continuum field
equations, and following DSX we define a class of gauges (conformally
Cartesian
harmonic gauges) that we use throughout the paper.  Section
\ref{sec:gaugefreedom} presents a simplified version of the theory of
post-1-Newtonian reference systems of Refs.\
\cite{dsxI,dsxII,dsxIII,Kopeikin0,Brumberg1989,Brumberg1991,Klioner}.
The key result of this section is the explicit parameterization
(\ref{eq:generalcoordtransform}) of the residual gauge freedom within
conformally Cartesian harmonic gauge in terms of a number of freely
specifiable functions of time and one harmonic function of time and space.

Section \ref{singlesys} is devoted to the definitions
of the mass multipole moments $M_L(t)$, current multipole moments
$S_L(t)$, gravitoelectric tidal moments $G_L(t)$ and gravitomagnetic
tidal moments $H_L(t)$
associated with a given object and a given conformally Cartesian, harmonic coordinate system.  These definitions are given in Sec.\
\ref{sec:pnsolns} in terms of the general solution
(\ref{basicmodelA}) -- (\ref{reducedzeta}) of the post-1-Newtonian
field equations in a vacuum region between two concentric coordinate
spheres that surround the object (the object's ``buffer region'').  Section
\ref{sec:momentsgaugetransformation} analyzes how
all of these moments transform under the class of allowed gauge
transformations discussed in Sec.\ \ref{sec:gaugefreedom}.
In Sec.\ \ref{sec:bodyadapted} we describe gauge specializations that
fix the gauge freedom completely and accordingly determine the
multipole and tidal moments uniquely.  We call the resulting
coordinate system a body-adapted coordinate system.  Section
\ref{sec:moments1} gives a definition of multipole moments and tidal
moments associated with a given object, a given worldline and a given
coordinate system.  These moments arise only in intermediate steps
in the derivations of this paper and not in our final results.
Finally, in Sec.\ \ref{sec:comparison} we compare the moment
definitions used here with other definitions in the literature.

Section \ref{eom} derives the law of motion for a single body,
that is, the relation between the second time derivative of its mass
dipole moment and its other multipole and tidal moments and
their time derivatives.  The assumptions and result are described in
Sec.\ \ref{sec:eomoverview}.  A general description of the
surface-integral method of derivation which we use is given in Sec.\
\ref{sec:method_of_derivation}.  In Sec.\ \ref{sec:post2} we give
some of the post-2-Newtonian vacuum field equations which are needed
for the derivation.  Section \ref{sec:nod} derives the single body
law of motion to Newtonian order, as a warm-up exercise.
Finally, the post-1-Newtonian derivation is given in Sec.\
\ref{sec:pnderiv}.  This derivation uses an idea due to Thorne and
Hartle \cite{thornehartle} to deduce the value of a complicated
surface integral from previous weak-field computations of DSX
\cite{dsxI,dsxII}.

Section \ref{manybody} lays the foundations for treating a system of
$N$ interacting,
finite-sized bodies.   Our assumptions are described and discussed in
Sec.\ \ref{sec:assumptions}.  In Sec.\ \ref{sec:body_frame_moments} we
define, for each body, a set of body-frame multipole and tidal moments
associated with that bodies' adapted coordinate system.  These are the
moments that would be measured by an observer in that bodies
local asymptotic rest frame.  Section \ref{sec:configvars} defines the
configuration variables that specify the location, orientation
etc.\ within the global coordinate system of the local asymptotic rest
frame which is attached to that body.  These variables include the
center of mass worldline and also the time functions and rotation
matrices discussed in Sec.\ \ref{sec:resultseom} above.
In Sec.\ \ref{sec:globalcoords} we define for each body multipole and
tidal moments associated with the global coordinate system.  These
quantities appear only in intermediate steps in our computations and
not in our final results.  The relation between the global-frame
multipole and tidal moments and the body-frame multipole and tidal
moments is computed in Sec.\ \ref{sec:cbf}.  Section \ref{sec:mdefs1}
defines the modified versions $\bodyM_L$ and $\bodyS_L$ of the
body-frame multipole moments, discussed in Sec.\ \ref{sec:resultseom}
above, which are defined with respect to a frame that is non-rotating with
respect to distant stars, and which are expressed as functions of the
global time coordinate.  These are the moments that appear in the
final equations of motion.

Finally, Sec.\ \ref{explicit} gives the derivation of the complete,
explicit translational equations of motion for the $N$ body system.

\subsection{Notations and conventions}
\label{sec:notation}

Throughout this paper we use geometric units in which $G=c=1$.  We use
the sign conventions of Misner, Thorne and Wheeler \cite{mtw}; in
particular we use the metric signature $(-,+,+,+)$.
Greek indices ($\mu$,$\nu$ etc...) run from 0 to 3 and denote
spacetime indices, while Roman indices ($a$, $b$, $i$,$j$, etc...) run
from 1 to 3
and denote spatial indices.
The spacetime coordinates will generically be denoted by $(x^0,x^i) =
(t,x^i)$.  Spatial indices are raised and lowered
using $\delta_{ij}$, and repeated spatial indices are contracted
regardless of whether they are covariant or contravariant indices.
We denote by $n^i$ the unit vector $x^i/r$, where $r = |{\bm{x}}| =
\sqrt{\delta_{ij} x^i x^j}$.

When dealing with sequences of spatial indices, we use the
multi-index notation introduced by Thorne \cite{thorne} as modified
slightly by Damour, Soffel and Xu \cite{dsxI}.
We use $L$ to denote the sequence of $l$ indices $a_1 a_2 \ldots a_l$,
so that for any $l$-index tensor\footnote{
Here by ``tensor'' we mean an object which transforms as a tensor
under the symmetry
group SO(3) of the zeroth order spatial metric $\delta_{ij}$, not a
spacetime tensor.} $T$ we have
\begin{equation}
T_L \equiv T_{a_1a_2... \,a_l} .
\end{equation}
If $l=0$, it is
understood that $T_L$ is a scalar. If $l<0$ then $T_L\equiv
0$.  We define $L-1$ to be the sequence of $l-1$ indices $a_1 a_2
\ldots a_{l-1}$, so that
\begin{equation}
T_{L-1} \equiv T_{a_1a_2... \,a_{l-1}}.
\end{equation}
If $l=0$, then by convention $T_{L-1} \equiv 0$.
We also define $N$ to be the sequence of $n$ indices $a_1 a_2
\ldots a_n$, and $L-N$ to be the sequence of $l-n$ indices $a_{n+1}
a_{n+2} \ldots a_l$, so that we can write a relation like
\begin{equation}
G_{a_1 ... a_l} = S_{a_1... \, a_n} T_{a_{n+1} ... \,a_l}
\end{equation}
as $G_L = S_N T_{L-N}$, for any tensors $G$, $S$ and $T$.
We define $K$, $P$ and $Q$ to be the sequences of
spatial indices $b_1 b_2 \ldots b_k$, $c_1 c_2 \ldots c_p$, and $d_1
d_2 \ldots d_q$, respectively.
Repeated multi-indices are subject to the Einstein summation
convention, as in $S_L T_L$.  We also use the notations
\begin{equation}
x^L \equiv x^{a_1a_2...\,a_l} \equiv x^{a_1}x^{a_2}...\,x^{a_l}
\end{equation}
and
\begin{equation}
\partial_L \equiv \partial_{a_1a_2...\,a_l} \equiv
\partial_{a_1}\partial_{a_2} ...\,\partial_{a_l}.
\end{equation}

We use angular brackets to denote the operation of taking the symmetric
trace-free (STF) part of a tensor. Thus for any tensor $T_L$, we define
\begin{equation}
T_{<L>} \equiv \text{STF}_L(T_L).
\end{equation}
where $\text{STF}_L$ means taking the symmetric trace-free projection on the
indices $L$. For example, if $l=2$, we have
\begin{eqnarray}\nn
T_{<L>} &=& T_{<a_1a_2>} \\\nn
&=& \frac{1}{2}\left(T_{a_1a_2} + T_{a_2a_1}\right) - \frac{1}{3}\delta_{a_1a_2}T_{jj}.
\end{eqnarray}
See Appendix \ref{stf} for the general definition of
STF projection, and for a collection of useful
relations involving STF tensors.

Throughout this paper, symbols will generally denote functions (as is
common in mathematics) rather than physical quantities (as is common
in physics).  For example, in Sec.\ \ref{manybody} we define a
mass multipole moment $M^A_L(s_A)$ which is a function of a time
coordinate $s_A$.  In that section we also use a different time
coordinate $t$.  Then, the notation $M^A_L(t)$ will always mean $M_L(s_A)$
evaluated at $s_a=t$, rather than $M^A_L[s_a(t)]$.

Finally, for the aid of the reader an index of symbols is provided in
Table \ref{table:symbols}.

\bigskip

%%%%%%%%%%%%%%%%%%%%%%%%%%%%%%%%%%%%%%%%%%%%%%%%%%%%%%%%%%%%%%%%%%%%%%%%%%%%%%%%%%%%%%%%%%%%%%%%%%%%%%%%%%%%%%%%%%%%%%%%%%%%%%%%%%%%%%%%%%%%%%%%%%%%%%%%%%%%%%

\section{Post-1-Newtonian continuum field equations and gauge freedom}
\label{coords}

In this section we summarize the form of the
post-1-Newtonian field equations that we use, and analyze the residual
gauge freedom left after imposing our gauge conditions.  Our notation
follows closely that of Weinberg \cite{weinberg}, though we relate our
conventions to those of DSX \cite{dsxI,dsxII,dsxIII,dsxIV}.

\subsection{Metric expansion and field equations}
\label{sec:cfe}

In the post-1-Newtonian approximation, one considers a one-parameter
family of solutions to Einstein's equations of the form
\begin{eqnarray}\nn
ds^2 &=& - [1 + 2\varepsilon^2\Phi + 2\varepsilon^4(\Phi^2 + \psi) +
O(\varepsilon^6)](dt/\varepsilon)^2 \\\nn
& & + [2\varepsilon^3\zeta_i + O(\varepsilon^5)]dx^i(dt/\varepsilon)
\\\label{metricA}
& & + [\delta_{ij} + \varepsilon^2 h_{ij} + O(\varepsilon^4)]dx^idx^j.
\end{eqnarray}
Here the Newtonian potential $\Phi$, the post-Newtonian potential
$\psi$, the gravitomagnetic potential $\zeta_i$ and the spatial metric
perturbation $h_{ij}$ are functions of
the coordinates $x^0=t$ and $x^i$, but are independent of the parameter
$\varepsilon$.
The corresponding expansion of the stress-energy tensor is
\begin{equation}
T^{\mu\nu} =
\varepsilon^4\left[\,^{\scriptscriptstyle \text{n}}\!T^{\mu\nu} +
  \varepsilon^2\,^{\scriptscriptstyle \text{pn}}\!T^{\mu\nu} + O(\varepsilon^4)
  \right],
\label{eq:st}
\end{equation}
where $\,^{\scriptscriptstyle \text{n}}\!T^{\mu\nu}$ is the
Newtonian-order piece and
$\,^{\scriptscriptstyle \text{pn}}\!T^{\mu\nu}$ is the
post-1-Newtonian order piece.  The post-Newtonian expansion parameter
$\varepsilon$ used here is equivalent to the expansion parameter
$c^{-1}$ used in some other treatments (we use units in which $G=c=1$).

We note that many presentations of the post-1-Newtonian equations use a
time coordinate ${\hat t}$ that is related to our time coordinate $t$ by
\begin{equation}
{\hat t} = \frac{t}{\varepsilon}.
\end{equation}
In $({\hat t},x^i)$ coordinate systems the various powers of
$\varepsilon$ that appear in the expansions (\ref{metricA})
of the metric and (\ref{eq:st}) of the
stress energy tensor differ from those given here.
The $({\hat t},x^i)$ coordinate systems have the advantage that the
pointwise limit as $\varepsilon\to0$ of the metric exists and is a
flat, Minkowski metric, but have the disadvantage that the potentials
$\Phi$, $\psi$
and $\zeta_i$ become $\varepsilon$-dependent.  That
$\varepsilon$-dependence is
usually accounted for by inserting an extra factor of $\varepsilon$
whenever a time derivative of a potential is taken.
By contrast, in the coordinates used here,
the potentials are independent of $\varepsilon$ and
no such extra factors of $\varepsilon$ are needed.

Assuming the validity of Einstein's equations for the metric
(\ref{metricA}) and stress energy tensor (\ref{eq:st}) for all
$\varepsilon$ in some open interval $0 < \varepsilon < \varepsilon_0$
then implies that the metric $\delta_{ij} + \varepsilon^2 (h_{ij} + 2
\Phi \delta_{ij})$ is flat to $O(\varepsilon^2)$ \cite{dsxI}.
Therefore one can always choose coordinate systems\footnote{This
 statement is always true locally, and is true globally if the spatial
 domain of the coordinates is simply connected.  In this paper the
 spatial coordinate domains will always be simply connected.}
 in which $h_{ij} =
- 2 \Phi \delta_{ij}$, for which metric expansion simplifies to
\begin{eqnarray}\nn
ds^2 &=& - [1 + 2\varepsilon^2\Phi + 2\varepsilon^4(\Phi^2 + \psi) +
O(\varepsilon^6)](dt/\varepsilon)^2 \\\nn
& & + [2\varepsilon^3\zeta_i + O(\varepsilon^5)]dx^i(dt/\varepsilon)
\\\label{metric}
& & + [\delta_{ij} -2\varepsilon^2\Phi \delta_{ij} +
O(\varepsilon^4)]dx^idx^j.
\label{metric0}
\end{eqnarray}
Such gauges are called conformally Cartesian \cite{dsxI}; we will
restrict attention to conformally Cartesian gauges throughout this
paper.

We will also assume
%for most of this paper
the harmonic gauge condition
\begin{equation}\label{Harmonicgauge}
\partial_\mu(\sqrt{-g}g^{\mu\nu}) = 0,
\end{equation}
which at post-1-Newtonian order and for conformally Cartesian gauges
reduces to
\begin{equation}\label{harmonicgauge}
4\frac{\partial\Phi}{\partial t} + \frac{\partial \zeta^i}{\partial x^i} = 0.
\end{equation}
Harmonic coordinate systems are usually conformally Cartesian, but one
can have local coordinate patches which are harmonic but not
conformally Cartesian.

The metric (\ref{metric}), stress-energy tensor
(\ref{eq:st}) and gauge condition (\ref{harmonicgauge})

imply the standard\footnote{
If we add $\varepsilon^2$ times Eq.\ (\protect{\ref{WfieldeqB}}) to Eq.\
(\protect{\ref{WfieldeqA}}) we can combine these two field equations
into the single wave equation
$
- \partial^2 w / \partial {\hat t}^2 + \nabla^2 w =
-4\pi\sigma + O(\varepsilon^4),
$
where $w = - \Phi - \varepsilon^2 \psi$ and
\begin{equation}
\sigma \equiv \,^{\scriptscriptstyle \text{n}}\!{T^{00}} + \varepsilon^2
\,^{\scriptscriptstyle \text{pn}}\!{T^{00}} + \varepsilon^2
\,^{\scriptscriptstyle \text{n}}\!{T^{jj}}.
\end{equation}
This is the notation used by DSX \cite{dsxI}.
The potential $w$ satisfies a flat-space wave equation, with
hidden $\varepsilon$ dependencies that come from the definition of
$w$.
For our purposes it will be more useful to keep the
  expansion in
$\varepsilon$ fully explicit by using the two elliptic equations
(\ref{WfieldeqA}) and (\ref{WfieldeqB}) for
$\Phi$ and $\psi$ instead of the above single hyperbolic equation
for $w$.} harmonic-gauge\footnote{In the
``standard post-Newtonian gauge'' \cite{Will} the gauge condition
(\protect{\ref{harmonicgauge}}) is replaced by $3 {\dot \Phi} +
 \bfnabla \cdot \bfzeta =0$, and the field equations
 (\protect{\ref{WfieldeqB}}) and (\protect{\ref{WfieldeqC}}) are
 replaced by $\nabla^2 \psi = 4\pi\,^{\scriptscriptstyle \text{pn}}\!{T^{00}} +
4 \pi \,^{\scriptscriptstyle \text{n}}\!{T^{jj}}$ and
$\nabla^2\zeta^i = 16\pi\,^{\scriptscriptstyle \text{n}}\!{T^{0i}} + {\dot
  \Phi}_{,i}$.}
field equations \cite{weinberg}
\begin{subequations}
\begin{eqnarray}\label{WfieldeqA}
\nabla^2\Phi &=& 4\pi \,^{\scriptscriptstyle \text{n}}\!{T^{00}} ,\\\label{WfieldeqB}
\nabla^2 \psi &=& \frac{\partial^2\Phi}{\partial t^2} + 4\pi\left(\,^{\scriptscriptstyle \text{pn}}\!{T^{00}} + \,^{\scriptscriptstyle \text{n}}\!{T^{jj}}\right),\\\label{WfieldeqC}
\nabla^2\zeta^i &=& 16\pi\,^{\scriptscriptstyle \text{n}}\!{T^{0i}}.
\end{eqnarray}
\end{subequations}
For most of this paper we will be concerned with the vacuum versions of
the field equations (\ref{WfieldeqA}) -- (\ref{WfieldeqC}).  These are
\begin{subequations}
\begin{eqnarray}\label{fieldeqA}
\nabla^2\Phi &=& 0 ,\\\label{fieldeqB}
\nabla^2 \psi &=& \frac{\partial^2\Phi}{\partial t^2} ,\\\label{fieldeqC}
\nabla^2\zeta^i &=& 0.
\end{eqnarray}
\end{subequations}

For later use we note the expansions of the electric and magnetic
components
$
{\cal E}_{ij} \equiv  R_{0i0j}
$
and
$
{\cal B}_{ij} \equiv - \frac{1}{2} \epsilon_{ikl} R_{kl0j}
$
of the curvature tensor.  In vacuum regions these tensors are
symmetric and traceless and can be expanded as
\be
{\cal E}_{ij} = \,^{\scriptscriptstyle \text{n}}\!{\cal E}_{ij} +
\ve^2
\,^{\scriptscriptstyle \text{pn}}\!{\cal E}_{ij} + O(\ve^4)
\label{eq:calEdef}
\ee
and
\be
{\cal B}_{ij} = \ve^2 \,^{\scriptscriptstyle \text{pn}}\!{\cal B}_{ij}
+ O(\ve^4).
\label{eq:calBdef}
\ee
Here $\,^{\scriptscriptstyle \text{n}}\!{\cal E}_{ij} =\Phi_{,ij}$
is the Newtonian electric tidal tensor, and
\be
\,^{\scriptscriptstyle \text{pn}}\!
{\cal E}_{ij} = {\dot \zeta}_{<i,j>} + 3 \Phi_{<,i}
\Phi_{,j>} + 4 \Phi \Phi_{,<ij>} + \psi_{,<ij>}
\ee
is the post-Newtonian electric tidal tensor.  The angular brackets
denote an STF projection, cf. Sec.\
\ref{sec:notation} above.  The
post-Newtonian magnetic tidal tensor is
\be
\,^{\scriptscriptstyle \text{pn}}\!
{\cal B}_{ij} = - \frac{1}{2} B_{(i,j)},
\ee
where
\be
{\bf B} \equiv {\bfnabla} \times \bfzeta.
\label{eq:Bdef}
\ee
is the so-called gravitomagnetic field.

For later use, we also note the definition of the gravitoelectric
field ${\bf E}$ used by DSX \cite{dsxI}:
\begin{equation}
\bm{E} \equiv -\nabla(\Phi + \varepsilon^2\psi) -
\varepsilon^2\dot{\bm{\zeta}}.
\label{eq:Edef}
\end{equation}

\subsection{Parameterization of residual gauge freedom in conformally
Cartesian harmonic gauge}
\label{sec:gaugefreedom}

In many applications of post-1-Newtonian theory, the elliptic
harmonic-gauge field equations
(\ref{WfieldeqA}) -- (\ref{WfieldeqC}) are valid in all of space.  In
such cases one normally solves the field equations by
imposing the boundary condition that all the potentials go to zero at
spatial infinity.  This boundary condition determines a unique
solution to the field equations and a unique choice of gauge.

However, in this paper we will be dealing with situations where
the field equations (\ref{WfieldeqA}) -- (\ref{WfieldeqC}) do not have
unique solutions, due to residual gauge freedom.
There are two reasons for this residual gauge freedom \cite{dsxI}.
First, even in cases where the field equations are
valid in all of space, there is in fact no physical reason for
imposing that the potentials go to zero at spatial infinity.
Instead, the physical boundary condition to impose is that the
components (\ref{eq:calEdef}) and (\ref{eq:calBdef})
of the Riemann curvature tensor go to zero at spatial infinity.
One then finds that there is a large class of solutions of the
harmonic gauge field equations, when the sources are fixed.
This non-uniqueness is present even at Newtonian order; there are many
solutions to the Newtonian Poisson equation (\ref{WfieldeqA}) with the
boundary condition $\Phi_{,ij} \to 0$ as $r \to \infty$.
In the Newtonian context, the new solutions are simply the standard
solution transformed to accelerated frames.  Similarly, in the
post-Newtonian context, the additional solutions correspond to the
original solution transformed to reference frames that are
accelerated, rotating or otherwise modified with respect to the
standard reference frame.

In Sec.\ \ref{sec:bodyadapted} below, when considering a
system of $N$ interacting
bodies, we will need to construct a coordinate system adapted to each
body.  Exploiting the additional freedom of allowing accelerated,
rotating coordinate systems will be crucial
for our construction of those adapted coordinate systems.

The second reason for
residual gauge freedom
in the situations considered in this paper is that we will
be considering spacetimes containing
strong-field regions in which the post-Newtonian approximation is not
valid.  Therefore, we must analyze the field equations
(\ref{WfieldeqA}) -- (\ref{WfieldeqC}) on some spatial region ${\cal
D}$ which is not all of space.  In this case, the boundary conditions
imposed on the potentials $\Phi$, $\psi$ and $\zeta_i$ on the boundary
$\partial {\cal D}$ of ${\cal D}$ influence the solution.
Part of the information inherent in those boundary conditions is
gauge, and part of the information is physical.

DSX \cite{dsxI}, Kopeikin \cite{Kopeikin0} and Klioner and Voinov
\cite{Klioner}
have derived a complete parameterization of the residual gauge
freedom present in some region ${\cal D}$ of space after the
conformally Cartesian condition has been imposed; see Ref.\
\cite{skpw} for a review.
In this subsection we give a simplified, streamlined version of the DSX
analysis in our somewhat different notation.  We also specialize the
DSX analysis by imposing in addition the harmonic gauge condition.

Our starting assumptions are as follows.  We assume the existence of
two different coordinate systems $(t,x^i)$
and $({\bar t},{\bar x}^i)$ on ${\cal D}$, each of which is
conformally Cartesian and harmonic.
We also assume
that both coordinate systems are such that the metric admits an expansion of
the form (\ref{metric}).  In particular, there exist potentials
${\bar \Phi}({\bar t},{\bar x}^j)$, ${\bar \psi}({\bar t},{\bar x}^j)$
and ${\bar \zeta}_i({\bar t},{\bar x}^j)$ such that
\begin{eqnarray}\nn
ds^2 &=& - [1 + 2\varepsilon^2\bar{\Phi} + 2\varepsilon^4(\bar{\Phi}^2 + \bar{\psi}) + O(\varepsilon^6)](d\bar{t}/\varepsilon)^2 \\\nn
& & + [2\varepsilon^3\bar{\zeta}_i +
  O(\varepsilon^5)]d\bar{x}^i(d\bar{t}/\varepsilon)
\\\label{transformedmetric}
& & + [ \delta_{ij} -2\varepsilon^2\bar{\Phi} \delta_{ij}+ O(\varepsilon^4)]d\bar{x}^id\bar{x}^j.
\end{eqnarray}
In Appendix \ref{app:gaugefreedom} we show that the most general
relation between the two coordinate systems that is
compatible with these assumptions is\footnote{Up to constant
displacements in time and time-independent spatial
rotations.  We also assume that the coordinate transformation is
orientation-preserving and time-orientation preserving.}
\begin{eqnarray}\nn
x^i &=& \bar{x}^i + z^i(\bar{t}) + \varepsilon^2h^i(\bar{t},\bar{x}^j)
+ O(\varepsilon^4)
, \\\label{coordinatetransformation}
t &=& \bar{t} + \varepsilon^2\alpha(\bar{t},\bar{x}^j) +
\varepsilon^4\beta(\bar{t},\bar{x}^j) + O(\varepsilon^6),
\label{eq:generalcoordtransform}
\end{eqnarray}
where
\begin{subequations}
\begin{eqnarray}\label{conshgA}
\alpha(\bar{t},\bar{x}^j) &=& \alpha_{\text{c}}(\bar{t}) + \bar{x}_i\dot{z}^i(\bar{t}) ,\\\nn
h^i(\bar{t},\bar{x}^j) &=& h_{\text{c}}^i(\bar{t}) +
\epsilon^{ijk}\bar{x}_jR_k(\bar{t}) +
\frac{1}{2}\ddot{z}^i(\bar{t})\bar{x}_j\bar{x}^j
- \bar{x}^i\dot{\alpha}_{\text{c}}(\bar{t})
\\\nn
& & - \bar{x}^i {\bar x}_j \ddot{z}^j(\bar{t}) +
\frac{1}{2}\bar{x}^i\dot{z}_j(\bar{t})\dot{z}^j(\bar{t}) +
\frac{1}{2}\dot{z}^i(\bar{t})\dot{z}^j(\bar{t})\bar{x}_j,\\\label{conshgB}
\end{eqnarray}
\begin{eqnarray}\nn
\beta(\bar{t},\bar{x}^j) &=&
\bar{x}_j\bar{x}^j\left[\frac{1}{10}\dddot{z}^k({\bar t})\bar{x}_k +
  \frac{1}{6}\ddot{\alpha}_{\text{c}}({\bar t})\right] +
\beta_{\text{h}}(\bar{t},\bar{x}^j), \\\label{conshgC}
\end{eqnarray}
\end{subequations}
and where overdots mean derivatives with respect to the time argument.

We next discuss the meaning of the various freely specifiable functions
$\alpha_{\text{c}}(\bar{t})$, $z^i(\bar{t})$,
$h^i_{\text{c}}(\bar{t})$, $R_k(\bar{t})$, and
$\beta_{\text{h}}(\bar{t},\bar{x}^j)$ that appear
in the coordinate transformation (\ref{coordinatetransformation}) --
(\ref{conshgC}). At Newtonian order there appears the function of time
$\alpha_{\text{c}}(\bar{t})$, which governs the
normalization of the time coordinate at $O(\varepsilon^2)$.
In standard treatments of Newtonian gravity, this freedom is fixed by
the usual assumption $\Phi \to 0$ as $r \to \infty$.
In the present context, however, this coordinate freedom is not fixed.
There also appears the spatial
3-vector $z^i(\bar{t})$, which parameterizes the
translational motion of the new frame with respect to the original
frame, to Newtonian order.
At post-1-Newtonian order, one has the
3-vector $h_{\text{c}}^i(\bar{t})$, which parameterizes
the post-Newtonian translational motion of the new frame with respect
to the original frame.  There also appears
the spatial 3-vector $R_k(\bar{t})$, whose time derivative is an
angular velocity that parameterizes the slow, post-Newtonian rotation
of the coordinate system.  Finally, there is the function
$\beta_{\text{h}}(\bar{t},\bar{x}^j)$ which governs the normalization
of the time coordinate at $O(\varepsilon^4)$.  This function is not
completely freely specifiable but must be harmonic, i.e.
\begin{equation}
\nabla^2 \beta_{\text{h}} = 0.
\end{equation}

It is straightforward to compute how the potentials transform by
combining the metric expansions (\ref{metric}) and
(\ref{transformedmetric}) with the coordinate transformation given by
Eqs.\ (\ref{coordinatetransformation}) -- (\ref{conshgC})
and using the tensor transformation law for the components of the metric. The
results are
\begin{subequations}
\begin{eqnarray}\label{transfA}
\bar{\Phi}(\bar{t},\bar{x}^j) &=& \hat{\Phi}(\bar{t},\bar{x}^j) +
\dot{\alpha} - \frac{1}{2}\dot{z}_j\dot{z}^j ,\\\nn
\bar{\zeta}_i(\bar{t},\bar{x}^j) &=& \hat{\zeta}_i(\bar{t},\bar{x}^j)
- [4\hat{\Phi}(\bar{t},\bar{x}^j) + \dot{\alpha}]\dot{z}_i
\\\label{transfB}
& & + \frac{\partial h_i}{\partial \bar{t}} + \frac{\partial
  h^j}{\partial \bar{x}^i}\dot{z}_j - \frac{\partial\beta}{\partial
  \bar{x}^i},
\end{eqnarray}
and
\begin{eqnarray}\nn
\bar{\psi}(\bar{t},\bar{x}^j) &=& \hat{\psi}(\bar{t},\bar{x}^j)  -
\hat{\zeta}_i(\bar{t},\bar{x}^j)\dot{z}^i + \alpha\frac{\partial
  \hat{\Phi}(\bar{t},\bar{x}^j)}{\partial \bar{t}} \\\nn
& & + 2\hat{\Phi}(\bar{t},\bar{x}^j)\dot{z}_i\dot{z}^i -
(\alpha\dot{z}^i - h^i)\frac{\partial
  \hat{\Phi}(\bar{t},\bar{x}^j)}{\partial \bar{x}^i} \\\nn
& & - \frac{1}{4}\left(\dot{z}^i\dot{z}_i\right)^2 +
\dot{\alpha}\dot{z}^i\dot{z}_i - \dot{z}^i\frac{\partial h_i}{\partial
  \bar{t}} + \frac{\partial \beta}{\partial \bar{t}} -
\frac{1}{2}\dot{\alpha}^2. \\\label{transfC}
\end{eqnarray}
\end{subequations}
Here the function ${\hat \Phi}({\bar t},{\bar x}^j)$ is defined as
\begin{equation}
{\hat \Phi}({\bar t},{\bar x}^j) = \Phi[{\bar t},{\bar x}^j + z^j({\bar
t})],
\label{eq:hatnotation}
\end{equation}
where the right hand side is $\Phi$ evaluated at the point $x^i =
{\bar x}^i + z^i({\bar t})$, $t = {\bar t}$.  We define the functions
${\hat \psi}({\bar t},{\bar x}^j)$ and ${\hat \zeta}^i({\bar t},{\bar
  x}^j)$ similarly.

The transformation laws (\ref{transfA}) -- (\ref{transfC}) are
expressed in terms of the functions $\alpha({\bar t},{\bar x}^j)$,
$\beta({\bar t},{\bar x}^j)$ and $h^i({\bar t},{\bar x}^j)$ defined by
Eqs.\ (\ref{coordinatetransformation}) -- (\ref{conshgC}).
More explicit versions of the transformation laws, in which they are
expressed in terms of the freely specifiable functions
$\alpha_{\text{c}}(\bar{t})$, $z^i(\bar{t})$,
$h^i_{\text{c}}(\bar{t})$, $R_k(\bar{t})$ and
$\beta_{\text{h}}(\bar{t},\bar{x}^j)$, can be obtained by substituting
the definitions (\ref{coordinatetransformation}) -- (\ref{conshgC})
into Eqs.\ (\ref{transfA}) -- (\ref{transfC}).

There are two special subgroups of the group
(\ref{coordinatetransformation}) of transformations
that will be of importance later.
The first subgroup applies when the spatial domain ${\cal D}$ is all
of space, and when in addition one imposes the boundary condition that
all the potentials vanish at spatial infinity.  In this case it is
easy to show that only uniform relative motion of the two frames is
allowed, ${\ddot z}^i = {\ddot h}^i_{\text{c}}=0$, that $\beta_\text{h}=0$,
$\alpha_{\text{c}}={\dot {\bf z}}^2/2$, and that $R^k$ is
constant.
These well-known ``post-Galilean'' transformations are discussed in,
for example, Sec.\ 39.9 of Ref.\ \cite{mtw}.

The second important subgroup is the subgroup parameterized by the
harmonic function $\beta_{\text{h}}$, for which $\alpha_{\text{c}} = z^i
= h_{\text{c}}^i = R^k =0$.  The corresponding coordinate
transformations are
\begin{equation}\label{1pntimegauge}
x^i = \bar{x}^i + O(\varepsilon^4)\,\,\,,\,\,\,t = \bar{t} +
\varepsilon^4\beta_{\text{h}} + O(\varepsilon^6),
\end{equation}
and the potentials transform according
to
\begin{equation}
 \bar{\Phi} = \Phi \,\,\,,\,\,\, \bar{\psi} = \psi +
\frac{\partial \beta_{\text{h}}}{\partial \bar{t}} \,\,\, , \,\,\,
\bar{\zeta^i} = \zeta^i - \frac{\partial \beta_{\text{h}}}{\partial
  \bar{x}^i}.
\end{equation}
One can show that the
Newtonian and post-Newtonian pieces of the connection coefficients
$\Gamma^\alpha_{\beta\gamma}$ and of the Riemann
tensor components $R_{\alpha\beta\gamma}^{\ \ \ \ \, \delta}$ are
invariant under this subgroup.
As noted by DSX \cite{dsxI}, this subgroup corresponds
to a gauge freedom in post-1-Newtonian theory analogous to that of
electromagnetism\footnote{If one requires only that the coordinate
systems be conformally Cartesian and not harmonic, then
the most general coordinate transformation is still given by Eqs.\
(\protect{\ref{coordinatetransformation}}) --
(\protect{\ref{conshgC}}), but with the modification that the function
$\beta_{\text{h}}$ can be arbitrary rather than being harmonic
\protect{\cite{dsxI}}.}.

%%%%%%%%%%%%%%%%%%%%%%%%%%%%%%%%%%%%%%%%%%%%%%%%%%%%%%%%%%%%%%%%%%%%%%%%%%%%%%%%%%%%%%%%%%%%%%%%%%%%%%%%%%%%%%%%%%%%%%%%%%%%%%%%%%%%%%%%%%%%%%%%%%%%%%%%%%%%%%

\bigskip
\section{Definitions of an object's mass and current multipole moments
and tidal moments}
\label{singlesys}

\subsection{Overview}
\label{sec:overview}

As discussed in the introduction, a crucial part of
the derivation of the equations of motion for strongly self-gravitating bodies
is the definition of an object's mass and current multipole moments and
also tidal moments in
terms of the behavior of the metric in a weak field
region surrounding the object.  In this section we discuss the definitions of these
quantities.

For orientation, we start by reviewing the definition and status of
multipole and tidal moments in Newtonian gravity.  Suppose that in
some reference frame $(t,x^i)$, there exists a region ${\cal D}$
between two spheres of the form
$$
r_- \le |{\bm{x}}| \le r_+,
$$
for some radii $r_-$ and $r_+$, in which there are no sources.  Then
the Newtonian potential
$\Phi$ satisfies the Laplace equation (\ref{fieldeqA}) in ${\cal D}$.
The general solution for $\Phi$ in ${\cal D}$ can then be written in
terms of a multipole expansion as
\begin{eqnarray}
\label{basicmodelA00}
\Phi(t,x^j) &=& \sum_{l=0}^\infty
\frac{(-1)^{l+1}}{l!}\nM_L(t)\partial_L\frac{1}{|\bm{x}|} -
\frac{1}{l!}\nG_L(t)x^{L}.\nn \\
\end{eqnarray}
Here $L$ denotes the multi-index $a_1 a_2 \ldots a_l$ and $x^L$
denotes the tensor $x^{a_1} x^{a_2}
\ldots x^{a_l}$, cf. Sec.\ \ref{sec:notation} above.
The quantity $\nM_L(t)$ is the $l$th Newtonian mass multipole moment
associated with the reference frame $(t,x^j)$
of the object or objects in the region $|{\bm{x}}| < r_-$.  [The
  superscript ``n'' in $\nM_L$
denotes ``Newtonian''.]  Similarly the quantity $\nG_L(t)$
is the $l$th Newtonian tidal moment associated with the reference
frame $(t,x^j)$ that acts on the region $r<r_-$
due to sources outside $r=r_+$, where $r= |{\bm{x}}|$.  The moments
$\nM_L$ and $\nG_L$ are both STF tensors.

The expansion (\ref{basicmodelA00}) can be taken as the definition of
the moments $\nM_L$ and $\nG_L$; it is possible to invert Eq.\
(\ref{basicmodelA00}) to obtain explicit
expressions for these moments in terms of surface integrals of
$\Phi$ in the domain ${\cal D}$ (see Appendix \ref{sec:welldefined}).
If we
additionally assume that the
Newtonian Poisson equation (\ref{WfieldeqA}) is valid everywhere in $r
< r_-$, then we obtain the conventional formula for the mass
multipole moments as an integral of the Newtonian mass density
$\,^{\scriptscriptstyle \text{n}}\!{T^{00}}$:
\begin{equation}
\nM_L(t) = \int_{r<r_-} \,
\,^{\scriptscriptstyle \text{n}}\!{T^{00}}(t,x^j) x^{<L>}
\,d^3 x.
\label{eq:integraldefine}
\end{equation}
Here the angular brackets denote an STF projection, cf. Sec.\
\ref{sec:notation} above.
As is well-known, the field-based definition (\ref{basicmodelA00}) of
the multipole moments is of greater generality than the integral-based
definition (\ref{eq:integraldefine}), since the former is applicable to
strong field sources that possess an asymptotic region in which the
Newtonian description is a good approximation.

Next, we recall that the moments $\nM_L(t)$ and $\nG_L(t)$ depend on
the choice of reference frame or coordinate system $(t,x^j)$.  They
change when one switches from one reference frame to another according
to ${\bm{x}} \to {\bm{x}} + {\bf z}(t)$ [cf. Eq.\
(\ref{eq:generalcoordtransform}) above, to Newtonian order].
This ambiguity is conventionally resolved in Newtonian physics by
specializing to the reference frame in which the origin of coordinates
coincides with the center of mass of the object, or, equivalently, in
which the mass dipole moment $\nM_i(t)$ vanishes.

Consider now the corresponding situation at post-1-Newtonian order.
A suitable generalization of the integral-based definition
(\ref{eq:integraldefine}) of multipole moments for an isolated system
was given by Blanchet and Damour \cite{bd}.  That definition was
generalized to the case of several interacting bodies by DSX \cite{dsxI}.
DSX also gave a field-based definition of multipole moments analogous
to the definition (\ref{basicmodelA00}) of $\nM_L$, and gave a
different type of definition of tidal moments.  In this section we
will review, simplify\footnote{The DSX definitions, given in Eqs.\
  (6.9a) -- (6.10b) of
Ref.\ \cite{dsxI}, involves a splitting of all the post-Newtonian
potentials into pieces associated with the individual objects.
Our simplified version of their definitions do not require any such
splitting.} and generalize the definitions of DSX
\cite{dsxI}.  Our analysis will be more general than theirs
because we will consider spacetimes where the post-1-Newtonian field
equations are not
satisfied everywhere, whereas DSX assumed the global validity of those
field equations.

As is the case at Newtonian order, the post-Newtonian multipole and
tidal moments associated with a body are not uniquely defined but
depend on the choice of reference frame or coordinate system.  This
ambiguity can be resolved, as in Newtonian theory, by making a
specific choice of canonical reference frame adapted to a given
body.   However, the freedom in choice of reference frame is much
larger at post-Newtonian order than at Newtonian order, cf.\ the
discussion in Sec.\ \ref{sec:gaugefreedom} above.  Therefore the
specialization to a body-adapted frame is more involved.

The remainder of this section is organized as follows.  In Sec.\
\ref{sec:pnsolns} we give a form of multipolar expansion of the
general solution of the post-1-Newtonian field equations that serves
to define the multipole and tidal moments associated with a given
body and with a given coordinate system.  Section
\ref{sec:momentsgaugetransformation} discusses the gauge
transformation properties of the moments, and Sec.\
\ref{sec:bodyadapted} defines the body-adapted gauge that fixes the
moments uniquely.
Finally in Sec.\ \ref{sec:moments1} we generalize our definition of
multipole moments to define moments associated with a given coordinate
system about a given specified worldline.

\subsection{Definition of mass and current multipole moments and tidal moments}
\label{sec:pnsolns}

To define the multipole moments of a body, we start by assuming the
existence of a local coordinate system $(t,x^j)$ with following
properties:
\begin{itemize}
\item The range of the coordinates contains the product of the open ball
\begin{equation}
| {\bm{x}} | < r_+,
\label{eq:domain0}
\end{equation}
where $r_+$ is some radius, with some open interval $(t_0,t_1)$ of time.

\item The vacuum post-1-Newtonian field equations (\ref{fieldeqA}) --
  (\ref{fieldeqC}) are valid in a
spatial region ${\cal D}$ of the form
\begin{equation}
r_- < |{\bm{x}}| < r_+,
\label{eq:domain1}
\end{equation}
for some non-zero radius $r_-$.
\end{itemize}

These assumptions allow us to define the multipole and tidal moments
of a body or bodies in the region $r < r_-$, where $r = |{\bm{x}}|$.
In parallel with the Newtonian case discussed above, the second
assumption allows for the possibility of strong field sources for
which the post-Newtonian approximation is not valid in the region $r <
r_-$.  When applying this definition to systems of several bodies, we
will choose both $r_-$ and $r_+$ to be of order the distance between the
bodies.

The first assumption, that the coordinates in the region ${\cal
D}$ can be extended into the interior to cover the body, is not
actually necessary for the definition of multipole moments.  However
it will be used later in the derivation of equations of motion so we
include it here.  Note that this first assumption does not exclude the
possibility that one or more black holes could reside in the region $r
< r_-$.   As pointed out by Thorne \cite{thornegw}, one merely needs to choose
as the $t=$ constant surfaces an appropriate set of time slices that
pass though the interior of the collapsing object(s) that form the
black hole(s).

Given these assumptions, the general solution of the vacuum field
equations (\ref{fieldeqA}) -- (\ref{fieldeqC}) in the region ${\cal
D}$ can be expanded in terms of STF tensors as
\begin{widetext}
\begin{subequations}
\begin{eqnarray}\label{basicmodelA}
\Phi(t,x^j) &=& \sum_{l=0}^\infty
\frac{(-1)^{l+1}}{l!}\nM_L(t)\partial_L\frac{1}{|\bm{x}|} -
\frac{1}{l!}\nG_L(t)x^L ,\\\nn
\psi(t,x^j) &=& \sum_{l=0}^\infty
\left\{\frac{(-1)^{l+1}}{l!}\left[\pnM_L(t)
+
\frac{(2l+1)}{(l+1)(2l+3)}\dot{\mu}_L(t)\right]\partial_L\frac{1}{|\bm{x}|}
\right.
\left.
+ \frac{(-1)^{l+1}}{l!}{\,}^{\scriptscriptstyle
  \text{n}}\!\ddot{M}_L(t)\partial_L \frac{|\bm{x}|}{2}\right. \\
\label{basicmodelB}
& & \left.- \frac{1}{l!}\left[\pnG_L(t) - \dot{\nu}_{L}(t)\right] x^L -
\frac{1}{l!}\frac{|\bm{x}|^2}{2(2l + 3)}\, ^{\scriptscriptstyle
  \text{n}}\!\ddot{G}_L(t) x^L \right\}, \\
\label{reducedzeta}
\zeta_i(t,x^j) &=& \sum_{l=0}^\infty
\left\{\frac{(-1)^{l+1}}{l!}\left[\frac{4}{l+1}{\,}^{\scriptscriptstyle
    \text{n}}\!\dot{M}_{iL}(t) - \frac{4l}{l+1}\epsilon_{ji<a_l}S_{L-1>j}(t)
  +  \frac{2l-1}{2l+1}\delta_{i<a_l}\mu_{L-1>}(t)\right]
\partial_L\frac{1}{|\bm{x}|}\right. \nn \\
& & \left.- \frac{1}{l!}\left[\nu_{iL}(t) +
  \frac{l}{l+1}\epsilon_{ji<a_l}H_{L-1>j}(t)-
  \frac{4(2l-1)}{2l+1}{\,}^{\scriptscriptstyle
    \text{n}}\!\dot{G}_{<L-1}(t)\delta_{a_l>i}\right] x^L \right\}.
\end{eqnarray}
\end{subequations}
\end{widetext}
The quantities that appear in these equations are the following.
First, there are the Newtonian mass multipole moments $\nM_L(t)$
and tidal moments $\nG_L(t)$ that were discussed in Sec.\
\ref{sec:overview}.  Second, there are post-Newtonian corrections
$\pnM_L(t)$ and $\pnG_L(t)$ to these quantities.  [The superscripts
``pn'' denote post-Newtonian.]  We shall call the quantities
\begin{equation}
M_L(t) \equiv \nM_L(t) + \varepsilon^2 \pnM_L(t)
\label{eq:totalMLdef}
\end{equation}
and
\begin{equation}
G_L(t) \equiv \nG_L(t) + \varepsilon^2 \pnG_L(t)
\label{eq:totalGLdef}
\end{equation}
the total mass multipole and tidal moments, respectively.
Third, there are current multipole moments
$S_L(t)$, and a new set of tidal moments $H_L(t)$ related to
gravitomagnetic forces.
Following DSX, we will refer to $H_L(t)$ and $G_L(t)$ as the
gravitomagnetic and gravitoelectric tidal moments, respectively.
Fourth, there are moments $\mu_L(t)$ and
$\nu_L(t)$ that contain information about the coordinate system being
used, but do not contain any gauge-invariant information about the
body.  We shall show below that it is always possible to find a gauge
in which $\mu_L(t) = \nu_L(t) = 0$.  We shall call these quantities
``gauge moments''.

All of the quantities $\nM_L$, $\nG_L$, $\pnM_L$, $\pnG_L$, $\mu_L$,
$\nu_L$, $S_L$ and $H_L$ are STF on all their indices.
As in the Newtonian case, the expansions (\ref{basicmodelA}) --
(\ref{reducedzeta}) can be taken
as the definition of all of these moments for a given coordinate
system; one can invert these expansions to obtain explicit
expressions for all the moments in terms of surface integrals of
various combinations of derivatives of the potentials [cf. Appendix
\ref{sec:welldefined}].
The moments
$\nM_L$, $\nG_L$, $\pnM_L$, $\pnG_L$, $\mu_L$ are defined for all $l
\ge 0$, while $\nu_L$, $S_L$ and $H_L$ are defined only for $l \ge 1$.
In Eq.\ (\ref{basicmodelB}) and throughout this paper, it is
understood that $\nu_L \equiv 0$ for $l=0$.

We shall call the pieces of the potentials that would diverge as
$|{\bm{x}}| \to \infty$ the {\it tidal pieces} of the potentials.
Correspondingly, we will use the phrase ``tidal moments'' to refer
to any of the moments $\nG_L$, $\pnG_L$, $H_L$ and $\nu_L$ that
appear in the coefficients of the growing terms in Eqs.\
(\ref{basicmodelA}) -- (\ref{reducedzeta}).
These moments encode the gravitational influence of other bodies on
the body in the region $|{\bm{x}}| < r_-$.
We shall call the remaining pieces of the potentials, which involve
the moments $\nM_L$, $\pnM_L$, $S_L$ and $\mu_L$, the {\it intrinsic
pieces}.

We now turn to the derivation of the expansions (\ref{basicmodelB}) and
(\ref{reducedzeta}).  We start by writing the general
solution in the region ${\cal D}$ of the Laplace equation
(\ref{fieldeqC}) for the gravitomagnetic potential as
\begin{eqnarray}
\label{basicmodelC}
\zeta_i(t,x^j) &=& \sum_{l=0}^\infty
\frac{(-1)^{l+1}}{l!}Z_{iL}(t)\partial_L\frac{1}{|\bm{x}|}  -
\frac{1}{l!}Y_{iL}(t)x^L. \nn \\
\end{eqnarray}
Here the quantities $Z_{iL}$ and $Y_{iL}$ are STF on their $L$ indices
only, and not on the $i$ index.
Next, we insert the expressions (\ref{basicmodelC}) and
(\ref{basicmodelA}) for the gravitomagnetic and Newtonian potentials
into the harmonic gauge condition (\ref{harmonicgauge}).
This gives the relations\footnote{Thus, the time independence of the
Newtonian mass monopole $\nM$ can be derived either from stress-energy
conservation at Newtonian order in the interior of the body (for a
weakly self-gravitating system), or from the validity of the
post-Newtonian vacuum field equations and harmonic gauge condition in
the far field of the body.  This type of phenomenon, where one
can avoid dealing with the interior physics by
going to one higher post-Newtonian order in the far field, is well
known in the literature on equations of motion, and will be
encountered again in Sec.\ \protect{\ref{eom}}.}
\begin{eqnarray}\label{hgA}
^{\scriptscriptstyle \text{n}}\!\dot{M} &=& 0 , \\\label{hgB}
Z_{<iL>} &=& \frac{4}{l+1}{\,}^{\scriptscriptstyle \text{n}}\!\dot{M}_{iL}
\end{eqnarray}
and
\begin{eqnarray}\label{hgC}
Y_{jjL} &=& -4{\,}^{\scriptscriptstyle \text{n}}\!\dot{G}_{L}.
\end{eqnarray}

Next, we use the identity given in Eq.\ (6.21) of Ref.\ \cite{dsxI}
that allows one to express in terms of irreducible, STF tensors any
tensor that is STF on all its indices except its first index.
Specifically, if $T_{iL}$ is any tensor satisfying $T_{iL} =
T_{i<L>}$, then we have
\begin{equation}
\label{reduction}
T_{iL} = T^{(+1)}_{iL} + \epsilon_{ji<a_l}T^{(0)}_{L-1>j} + \delta_{i<a_l}T^{(-1)}_{L-1>},
\end{equation}
where
\begin{subequations}
\begin{eqnarray}
T^{(+1)}_{iL} &\equiv& T_{<iL>}, \\
T^{(0)}_L &\equiv& \frac{l}{l+1} T_{jk<L-1} \epsilon_{a_l>jk}
\end{eqnarray}
and
\begin{eqnarray}
T^{(-1)}_{L-1} &\equiv& \frac{2l-1}{2l+1}T_{jjL-1}.
\end{eqnarray}
\end{subequations}
In order to apply this identity to the tensors $Z_{iL}$ and $Y_{iL}$,
we define
\begin{subequations}
\begin{eqnarray}
\label{eq:SLdef}
S_L &\equiv& - \frac{1}{4} Z_{jk<L-1} \epsilon_{a_l>jk}, \\
\label{eq:HLdef}
H_L &\equiv&  Y_{jk<L-1} \epsilon_{a_l>jk}, \\
\label{eq:nuLdef}
\nu_L &\equiv&  Y_{<L>}
\end{eqnarray}
and
\begin{equation}
\label{eq:muLdef}
\mu_L \equiv Z_{jjL}.
\end{equation}
\end{subequations}
We now insert the relations (\ref{hgB}) and (\ref{hgC}) and the
definitions (\ref{eq:SLdef}) -- (\ref{eq:muLdef}) into the general
identity (\ref{reduction}) to obtain
\begin{equation}
Z_{iL} =
\frac{4}{l+1}{\,}^{\scriptscriptstyle
    \text{n}}\!\dot{M}_{iL} - \frac{4l}{l+1}\epsilon_{ji<a_l}S_{L-1>j}
  +  \frac{2l-1}{2l+1}\delta_{i<a_l}\mu_{L-1>}
\label{eq:ZiLdecompos}
\end{equation}
and
\begin{equation}
Y_{iL} = \nu_{iL} +
  \frac{l}{l+1}\epsilon_{ji<a_l}H_{L-1>j}-
  \frac{4(2l-1)}{2l+1}{\,}^{\scriptscriptstyle
    \text{n}}\!\dot{G}_{<L-1} \delta_{a_l>i}.
\label{eq:YiLdecompos}
\end{equation}
Inserting these formulae into the expansion (\ref{basicmodelC}) yields
Eq.\ (\ref{reducedzeta}).

We remark that the parameterization (\ref{basicmodelC}) of the
gravitomagnetic potential in terms of the tensors $Y_{iL}$ and $Z_{iL}$
will frequently be more convenient to use in
our computations below than the fully STF parameterization
(\ref{reducedzeta}).

Consider now the post-Newtonian potential $\psi$.  We can write the
general solution in the region ${\cal D}$ of the vacuum field equation
(\ref{fieldeqB}) as the sum
$
\psi = \psi_\text{p} + \psi_\text{h}
$
of a particular solution $\psi_\text{p}$ of the inhomogeneous equation
and a general solution $\psi_\text{h}$ of the homogeneous equation $\nabla^2
\psi =0$. A particular solution can be obtained by inspection,
using the expansion (\ref{basicmodelA}) of the Newtonian potential:
\begin{eqnarray}
\psi_\text{p} &=& \sum_{l=0}^\infty
\left[
 \frac{(-1)^{l+1}}{l!}{\,}^{\scriptscriptstyle
  \text{n}}\!\ddot{M}_L \partial_L \frac{|\bm{x}|}{2}
-\frac{1}{l!}\frac{|\bm{x}|^2}{2(2l + 3)}\, ^{\scriptscriptstyle
  \text{n}}\!\ddot{G}_L x^L \right]. \nn \\
\label{eq:psipdef}
\end{eqnarray}
The homogeneous solution can be written in a form paralleling the
expansion (\ref{basicmodelA}) of the Newtonian potential:
\begin{eqnarray}
\psi_\text{h} &=& \sum_{l=0}^\infty
\frac{(-1)^{l+1}}{l!}
{\,^{\scriptscriptstyle \text{pn}}\!{\tilde M}}_L\partial_L\frac{1}{|\bm{x}|} -
\frac{1}{l!}
{\,^{\scriptscriptstyle \text{pn}}\!{\tilde G}}_Lx^L.\ \ \
\label{eq:psihdef}
\end{eqnarray}
Next we define
\begin{equation}
\pnM_L(t) = {\,^{\scriptscriptstyle \text{pn}}\!{\tilde M}}_L(t)
- \frac{(2l+1)}{(l+1)(2l+3)}\dot{\mu}_L(t)
\label{eq:pnMdef}
\end{equation}
and
\begin{equation}
\pnG_L(t) = {\,^{\scriptscriptstyle \text{pn}}\!{\tilde G}}_L(t)
+ {\dot \nu}_L(t).
\label{eq:pnGdef}
\end{equation}
Inserting the definitions (\ref{eq:pnMdef}) and (\ref{eq:pnGdef}) into
the homogeneous solution (\ref{eq:psihdef}) and adding the particular solution
(\ref{eq:psipdef}) yields Eq.\ (\ref{basicmodelB}).
The reason for choosing the particular parameterization given by Eqs.\
(\ref{eq:pnMdef}) and (\ref{eq:pnGdef}) is so that the moments
$\pnM_L(t)$ and $\pnG_L(t)$ be invariant\footnote{Except for $\pnG_L$
  for $l=0$ which is not invariant.} under the subclass
(\ref{1pntimegauge}) of gauge transformations; see Sec.\
\ref{sec:momentsgaugetransformation} below for more details.

We now specialize to the situation, considered by DSX,
where the post-1-Newtonian field equations with sources (\ref{WfieldeqA}) --
(\ref{WfieldeqC}) are assumed to hold for all $r <
r_+$, i.e. in the interior of the body.  In this special case, we
now show that our definitions of the quantities $\pnM_L$,
$\pnG_L$, $H_L$ and $S_L$ are equivalent to the DSX definitions
of these quantities.

Consider first the multipole moments $\pnM_L$
and $S_L$.  Equations (6.9a) and (6.9b) of DSX \cite{dsxI}
define the moments $M_L(t)$ and $S_L(t)$ in terms of an expansion of
``locally generated'' pieces of the potentials.  Their splitting of
the potentials into ``locally
generated'' and ``external'' pieces is defined by their Eq.\ (4.5),
and is easily seen to be equivalent to the splitting which we
discussed above of our expansions (\ref{basicmodelA}) --
(\ref{reducedzeta}) into intrinsic terms and tidal terms.
Therefore it is sufficient to show that the DSX expansions (6.9a) and
(6.9b) coincide with the tidal terms in our Eqs.\ (\ref{basicmodelA})
-- (\ref{reducedzeta}).  This follows from the definition
(\ref{eq:totalMLdef}) and the relations $W = - \Phi - \varepsilon^2
\psi$ and $W_i = - \zeta_i/4$ between the DSX potentials $(W,W^i)$ and
our potentials $\Phi$, $\psi$ and $\zeta^i$.

Consider next the gravitoelectric and gravitomagnetic tidal moments
$G_L$ and $H_L$.  These are defined by DSX in terms of STF
projections of gradients of the external (or tidal) pieces of the
gravitoelectric and gravitomagnetic fields evaluated at the origin of
spatial coordinates, cf. Eq.\ (6.13) of Ref.\ \cite{dsxI}.
Inserting our expansions (\ref{basicmodelA}) --
(\ref{reducedzeta}) into the definitions (\ref{eq:Bdef}) and
(\ref{eq:Edef}) of the gravitomagnetic and gravitoelectric fields
yields for the tidal pieces (denoted by a superscript T) of these
quantities
\begin{equation}
\label{bfield}
B_i^T = \sum_{l=0}^\infty \frac{1}{l!}\left[ H_{iL}x^{<L>} +
  \frac{4l}{l+1}\epsilon_{ija_l}{\,}^{\scriptscriptstyle
    \text{n}}\!\dot{G}_{jL-1}x^{<L>} \right]
\end{equation}
and
\begin{eqnarray}\nn
E_i^T &=& \sum_{l=0}^\infty \frac{1}{l!} \bigg\{ (\nG_{iL} +
\varepsilon^2 \pnG_{iL})x^{<L>}
\\ \nn  & &
+ \varepsilon^2
\bigg[\frac{|\bm{x}|^2}{2(2l+3)} x^{<L>}{\,}^{\scriptscriptstyle
    \text{n}}\!\ddot{G}_{iL}
-\frac{7l - 4}{2l+1}x^{<iL-1>}{\,}^{\scriptscriptstyle
    \text{n}}\!\ddot{G}_{L-1}
\\  & &
  + \frac{l}{l+1}\epsilon_{ijk}x^{<jL-1>}\dot{H}_{kL-1}
  \bigg] \bigg\}. \label{efield}
\end{eqnarray}
These expressions agree to $O(\varepsilon^2)$ with corresponding
expansions derived by DSX [Eqs.\ (6.23) of Ref.\ \cite{dsxI}],
when the definition (\ref{eq:totalGLdef}) is used.  It follows that
the two definitions of tidal moments are equivalent.

Thus, our definition of both the multipole moments $\nM_L$, $\pnM_L$,
$S_L$ and the tidal moments $\nG_L$, $\pnG_L$ and $H_L$ in terms of
the general solution (\ref{basicmodelA}) -- (\ref{reducedzeta}) of the
post-1-Newtonian vacuum field equations unifies and simplifies the two
different types of definition given by DSX.
Our definitions also generalize the DSX definitions to strong field
sources.

Finally, we note that in the case considered by DSX, one can
alternatively define the post-Newtonian multipole moments by integrals
over the source that are analogous to the Newtonian integral
(\ref{eq:integraldefine}) \cite{dsxI}.  Translating
Eqs.\ (6.11) of Ref.\ \cite{dsxI} into our notation gives for these
integrals
\begin{equation}
S_L = \int_{r<r_-} \,
\epsilon^{jk<a_l} x^{L-1>j} \, \,^{\scriptscriptstyle \text{n}}\!{T^{0k}}
\, d^3 x
\label{eq:Sintegraldefine}
\end{equation}
and
\begin{eqnarray}
\!\pnM_L &=& \!\int_{r<r_-} \!\bigg\{ \bigg[
\,^{\scriptscriptstyle \text{pn}}\!{T^{00}} +
\,^{\scriptscriptstyle \text{n}}\!{T^{jj}} + \frac{x^j x^j}{2 (2l+3)}
\,^{\scriptscriptstyle \text{n}}\!{T^{00}}_{\!\!,00} \bigg]
 x^{<L>} \nn \\
&& - \frac{4 (2l+1)}{(l+1)(2l+3)}
\,^{\scriptscriptstyle \text{n}}\!{T^{0j}}_{,0} x^{<jL>} \bigg\}
\,d^3 x.
\label{eq:pnMintegraldefine}
\end{eqnarray}

\subsection{Gauge transformation properties of the moments}
\label{sec:momentsgaugetransformation}

In this section we compute how the various moments transform under
the general transformation (\ref{coordinatetransformation})
from an original coordinate system $(t,x^j)$ and a new coordinate
system $({\bar t},{\bar x}^j)$.
Under that transformation, the spatial domain ${\cal D}$ defined by
Eq.\ (\ref{eq:domain1}) is mapped onto the domain
\begin{equation}
r_- < | {\bar {\bm x}} + {\bm z}({\bar t})| < r_+,
\end{equation}
to zeroth order in $\varepsilon$.  We therefore restrict the set of
coordinate transformations to those that satisfy
\begin{equation}
\max_{\bar{t}} |\bm{z}(\bar{t})| < \frac{r_+ - r_-}{2}.
\label{eq:restriction}
\end{equation}
This restriction ensures that the image ${\bar {\cal D}}$
of the domain ${\cal D}$ contains a
non-empty region of the form ${\bar r}_- < | {\bar
  {\bm x}} | < {\bar
r}_+$ for some radii ${\bar r}_-$, ${\bar r}_+$ with ${\bar r}_- > 0$,
and thus allows us to define multipole and tidal moments in the new
coordinate system.
We also parameterize the freely
specifiable harmonic function $\beta_{\text{h}}$ as
\begin{equation}
\beta_\text{h} = \left\{ \begin{array}{ll} {\displaystyle
\sum_{l=0}^\infty
\frac{(-1)^{l+1}}{l!}\lambda_L(\bar{t})\partial_L\frac{1}{|\bar{\bm{x}}|}
- \frac{1}{l!}\tau_L(\bar{t})\bar{x}^L,} & {\mbox\ {\rm in}\ } {\bar {\cal D}}
\\
\\
\mbox{arbitrary smooth function} & {\mbox\ {\rm in}\ } {\bar {\cal
    D}}_{\rm int}, \\
\
\end{array} \right. \\
\\
\label{eq:lambdaLdef}
\end{equation}
where ${\bar {\cal D}}_{\rm int}$ is the image of the domain
$0 \le r < r_-$.  Here the quantities $\lambda_L(\bar{t})$ and
$\tau_L(\bar{t})$ are STF on all their indices.
This choice guarantees that the coordinate transformation
(\ref{coordinatetransformation}) is defined and smooth on the entire
domain $0 \le r < r_+$, while maintaining the harmonic property of
the coordinates in the domain $r_- < r < r_+$ where the
post-1-Newtonian equations are valid.

In the barred coordinate system we define the
transformed moments
$\barnM_L({\bar t})$, $\barnG_L({\bar t})$, $\barpnM_L({\bar t})$,
$\barpnG_L({\bar t})$, ${\bar Y}_{iL}({\bar t})$, ${\bar Z}_{iL}({\bar
t})$, ${\bar S}_L({\bar t})$, ${\bar H}_L({\bar t})$, ${\bar
\nu}_L({\bar t})$ and ${\bar \mu}_L({\bar t})$ by the following barred
versions of Eqs.\ (\ref{basicmodelA}), (\ref{basicmodelB}),
(\ref{basicmodelC}), (\ref{eq:ZiLdecompos}) and
(\ref{eq:YiLdecompos}):
\begin{widetext}
\begin{subequations}
\begin{eqnarray}\label{basicmodelD}
\bar{\Phi}(\bar{t},\bar{x}^j) &=& \sum_{l=0}^\infty
\frac{(-1)^{l+1}}{l!}{\,}^{\scriptscriptstyle
  \text{n}}\!\bar{M}_L(\bar{t})\partial_L\frac{1}{|\bar{\bm{x}}|} -
\frac{1}{l!}{\,}^{\scriptscriptstyle
  \text{n}}\!\bar{G}_L(\bar{t})\bar{x}^L ,\\\nn
\bar{\psi}(\bar{t},\bar{x}^j) &=& \sum_{l=0}^\infty
\frac{(-1)^{l+1}}{l!}\left[{\,}^{\scriptscriptstyle
  \text{pn}}\!\bar{M}_L(\bar{t}) +
\frac{(2l+1)}{(l+1)(2l+3)}
\dot{\bar{\mu}}_L(\bar{t})\right] \partial_L\frac{1}{|\bar{\bm{x}}|}
+ \frac{(-1)^{l+1}}{l!}{\,}^{\scriptscriptstyle
  \text{n}}\!\ddot{\bar{M}}_L(\bar{t})\partial_L\frac{|\bar{\bm{x}}|}{2}
 \\\label{basicmodelE}
& & - \frac{1}{l!} \left[{\,}^{\scriptscriptstyle
    \text{pn}}\!\bar{G}_L(\bar{t}) -
  \dot{\bar{\nu}}_L(\bar{t})\right] \bar{x}^L -
\frac{1}{l!}\frac{|\bar{\bm{x}}|^2}{2(2l + 3)}{\,}^{\scriptscriptstyle
  \text{n}}\!\ddot{\bar{G}}_L(\bar{t})\bar{x}^L
,\\\label{basicmodelF}
\bar{\zeta}_i(\bar{t},\bar{x}^j) &=& \sum_{l=0}^\infty
\frac{(-1)^{l+1}}{l!}\bar{Z}_{iL}(\bar{t})\partial_L\frac{1}{|\bar{\bm{x}}|}
- \frac{1}{l!}\bar{Y}_{iL}(\bar{t})\bar{x}^{<L>}, \\
{\bar Z}_{iL}({\bar t}) &=&
\frac{4}{l+1}{\,}^{\scriptscriptstyle
    \text{n}}\!\dot{{\bar M}}_{iL}({\bar t}) -
\frac{4l}{l+1}\epsilon_{ji<a_l} {\bar S}_{L-1>j}({\bar t})
  +  \frac{2l-1}{2l+1}\delta_{i<a_l}{\bar \mu}_{L-1>}({\bar t}),
\label{eq:barZiLdecompos}
\end{eqnarray}
and
\begin{equation}
{\bar Y}_{iL}({\bar t}) = {\bar \nu}_{iL}({\bar t}) +
  \frac{l}{l+1}\epsilon_{ji<a_l}{\bar H}_{L-1>j}({\bar t}) -
  \frac{4(2l-1)}{2l+1}{\,}^{\scriptscriptstyle
    \text{n}}\!\dot{{\bar G}}_{<L-1}({\bar t}) \delta_{a_l>i}.
\label{eq:barYiLdecompos}
\end{equation}
\end{subequations}
\end{widetext}
As before dots denote derivatives with respect to the time argument.
By substituting
the multipole expansions (\ref{basicmodelA}) --- (\ref{reducedzeta}) of
the original potentials into the transformation formulae
(\ref{transfA}) --- (\ref{transfC}), and comparing the results with the
multipole expansions (\ref{basicmodelD}) --- (\ref{basicmodelF}), we can
derive transformation laws for the various moments.

To illustrate this procedure, we begin with the transformation of the
Newtonian potential. We obtain
\begin{eqnarray}\nn
 \bar{\Phi}(\bar{t},\bar{x}^k) &=& \hat{\Phi}(\bar{t},\bar{x}^k) +
\ddot{z}_i(\bar{t})\bar{x}^i -
\frac{1}{2}\dot{z}_i(\bar{t})\dot{z}^i(\bar{t}) +
\dot{\alpha}_{\text{c}}(\bar{t})\\\nn
&=&
\dot{\alpha}_{\text{c}}(\bar{t}) +
\sum_{l=0}^\infty
\frac{(-1)^{l+1}}{l!}\nM_L(\bar{t})\partial_L\frac{1}{|\bar{\bm{x}} +
  \bm{z}(\bar{t})|}
\nn \\ &&
- \frac{1}{l!}\nG_L({\bar t})[\bar{x} +
  z(\bar{t})]^L + \ddot{z}_i(\bar{t})\bar{x}^i -
\frac{1}{2}\dot{z}^i(\bar{t})\dot{z}_i(\bar{t}). \nn \\
\label{barphi}
\end{eqnarray}
Next we use the expansion
\begin{equation}
\frac{1}{|\bar{\bm{x}} + \bm{z}|} = \sum_{n=0}^\infty
\frac{1}{n!}z^N\partial_N\frac{1}{|\bar{\bm{x}}|}
\end{equation}
which is valid for
$|{\bar {\bm x}}| > |{\bm z}|$ [cf. Eq.\ (\ref{eq:restriction})
above], and we compare with the multipole expansion (\ref{basicmodelD}).
This gives the well known transformation laws for the monopole and dipole
\begin{equation}
{\,}^{\scriptscriptstyle \text{n}}\!\bar{M} = \nM,
\end{equation}
and
\begin{equation}
{\,}^{\scriptscriptstyle \text{n}}\!\bar{M}_i = \nM_i - \nM z_i,
\label{eq:dipoletransformnewtonian}
\end{equation}
together with the general result
\begin{equation}\label{newM}
{\,}^{\scriptscriptstyle \text{n}}\!\bar{M}_L =
\sum_{n=0}^l(-1)^{n+l}\frac{l!}{n!(l-n)!}\nM_{<N}z_{L-N>}.
\end{equation}
Here $\nM_{<N} z_{L-N>}$ denotes $\nM_{<a_1 \ldots a_n} z_{a_{n+1}} z_{a_{n+2}}
\ldots z_{a_l>}$, cf.\ Sec.\ \ref{sec:notation} above.
Similarly we obtain from Eq.\ (\ref{barphi}) the
transformation law for the Newtonian gravitoelectric tidal moments
\begin{equation}\label{newG}
{\,}^{\scriptscriptstyle \text{n}}\!\bar{G}_L =
\sum_{k=0}^\infty\frac{1}{k!}\nG_{LK}z^K - l!\Lambda^{\Phi}_L.
\end{equation}
Here
\begin{subequations}
\begin{eqnarray}\label{lambdaphizero}
\Lambda^{\Phi} &=& -\frac{1}{2}\dot{z}_j\dot{z}_j +
\dot{\alpha}_{\text{c}}, \\\label{lambdaphii}
\Lambda^{\Phi}_i &=&  \ddot{z}_i,
\end{eqnarray}
\end{subequations}
and $\Lambda^{\Phi}_L = 0$ for $l \geq 1$.
Following DSX \cite{dsxI}, we call the quantities
$\Lambda^{\Phi}_L$ {\it inertial moments}.

We similarly compute the transformation laws for the multipole and
tidal moments $Z_{iL}$ and $Y_{iL}$.
We insert the expansion (\ref{basicmodelC}) into
the transformation law (\ref{transfB}) and use the formulae
(\ref{conshgA}) -- (\ref{conshgC}) for the
quantities $\alpha$, $h^i$ and $\beta$.  We also use the
parameterization (\ref{eq:lambdaLdef})
of the harmonic function $\beta_\text{h}$.
We then equate the result to the multipole
expansion (\ref{basicmodelF}) and obtain the transformation laws

\begin{subequations}
\begin{eqnarray}\nn
\bar{Z}_{iL} &=& \sum_{n=0}^l(-1)^{n+l}\frac{l!}{n!(l-n)!}\left(Z_{i<N}z_{L-N>} \right. \\\label{newZ}
& & - 4\dot{z}_i\nM_{<N}z_{L-N>}) + l\delta_{i<a_l}\lambda_{L-1>}, \\\nn
\bar{Y}_{iL} &=& \sum_{k=0}^\infty\frac{1}{k!}(Y_{iLK} -
4\dot{z}_i\nG_{LK})z^K - \tau_{iL}
-l!\Lambda^{\bm{\zeta}}_{iL}.\\\label{newY}
\end{eqnarray}
\end{subequations}
Here
\begin{subequations}
\begin{eqnarray}
\label{gravitoinertialA}
\Lambda^{\bm{\zeta}}_i &=& \dot{h}_{\text{c}}^i +
\dot{z}_i\dot{z}_j\dot{z}_j - \epsilon_{ijk}\dot{z}_jR_k  -
2\dot{\alpha}_\text{c}\dot{z}_i, \\
\label{Lambdaijdef}
\Lambda^{\bm{\zeta}}_{ij} &=& -\dot{z}_i\ddot{z}_j
-\dot{z}_{(i}\ddot{z}_{j)} + \epsilon_{ijk}\dot{R}_k
\label{gravitoinertialB}
+ 2\delta_{ij}\dot{z}_k\ddot{z}_k  -
\frac{4}{3}\delta_{ij}\ddot{\alpha}_\text{c}, \nn \\
\\
\label{gravitoinertialC}
\Lambda^{\bm{\zeta}}_{ijk} &=& -\frac{6}{5}\delta_{i<j}\dddot{z}_{k>},
\end{eqnarray}
\end{subequations}
and all the other gravitomagnetic inertial moments
$\Lambda^{\bm{\zeta}}_{iL}$ are zero.

The transformation laws for the fully STF moments $S_L$, $H_L$,
$\mu_L$ and $\nu_L$ parameterizing the
gravitomagnetic potential [cf. Eq.\ (\ref{reducedzeta}) above] are obtained
by splitting the expressions (\ref{newZ}) and (\ref{newY})
for ${\bar Z}_{iL}$ and ${\bar Y}_{iL}$ into their fully
STF pieces using the identities (\ref{reduction}), (\ref{eq:peeling}),
(\ref{eq:ident14}) and (\ref{eq:ident15}).
The results are
\begin{widetext}
\begin{subequations}
\begin{eqnarray}
{\bar S}_L &=& \sum_{n=0}^l (-1)^{n+l} \frac{l!}{n! (l-n)!} \bigg[
\frac{n(l+1)}{l(n+1)} S_{<N} z_{L-N>} - \frac{(l-n)}{l(n+1)} z_s
\ndotM_{r<N} z_{(L-1)-N} \epsilon_{a_l>rs} \nn \\
&&+ \frac{n}{l} {\dot z}_r \nM_{s<N-1} z_{(L-1) - (N-1)}
\epsilon_{a_l>rs}
+ \frac{l-n}{l} {\dot z}_r z_s \nM_{<N} z_{(L-1)-N} \epsilon_{a_l>rs}
\bigg],
\label{eq:Stransform}
\end{eqnarray}
\begin{eqnarray}
{\bar \mu}_L &=& \sum_{n=0}^{l+1} (-1)^{n+l+1} \frac{(l+1)!}{n! (l+1-n)!}
\bigg[ \left[\frac{n}{l+1} + \frac{l+1-n}{(l+1)(2n+1)}\left(2n-1 - \frac{2}{2l+1}\right)\right]\mu_{<N-1} z_{L-(N-1)>} \nn \\
&& + \frac{4(l+1-n)(2l-2n+1)}{(l+1)(2l+1)(n+1)} z_j
  \ndotM_{j<N} z_{L-N>} - \frac{4n(l+1-n)(2l+3)}{(l+1)(n+1)(2l+1)} z_j S_{p<N-1} z_{(L-1)-(N-1)} \epsilon_{a_l>pj} \nn \\
&& -\frac{4n}{l+1} {\dot z}_j \nM_{j<N-1} z_{L-(N-1)>} - \frac{4(l+1-n)}{l+1} z_j {\dot z}_j \nM_{<N} z_{L-N>} \nn \\
&& + \frac{8n(l+1-n)}{(l+1)(2l+1)} z_j \nM_{j<N-1} z_{(L-1) - (N-1)} {\dot
  z}_{a_l>} + \frac{4(l+1-n)(l-n)}{(l+1)(2l+1)} z_j z_j \nM_{<N} z_{(L-1)-N} {\dot z}_{a_l>} \nn \\
&& - \frac{4(l+1-n)(l-n)}{(l+1)(2l+1)(n+1)} z_j z_j \ndotM_{<N+1} z_{L-(N+1)>} \bigg] + \frac{(l+1)(2l+3)}{2l+1} \lambda_L,
\label{eq:mutransform}
\end{eqnarray}
\begin{eqnarray}
{\bar H}_L &=& \sum_{k=0}^\infty \frac{1}{k!}\bigg[
H_{LK} z^K + 4 \nG_{iK<L-1} \epsilon_{a_l>ij} {\dot z}^j z^K
+ \frac{8k}{2l + 2k + 1} \ndotG_{iK-1<L-1} \epsilon_{a_l>ij} z^j z^{K-1}
\bigg] \nn \\
&& - l! \,
\Lambda^{\bm{\zeta}}_{ij<L-1} \epsilon_{a_l>ij},
\label{eq:Htransform}
\end{eqnarray}
and
\begin{eqnarray}
{\bar \nu}_{iL} &=&- \tau_{iL} - l! \, \Lambda^{\bm{\zeta}}_{<iL>} +
\sum_{k=0}^\infty \bigg[ \nu_{iLK}z^K -4 {\dot z}_{<i} \nG_{L>K} z^K
+ \frac{k}{l+k+1} z^{j} \epsilon_{jm<i} H_{L>mK-1} z^{K-1} \nn \\ \nn
&& -\frac{4k(2l+2k-1)}{(l+k)(2l+2k+1)} z_{<i} \ndotG_{L>K-1} z^{K-1}
+ \frac{4k(k-1)}{(l+k)(2l+2k+1)} \ndotG_{iLK-2} z^{K-2} (z_j z_j) \\
&& + \frac{4kl}{(l+k)(2l+2k+1)}\ndotG_{K-1<iL-1} z_{a_l>} z^{K-1} \bigg].
\label{eq:nutransform}
\end{eqnarray}
\end{subequations}
\end{widetext}

It is also possible to derive from the expressions (\ref{newZ}) and
(\ref{newY}) transformation laws for the moments
${\,}^{\scriptscriptstyle \text{n}}\!\dot{M}_{iL}$
and
${\,}^{\scriptscriptstyle \text{n}}\!\dot{G}_{iL}$
that enter into the fully STF parameterizations (\ref{eq:ZiLdecompos}) and
(\ref{eq:YiLdecompos}) of $Z_{iL}$ and $Y_{iL}$.  The resulting
transformation laws are consistent with the transformation laws
(\ref{newM}) and (\ref{newG}) derived earlier for $\nM_L$ and $\nG_L$; this
consistency is an important check of the formalism.

Finally we turn to the moments $\pnM_L$ and $\pnG_L$ parameterizing
the post-Newtonian potential $\psi$.  The transformation laws for
these moments are by far the most tedious to compute.  The results
are
\begin{widetext}
\begin{eqnarray}\label{newpnM}
{\,}^{\scriptscriptstyle \text{pn}}\!\bar{M}_L &=&
-\frac{2l+1}{(l+1)(2l+3)}\dot{\bar{\mu}}_L + {\dot \lambda}_L +
\sum_{n=0}^l(-1)^{n+l}\frac{l!}{n!(l-n)!}\sigma_{<N}z_{L-N>},
\end{eqnarray}
where
\begin{subequations}
\begin{eqnarray}\nn
\sigma_L &=& \pnM_L + \frac{2l+1}{(l+1)(2l+3)}\dot{\mu}_L - \dot{z}_jZ_{jL} + 2\dot{z}_j\dot{z}_j\nM_L + (\alpha_\text{c} - z_j\dot{z}_j){\,}^{\scriptscriptstyle \text{n}}\!\dot{M}_L
- l U_{<a_l}\nM_{L-1>} \\\label{sigmaL}
& & + \left(\frac{2l+1}{2l+3}\right)\left[(l+2)\ddot{z}_j\nM_{jL} - U_{jj}\nM_L - l\nM_{j<L-1}U_{a_l>j} + \frac{2l}{2l+1}U_{j<a_l}\nM_{L-1>j} + \dot{z}_j{\,}^{\scriptscriptstyle \text{n}}\!\dot{M}_{jL}\right], \\
U_i &=& h^i_{\text{c}} - z_jV_{ij} - z_jz_kV_{ijk},\\
U_{ij} &=& V_{ij} - 2z_kV_{ijk} ,\\
V_{ij} &=& \epsilon_{ijk}R^k - \delta_{ij}\dot{\alpha}_\text{c} + \frac{1}{2}\delta_{ij}\dot{z}_k\dot{z}_k + \frac{1}{2}\dot{z}_i\dot{z}_j ,\\
V_{ijk} &=& -\frac{1}{2}\delta_{ij}\ddot{z}_k + \ddot{z}_{[i}\delta_{j]k},
\end{eqnarray}
\end{subequations}
and
\begin{eqnarray}\label{newpnG}
{\,}^{\scriptscriptstyle \text{pn}}\!\bar{G}_L &=& \dot{\bar{\nu}}_{L} + \sum_{k=0}^\infty \frac{1}{k!}(\rho_{LK} + lz_{<a_l}\rho^\prime_{L-1>K}) z^K  + \dot{\tau}_L- l!\Lambda^{\psi_\text{h}}_L,
\end{eqnarray}
where
\begin{subequations}
\begin{eqnarray}\nn
\rho_L &=& \pnG_L - \dot{\nu}_{L} - \dot{z}_iY_{iL} + 2\dot{z}_j\dot{z}_j\nG_L + U_j\nG_{jL} + lU_{j<a_l}\nG_{L-1>j} \\
& & -l(l-1)\ddot{z}_{<a_l}\nG_{L-1>} + l\dot{z}_{<a_l}{\,}^{\scriptscriptstyle \text{n}}\!\dot{G}_{L-1>} + (\alpha_{\text{c}} - \dot{z}_j z^j){\,}^{\scriptscriptstyle \text{n}}\!\dot{G}_L + \frac{z_jz^j}{2(2l+3)}\rho^\prime_L ,\\
\rho^\prime_L &=& {\,}^{\scriptscriptstyle \text{n}}\!\ddot{G}_L +
2\dot{z}_j{\,}^{\scriptscriptstyle \text{n}}\!\dot{G}_{jL} +
\dot{z}_j {\dot z}_k \nG_L + \ddot{z}_j\nG_{jL}, \\
\Lambda^{\psi_\text{h}} &=& -\dot{z}_j\dot{h}^j_\text{c} - \frac{1}{4}(\dot{z}_j\dot{z}_j)^2 - \frac{1}{2}(\dot{\alpha}_\text{c})^2 + \dot{\alpha}_\text{c}\dot{z}_j\dot{z}_j ,\\
\Lambda^{\psi_\text{h}}_i &=& \epsilon_{ijk}\dot{z}_j\dot{R}_k + \frac{1}{2}\ddot{z}_i\dot{z}_j\dot{z}_j - \frac{3}{2}\dot{z}_i\ddot{z}_j\dot{z}_j - \dot{\alpha}_\text{c}\ddot{z}_i + \ddot{\alpha}_\text{c}\dot{z}_i ,\\
\Lambda^{\psi_\text{h}}_{jk} &=&
-\frac{1}{2}\ddot{z}_{<j}\ddot{z}_{k>} + \dot{z}_{<j}\dddot{z}_{k>},
\end{eqnarray}
\end{subequations}
\end{widetext}
with all the other $\Lambda^{\psi_{\text{h}}}_L$ being zero.

Note that the expression (\ref{newpnM}) for the transformed moment
$\barpnM_L$ depends on the transformed moment $\dot{\bar{\mu}}_L$.
However, using the transformation law
(\ref{eq:mutransform}) for $\mu_L$ we can write the expression (\ref{newpnM})
entirely in terms of the untransformed moments, and the dependence on
$\mu_L$ cancels out.  Similarly the expression (\ref{newpnG}) for the
transformed moment $\barpnG_L$ depends on the transformed moment
${\dot {\bar \nu}}_L$, but that dependence can be eliminated using the
transformation law (\ref{eq:nutransform}) for $\nu_L$.

Finally we note that the left hand sides of Eqs.\ (\ref{newM}),
(\ref{newG}), (\ref{newZ}), (\ref{newY}),
(\ref{eq:Stransform}) -- (\ref{newpnM}) and (\ref{newpnG}) are
functions of the new time coordinate ${\bar t}$.
The right hand sides are expressed as functions
of ${\bar t}$ by evaluating the untransformed (unbarred) moments,
which are functions of $t$, at $t={\bar t}$.  This replacement of $t$
by ${\bar t}$ in the arguments of the untransformed moments is implicit in
the transformation laws (\ref{transfA}) -- (\ref{transfC}) for the
potentials\footnote{Note in particular that we do {\it not} evaluate the
untransformed moments at $t = t({\bar t})$, where $t({\bar t})$ is the
function $t({\bar t},{\bar x}^j)$ of the coordinate transformation
(\protect{\ref{coordinatetransformation}}) evaluated at ${\bar x}^j=0$.   The
corresponding correction terms have already been included in the
derivation of Eqs.\ (\protect{\ref{transfA}}) -- (\protect{\ref{transfC}})}.

\subsection{Specialization to body-adapted gauge}
\label{sec:bodyadapted}

The construction in Sec.\ \ref{sec:pnsolns} above defined the mass
multipole moments
$\nM_L(t)$, $\pnM_L(t)$, current multipole moments $S_L(t)$,
gravitoelectric tidal moments $\nG_L(t)$, $\pnG_L(t)$ and
gravitomagnetic tidal moments $H_L(t)$ that are associated with a
given body and with a given choice of coordinate system.
In this section we discuss how to obtain a unique set of multipole and
tidal moments associated with a given body by specializing to a
coordinate system that is adapted to the body in a certain way.  The
particular coordinate system defined here is relatively well known;
see for example Ref.\ \cite{dsxI}.  We shall call it the
``body-adapted gauge''.

We start by reviewing the well-known construction at Newtonian order.
From the transformation laws (\ref{eq:dipoletransformnewtonian}) and
(\ref{newG}) for $\nM_i$ and $\nG_L$, we see that
the freedom carried by the Newtonian-order worldline $\bm{z}({\bar
t})$ could be used in either of two ways.  One can set to zero
either the mass dipole moment $\barnM_i({\bar t})$ or the $l=1$
gravitoelectric tidal moment ${\bar G}_i({\bar t})$.
The second choice is not very useful, as it makes the worldline of the
origin of spatial coordinates follow the tidal field instead of following the
body.  The first choice is the conventional and useful choice;
choosing $z_i(\bar{t}) = \nM_i(\bar{t})/\nM$ achieves
\begin{equation}
\barnM_i = 0
\label{eq:bc1}
\end{equation}
and the coordinates are then mass-centered to Newtonian accuracy.
Normally such mass-centering would mean that the origin of coordinates
coincides with
the body's center of mass.  In the present context, however, this
conclusion is not valid, since the origin of coordinates is in the
strong-field region $r<r_-$ of space where the Newtonian equations are
not necessarily valid.  Moreover the definition of the center-of-mass
in the present context is somewhat subtle; we defer
to Sec.\ \ref{sec:configvars} the discussion of this definition.

The second gauge specialization we make at Newtonian order is
to set
\begin{equation}
\barnG =0
\label{eq:bc2}
\end{equation}
using the choice
\begin{equation}
\alpha_\text{c}({\bar t}) = \int d\bar{t}\left[\frac{1}{2}\dot{z}_j\dot{z}_j +
  \sum_{l=0}^\infty\nG_Lz^L\right];
\label{eq:alphac_choice}
\end{equation}
cf. Eqs.\ (\ref{newG}) and (\ref{lambdaphizero}) above. In Newtonian physics,
this choice simply corresponds to adjusting the zero of the gravitational
potential, which does not affect the dynamics.

We now turn to a discussion of the gauge specialization at
post-Newtonian order.  Without loss of generality we can assume that
we have already achieved the Newtonian conditions $\nM_i = \nG =0$,
and so we can specialize to the purely post-Newtonian subgroup of the
coordinate transformations
which is characterized by $z^i({\bar t}) =
\alpha_{\text{c}}({\bar t})=0$.  That subgroup is parameterized by the
functions $h^i_\text{c}({\bar t})$, $R^k({\bar t})$ and
by the STF tensors $\lambda_L({\bar t})$ and $\tau_L({\bar t})$ that
define the harmonic function $\beta_\text{h}$ via Eq.\
(\ref{eq:lambdaLdef}).

We first discuss the gravitoelectric sector.  For the purely
post-Newtonian coordinate transformations the transformation law
(\ref{newpnM}) for the post-Newtonian mass multipole simplifies
considerably to
\begin{equation}
\barpnM_L = \pnM_L - l \nM_{<L-1} h^{\text{c}}_{a_l>} - l \nM_{j<L-1}
\epsilon_{a_l>jk} R^k.
\label{eq:pnMtransform1}
\end{equation}
If we specialize to the mass dipole, the last term in Eq.\
(\ref{eq:pnMtransform1}) vanishes since the coordinates are
already mass-centered to Newtonian order.  This gives
\begin{equation}
\ \ \ \ \ \ \ \  \barpnM_i = \pnM_i - \nM  h^{\text{c}}_i.
\end{equation}
Therefore, as in the Newtonian case, we can use the translational freedom
encoded in the function $h^{\text{c}}_i$ to mass-center the
coordinates by making
\begin{equation}
\barpnM_i =0.
\label{eq:bc3}
\end{equation}
The transformation law (\ref{newpnG}) for the gravitoelectric tidal
moments similarly simplifies to
\begin{equation}
\barpnG_L = \pnG_L + h^\text{c}_j \nG_{jL} - l \nG_{j<L-1}
\epsilon_{a_l>jk} R^k + \Lambda^G_L,
\label{eq:pnGtransform1}
\end{equation}
where the quantities $\Lambda^G_L$ are given by
$\Lambda^G = {\dot \tau}$, $\Lambda^G_i = - {\ddot h}^\text{c}_i$,
and $\Lambda^G_L =0$ for $l \ge 2$.  By choosing $\tau({\bar t})$
suitably we can make
\begin{equation}
\barpnG =0,
\label{eq:bc4}
\end{equation}
which is analogous to the Newtonian condition (\ref{eq:bc2}).

Consider next the gravitomagnetic sector.
For purely post-Newtonian coordinate transformations, the
transformation laws (\ref{eq:Stransform}) and (\ref{eq:Htransform})
for the current multipole moments $S_L$ and gravitomagnetic tidal
moments $H_L$ simplify to
\begin{eqnarray}
\label{eq:Stransform1}
{\bar S}_L &=& S_L, \\
{\bar H}_L &=& H_L + \Lambda^H_L,
\label{eq:Htransform1}
\end{eqnarray}
where $\Lambda^H_i = - 2 {\dot R}_i$ and $\Lambda^H_L =0$ for $l \ge
2$.  Therefore we can choose to make
\begin{equation}
\bar{H}_i = 0
\label{eq:bc5}
\end{equation}
by an appropriate choice of the angular velocity ${\dot R}^k$, as
noted by DSX \cite{dsxI}.
This gauge specialization makes
the $l=0$ part of the tidal piece $\bm{B}^T$ of the gravitomagnetic field
vanish.  The resulting coordinate system slowly rotates relative to distant
stars, in such a way that the
leading order Coriolis acceleration due to the tidal gravitomagnetic
field is effaced.

At this stage, the remaining coordinate freedom is parameterized by
the STF tensors $\lambda_L$ for $l \ge 0$ and $\tau_L$ for $l \ge 1$,
which appear in
the formula (\ref{eq:lambdaLdef}) for the
harmonic function $\beta_\text{h}$.
Note that these tensors do not enter into the transformation laws
(\ref{eq:pnMtransform1}), (\ref{eq:pnGtransform1}),
(\ref{eq:Stransform1}) and (\ref{eq:Htransform1}) for the moments
$\pnM_L$, $\pnG_L$, $S_L$ and $H_L$
\footnote{Achieving this gauge invariance was the reason for picking
the particular choices of parameterization of Eqs.\
(\protect{\ref{eq:pnMdef}}) and (\protect{\ref{eq:pnGdef}}) above.}.
Therefore the values of these
moments will be the same in all coordinate systems that satisfy the
conditions specified so far, Eqs.\ (\ref{eq:bc1}), (\ref{eq:bc2}),
(\ref{eq:bc3}), (\ref{eq:bc4}) and (\ref{eq:bc5}).  In other words,
these moments (as well as the Newtonian moments $\nM_L$ and $\nG_L$) are
already uniquely defined by the conditions we have
specified so far.  Nevertheless, it is useful for some purposes to fix
the remaining gauge freedom.  To do this we consider the gauge moments
$\mu_L$ and $\nu_L$.  For purely post-Newtonian transformations the
transformation laws (\ref{eq:mutransform}) and (\ref{eq:nutransform})
for these moments reduce to
\begin{eqnarray}
\label{eq:mutransform1}
{\bar \mu}_L &=& \mu_L + \frac{(l+1)(2l+3)}{2l+1} \lambda_L, \\
{\bar \nu}_L &=& \nu_L - \tau_L + \Lambda^\nu_L,\ \ \ \ l \ge 1,
\label{eq:nutransform1}
\end{eqnarray}
where $\Lambda^\nu_i = - {\dot h}^\text{c}_i$ and $\Lambda^\nu_L =0$
for $l \ge 2$.  It follows that we can make
\begin{equation}
{\bar \nu}_L =0
\label{eq:bc6}
\end{equation}
by choosing $\tau_L$ suitably, for all $l \ge 1$.  Similarly we can
make
\begin{equation}
{\bar \mu}_L=0
\label{eq:bc7}
\end{equation}
by choosing $\lambda_L$ suitably, for all $l \ge 0$.
We shall call the unique coordinate system\footnote{Up to constant
displacements in time and time-independent spatial
rotations, cf. Sec.\ \ref{sec:gaugefreedom} above.} that
achieves all of the
conditions (\ref{eq:bc1}), (\ref{eq:bc2}),
(\ref{eq:bc3}), (\ref{eq:bc4}), (\ref{eq:bc5}), (\ref{eq:bc6}) and
(\ref{eq:bc7}) the {\it body-adapted} coordinate system\footnote{This
gauge is called the skeletonized-body harmonic gauge by DSX
\protect{\cite{dsxI}}.}.

To summarize, we have demonstrated that there is enough coordinate
freedom to accomplish the following:
\begin{enumerate}
\item{Mass-center the coordinate system to post-1-Newtonian accuracy by setting
  $\nM_i = \pnM_i = 0$.}
\item{Set to zero the $l=0$ pieces $\nG$, $\pnG$ of the
  tidal pieces of the potentials $\Phi$ and $\psi$}
\item{Set to zero the gravitomagnetic tidal moment $H_i$.}
\item{Set to zero the all the gauge moments $\nu_L$ and $\mu_L$.}
\end{enumerate}
The role of each free function in the coordinate transformation
(\ref{eq:generalcoordtransform}) in the derivation of the above
conditions is recapitulated in table \ref{gaugefreedom}.

We note that our definition of the body-frame coordinates or local
asymptotic rest frame differs slightly from that of Thorne and Hartle
\cite{thornehartle}.   We require that the mass dipole $M_i$ should vanish
for all time, whereas Thorne and Hartle instead require
the gravitoelectric tidal moment $G_i$ should vanish for all time.  At a
given initial instant, they also demand that $M_i$ and ${\dot M}_i$
vanish, while ${\ddot M}_i$ can be nonvanishing.

We next consider the special case treated by DSX, where the
post-1-Newtonian field
equations (\ref{WfieldeqA}) -- (\ref{WfieldeqC}) are assumed to hold
all the way down to $r=0$.  In this case the gauge freedom is somewhat
reduced, since the function $\beta_\text{h}$ must now be both harmonic
and smooth for all $r < r_+$.  This requirement eliminates the terms
parameterized
by $\lambda_L$ in Eq.\ (\ref{eq:lambdaLdef}), as those terms diverge
at $r=0$.  Therefore we no longer have sufficient gauge freedom to set to
zero the gauge-moments $\mu_L$ via Eq.\ (\ref{eq:mutransform1}).  We
still obtain a unique coordinate system by imposing the remaining
requirements (\ref{eq:bc1}), (\ref{eq:bc2}),
(\ref{eq:bc3}), (\ref{eq:bc4}), (\ref{eq:bc5}) and (\ref{eq:bc6}), but
the $\mu_L$ moments will now in general be nonvanishing.
This modified version of the body-adapted gauge will be important in
Sec.\ \ref{eom} below.\\

\begin{table}
\caption{\label{gaugefreedom} Free functions in the coordinate
transformation and their role in defining the body-adapted harmonic
gauge.}
\begin{ruledtabular}
\begin{tabular}{cc}
Free function & Role \\
\hline \\
$z^i(\bar{t})$ & sets ${\,}^{\scriptscriptstyle \text{n}}\!\bar{M}_i = 0$ \\
$\alpha_{\text{c}}(\bar{t})$ & sets ${\,}^{\scriptscriptstyle \text{n}}\!\bar{G} = 0$ \\
$h^i_{\text{c}}(\bar{t})$ & sets ${\,}^{\scriptscriptstyle \text{pn}}\!\bar{M}_i = 0$ \\
$\tau(\bar{t})$ & sets ${\,}^{\scriptscriptstyle \text{pn}}\!\bar{G} =
0$  \\
$R_k(\bar{t})$ & sets $\bar{H}_i = 0$ \\
$\tau_{L}(\bar{t}),\ \ \ l\ge1$ & sets ${\bar \nu}_L = 0$  \\
$\lambda_{L}(\bar{t}),\ \ \ l\ge0$ & sets ${\bar \mu}_L = 0$
\end{tabular}
\end{ruledtabular}
\end{table}

\subsection{Definition of multipole and tidal moments about a given worldline}
\label{sec:moments1}

In Sec.\ \ref{sec:pnsolns} above we defined the multipole and tidal
moments of a body associated with a given coordinate system.  Those
moments can be interpreted as being moments about the origin
$\bm{x}=0$ of that coordinate system.  In this section, we generalize
that definition to define tidal and multipole moments about a
specified worldline\footnote{By a worldline we mean simply
a function $x^i = z^i(t)$ which transforms appropriately under the
group (\protect{\ref{eq:generalcoordtransform}}) of coordinate
transformations.
If $z^i(t)$ lies outside of the domain of definition of the
coordinates then it does not correspond to an actual worldline in
spacetime.  See Sec.\ \protect{\ref{sec:configvars}} below for further
discussion of
this point.} $x^i = z^i(t)$ and associated with a given
coordinate system $(t,x^i)$.  This more general definition will be
used in Sec.\ \ref{sec:globalcoords} below.

We assume that the vacuum
post-1-Newtonian field equations are satisfied in a region of the form
$ r_- < | \bm{x} - \bm{z}(t) | < r_+$.
Our definition of the moments is given by the following multipole
expansions of the potentials $\Phi$, $\zeta^i$ and $\psi$ in this region:
\begin{widetext}
\begin{subequations}
\begin{eqnarray}\label{basicmodelA0}
\Phi(t,x^j) &=& \sum_{l=0}^\infty
\frac{(-1)^{l+1}}{l!}\nM_L(t)\partial_L\frac{1}{|\bm{x} - \bm{z}(t)|} -
\frac{1}{l!}\nG_L(t) [x - z(t)]^L ,\\\nn
\psi(t,x^j) &=& \sum_{l=0}^\infty
\left(\frac{(-1)^{l+1}}{l!}\left\{ \pnM_L(t)
\partial_L\frac{1}{|\bm{x} - \bm{z}(t)|}
+ \frac{(2l+1)}{(l+1)(2l+3)} \frac{\partial}{\partial t} \left[ \mu_L(t)
\partial_L\frac{1}{|\bm{x} - \bm{z}(t)|} \right]
\right\}
\right.
\nonumber \\ \nn \mbox{} &&
\left.
+ \frac{(-1)^{l+1}}{l!} \frac{\partial^2}{\partial t^2} \left[ \nM_L
(t)\partial_L \frac{|\bm{x} - \bm{z}(t)|}{2} \right]
- \frac{1}{l!} \pnG_L(t) [x-z(t)]^L
\right. \\
\label{basicmodelB0}
& & \left.
+ \frac{1}{l!}
\frac{\partial}{\partial t} \left\{ \nu_{L}(t) [x - z(t)]^L \right\}
-\frac{1}{2 l! (2l + 3)}
\frac{\partial^2}{\partial t^2} \left\{ \nG_L(t) [x-z(t)]^L |\bm{x}-{\bm
  z}(t)|^2 \right\}
\right), \\
\label{reducedzeta0}
\zeta_i(t,x^j) &=& \sum_{l=0}^\infty
\frac{(-1)^{l+1}}{l!}Z_{iL}\partial_L\frac{1}{|\bm{x} -
\bm{z}(t)|} - \frac{1}{l!}Y_{iL}[x - z(t)]^L,
\end{eqnarray}
where
\begin{equation}
Z_{iL} =
\frac{4}{l+1}{\,}^{\scriptscriptstyle
    \text{n}}\!\dot{M}_{iL} - \frac{4l}{l+1}\epsilon_{ji<a_l}S_{L-1>j}
  +  \frac{2l-1}{2l+1}\delta_{i<a_l}\mu_{L-1>}
+ 4 {\dot z}_{<i} \nM_{L>},
\label{eq:Z1def}
\end{equation}
and
\begin{equation}
Y_{iL} = \nu_{iL} +
  \frac{l}{l+1}\epsilon_{ji<a_l}H_{L-1>j}-
  \frac{4(2l-1)}{2l+1}{\,}^{\scriptscriptstyle
    \text{n}}\!\dot{G}_{<L-1} \delta_{a_l>i}
+ \frac{4(2l+1)}{2l+3} \nG_{<L} \delta_{k>i} {\dot z}^k.
\label{eq:Y1def}
\end{equation}
\end{subequations}
\end{widetext}
The expansions (\ref{basicmodelA0}) and (\ref{basicmodelB0})
have the same form as Eqs.\ (\ref{basicmodelA}) and
(\ref{basicmodelB}) except that $\bm{x}$ is replaced everywhere by
$\bm{x} - {\bm z}(t)$, and the various time derivative operators
are allowed to act on the factors of ${\bm x} - {\bm z}(t)$ as well as
on the moments.
Note that the formulae (\ref{eq:Z1def}) and
(\ref{eq:Y1def}) for $Z_{iL}$ and $Y_{iL}$ contain extra terms
involving ${\dot z}^i$ compared to the original formulae
(\ref{eq:ZiLdecompos}) and (\ref{eq:YiLdecompos}).

We now discuss the derivation of the expansions (\ref{basicmodelA0})
-- (\ref{reducedzeta0}).  First, one can verify that these expansions
satisfy the field equations (\ref{fieldeqA}) -- (\ref{fieldeqC}).
Next, inserting the
expansions (\ref{basicmodelA0}) and (\ref{reducedzeta0}) into the
gauge condition (\ref{harmonicgauge}) yields the Newtonian
conservation of mass equation (\ref{hgA}) as before, and also the
following replacements for Eqs.\ (\ref{hgB}) and (\ref{hgC}):
\begin{eqnarray}
\label{hgBnew}
Z_{<iL>} &=& \frac{4}{l+1} \ndotM_{iL} + 4 {\dot z}_{<i} \nM_{L>}
\end{eqnarray}
and
\begin{eqnarray}\label{hgCnew}
Y_{jjL} &=& -4 \ndotG_{L} + 4 {\dot z}_j \nG_{jL}.
\end{eqnarray}
If we now define the moments $S_L$, $H_L$, $\mu_L$ and $\nu_L$ in terms of
$Z_{iL}$ and $Y_{iL}$ using the
same formulae (\ref{eq:SLdef}) -- (\ref{eq:muLdef}) as before and
use the decomposition identity (\ref{reduction}), we obtain Eqs.\
(\ref{eq:Z1def}) and (\ref{eq:Y1def}).  Finally, the expression
(\ref{basicmodelB0}) for $\psi$ is chosen so that
the moments $\nM_L$, $\pnM_L$, $\nG_L$,
$\pnG_L$, $H_L$ and $S_L$ defined by these expansions are invariant
under the group (\ref{1pntimegauge}) of gauge transformations.  This
invariance
can be verified by using a parameterization of $\beta_\text{h}$ of the
form (\ref{eq:lambdaLdef}) with $\bar{\bm{x}}$ replaced by
$\bar{\bm{x}} - {\bm z}({\bar t})$.

For later computations it is useful
to expand the time derivatives in Eq.\ (\ref{basicmodelB0})
and express
results in terms of STF tensors.  This computation gives
\begin{widetext}
\begin{subequations}
\begin{eqnarray}\nn
\psi &=& \sum_{l=0}^\infty
\left\{\frac{(-1)^{l+1}}{l!}\left[N_L
\partial_L\frac{1}{|\bm{x} - \bm{z}(t)|}  + P_L
\partial_L\frac{|\bm{x} - \bm{z}(t)|} {2} \right]  -
\frac{1}{l!}\left[F_L[x - z(t)]^L +
J_L\frac{|\bm{x} - \bm{z}(t)|^2 [x -
z(t)]^L}{2(2l+3)}\right]\right\}, \\\label{eq:tidalA}
\end{eqnarray}
where
\begin{eqnarray}
N_L &=& \pnM_L + \frac{2l+1}{(l+1)(2l+3)} {\dot \mu}_L +
\frac{2l-1}{2l+1} \mu_{<L-1} {\dot z}_{a_l>} \nn \\
\mbox{} &&
+ \frac{1}{2l+3}\left[ \ddot{z}_j \nM_{jL} +
\dot{z}_j \dot{z}_j\,\nM_L +
2\dot{z}_j \ndotM_{jL} + 2l\dot{z}_j
\dot{z}_{<a_l}\nM_{L-1>j}\right],\label{eq:Ndef}
\end{eqnarray}
\begin{eqnarray}
P_L &=&
\nddotM_L +
2l\dot{z}_{<a_l} \ndotM_{L-1>} +
l(l-1)\dot{z}_{<a_l} \dot{z}_{a_{l-1}}\nM_{L-2>} +
l\ddot{z}_{<a_l}\nM_{L-1>},
\label{eq:Pdef}
\end{eqnarray}
\begin{eqnarray}
F_L &=& \pnG_L - {\dot \nu}_L + \nu_{Lj} {\dot z}^j - \frac{2l}{2l+1}
\ndotG_{<L-1} {\dot z}_{a_l>}
+ \frac{2l}{2l + 3} {\dot z}^k \nG_{k<L-1} {\dot z}_{a_l>}
\nn \\
\mbox{} && + \frac{1}{2l+3} \nG_L {\dot z}^j {\dot z}^j
- \frac{l}{2l+1} \nG_{<L-1} {\ddot z}_{a_l>},
\label{eq:Fdef}
\end{eqnarray}
and

\begin{eqnarray}
J_L &=& \nddotG_L - 2 \ndotG_{Lj} {\dot z}^j + \nG_{Ljk} {\dot z}^j
{\dot z}^k - \nG_{Lk} {\ddot z}^k.\ \ \ \
\label{eq:Jdef}
\end{eqnarray}
\end{subequations}
\end{widetext}
All of the tensors $N_L$, $P_L$, $F_L$ and $J_L$ are STF,
while the tensors $Z_{iL}$ and $Y_{iL}$ are STF only on their last $l$
indices, as before.

Finally, we note that there is no natural, unique definition of
multipole and tidal moments about a given worldline associated with a
given coordinate system.  We have chosen a particular definition, but
there are other definitions that are equally valid.  For example, if we
replace Eq.\ (\ref{eq:Z1def}) with
\begin{eqnarray}
Z_{iL} &=&
\frac{4}{l+1}{\,}^{\scriptscriptstyle
    \text{n}}\!\dot{M}_{iL} - \frac{4l}{l+1}\epsilon_{ji<a_l}S_{L-1>j}
  +  4 {\dot z}_i \nM \delta_{l0}
\nonumber \\
\mbox{} &&
+ \frac{2l-1}{2l+1}\delta_{i<a_l}\mu_{L-1>} + 4 \nM_{i<L-1} {\dot z}_{a_l>},
\label{eq:Z1defalternative}
\end{eqnarray}
then field equations and harmonic gauge condition are still satisfied.
However, while the moments defined by this equation are still
invariant under the group (\ref{1pntimegauge}) of gauge
transformations, they differ from the moments defined by Eq.\ (\ref{eq:Z1def}).
By contrast, the definitions of multipole and tidal
moments about the origin of coordinates discussed in Sec.\ \ref{sec:pnsolns}
above are essentially unique.
However, this lack of uniqueness will be unimportant for our
purposes, since multipole and tidal moments about a given
worldline will appear only in intermediate steps in our computations and
not in our final results.

\subsection{Comparison with other definitions of multipole and tidal
moments in the literature}
\label{sec:comparison}

As we have discussed, our multipole moments coincide with those of
Blanchet and Damour \cite{bd} and of DSX \cite{dsxI,dsxII,dsxIII} for
weakly self-gravitating bodies.  For isolated systems these moments agree
to leading order with the unique moments that can be defined for
stationary systems in general relativity \cite{stationary}, and with
the asymptotic radiative multipole moments of Thorne \cite{thorne}, in
the appropriate limits, as noted by DSX \cite{dsxI}.

For non-isolated systems, there is another notion of mass and current multipole
moments, defined in terms of the metric in a buffer region
surrounding a body, due to Thorne and Hartle \cite{thornehartle}.
These moments are defined in full general relativity by the same type
of surface integrals as used here [cf.\ Appendix \ref{sec:welldefined}],
but applied to the full metric rather than to the post-Newtonian
potentials.  These moments depend on the choice of coordinate system
used to evaluate the surface integrals, but the magnitude of the
resulting ambiguities can be estimated and in many applications are
small enough to be unimportant \cite{thornehartle}.  By contrast, the
multipole moments used here are defined only the context of
post-1-Newtonian theory, but are unique.

Our tidal moments also coincide with those of DSX for weakly
self-gravitating bodies.  They also appear to coincide with the tidal
moments defined by Suen in the context of stationary systems in full
general relativity \cite{Suen}.

\section{Post-1-Newtonian laws of motion: a single body}
\label{eom}

\subsection{Overview}
\label{sec:eomoverview}

As discussed in the introduction, the derivation of equations of motion for
several interacting bodies can be divided into two pieces:
\begin{itemize}
\item A derivation of a formula\footnote{DSX call this formula the
  ``law of motion'' \protect{\cite{dsxI}}.}, for any given body, of the second time
derivative of its mass dipole moment in terms of its other multipole
and tidal moments and their time derivatives.
\item A derivation of the relation between the tidal moments acting on
each body and the multipole moments and center-of-mass worldlines of
all the other bodies.
\end{itemize}
In this section we will carry out the first of these tasks.  The
second task will be the subject of Secs.\ \ref{manybody} and \ref{explicit}
below.

We start by describing the assumptions and the result.
As in Sec.\ \ref{sec:pnsolns} above, we assume
existence of a local coordinate system $(t,x^j)$ with following
properties:
\begin{itemize}
\item The range of the coordinates contains the product of the open ball
\begin{equation}
| {\bm{x}} | < r_+,
\label{eq:domain00}
\end{equation}
where $r_+$ is some radius, with some open interval $(t_0,t_1)$ of time.
\item On the spatial region ${\cal D}$ given by $r_- < r < r_+$, for
some $r_->0$, the coordinates are conformally Cartesian and harmonic.
Also the vacuum Einstein
equations are valid on ${\cal D}$ for a
one-parameter family of metrics of the form of Eq.\
(\ref{eq:metricP2N}) below.
Essentially this says that the
Newtonian, post-1-Newtonian and post-2-Newtonian vacuum field
equations are valid on ${\cal D}$.
\end{itemize}
The reason for imposing the post-2-Newtonian field equations in
addition to the post-1-Newtonian field equations is discussed below.

These assumptions allow us to define the multipole moments
$\nM_L(t)$, $\pnM_L(t)$, $S_L(t)$ which characterize the sources in
the region $r<r_-$, as well as the tidal moments $\nG_L(t)$,
$\pnG_L(t)$ and $H_L(t)$, as discussed in the
previous section.  They also imply
some formulae relating time derivatives of the multipole moments.
At Newtonian order these formulae are
\begin{subequations}
\begin{eqnarray}
\label{eq:masscons}
\ndotM(t) &=&0,\\
{\,}^{\scriptscriptstyle \text{n}}\!\ddot{M}_i(t) &=& \sum_{l=0}^\infty
\frac{1}{l!}\nM_L(t) \nG_{iL}(t).
\label{newtonianlom}
\end{eqnarray}
\end{subequations}
Here the first formula is just the Newtonian conservation of mass
derived earlier [Eq.\ (\ref{hgA}) above].  The second formula
equates the acceleration of the center of mass to the
acceleration produced by the external tidal moments coupling with the
multipole moments of the system.  This formula is usually derived by
integrating the Newtonian stress-energy conservation equations over
the interior of the body.  Here, however, we do not assume the
validity of the Newtonian equations in the interior.
Equation (\ref{newtonianlom}) contains all the information one needs
in order to derive the explicit coupled equations of motion for the center of
mass worldlines of each body in an $N$-body system.  We will describe
this derivation later in Sec.\ \ref{explicit}.

At post-1-Newtonian order, the formulae analogous to Eqs.\
(\ref{eq:masscons}) and
(\ref{newtonianlom}) are
\begin{widetext}
\begin{subequations}
\begin{eqnarray}
\label{eq:mainresult0}
\pndotM &=& - \sum_{l=0}^\infty \frac{1}{l!} \left[ (l+1) \nM_L
  \ndotG_L + l \ndotM_L \nG_L \right], \label{eq:pnmasscons} \\
\pnddotM_i&=&
\sum_{l=0}^\infty\frac{1}{l!}\left[\pnM_L\nG_{iL} +
  \nM_L\pnG_{iL} + \frac{l}{l+1}S_L H_{iL} +
  \frac{1}{l+2}\epsilon_{ijk}\nM_{jL}\dot{H}_{kL} +
  \frac{1}{l+1}\epsilon_{ijk}{\,}^{\scriptscriptstyle
    \text{n}}\!\dot{M}_{jL}H_{kL}
\right. \nn \\\nn  & & \left.
- \frac{4(l+1)}{(l+2)^2}\epsilon_{ijk}S_{jL}{\,}^{\scriptscriptstyle
    \text{n}}\!\dot{G}_{kL}
- \frac{4}{l+2}\epsilon_{ijk}\dot{S}_{jL}\nG_{kL}
\right. \nn \\  & & \left.
  -\frac{2l^3 +7l^2 +15l +
    6}{(l+1)(2l+3)}\nM_{iL}{\,}^{\scriptscriptstyle
    \text{n}}\!\ddot{G}_L - \frac{2l^3 + 5l^2 +12l +
    5}{(l+1)^2}{\,}^{\scriptscriptstyle
    \text{n}}\!\dot{M}_{iL}{\,}^{\scriptscriptstyle
    \text{n}}\!\dot{G}_L - \frac{l^2 + l +
    4}{l+1}{\,}^{\scriptscriptstyle \text{n}}\!\ddot{M}_{iL}\nG_L
  \right],
\label{mainresult} \\
{\dot S}_i &=& \sum_{l=0}^\infty \frac{1}{l!} \epsilon_{ijk} \nM_{jL} \nG_{kL}.
\label{eq:spinresult}
\end{eqnarray}
\end{subequations}
\end{widetext}
For bodies in which the post-Newtonian field equations are valid
everywhere, DSX derived the formulae (\ref{eq:mainresult0}) --
(\ref{eq:spinresult}) by using Newtonian\footnote{As is well known,
the derivation of the spin evolution equation (\ref{eq:spinresult})
requires only Newtonian-order stress-energy conservation.}
and post-Newtonian stress-energy
conservation in the interior of the body \cite{dsxII}.  In this
section we will derive Eq.\ (\ref{mainresult}) from the assumptions listed
above.  The formulae (\ref{eq:mainresult0}) and (\ref{eq:spinresult})
giving the time evolution of the mass monopole and the spin
will be derived from the same assumptions in the second paper in this
series \cite{Racine}.

Note that the formulae (\ref{eq:mainresult0}) -- (\ref{eq:spinresult}) are
valid for all coordinate systems satisfying the assumptions listed above, not
just for the body-adapted coordinate system discussed in Sec.\
\ref{sec:bodyadapted} above.

\subsection{Method of derivation}
\label{sec:method_of_derivation}

We now turn to a description of the surface-integral method of
derivation that we use.   For this description we return, temporarily,
to the context of the full, nonlinear equations of general
relativity.  The method is well-known and is described in Landau and Lifshitz
\cite{landaulifshitz} and in Misner, Thorne and Wheeler \cite{mtw}.
It has been previously applied to the derivation of laws of motion by
Thorne and Hartle \cite{thornehartle}.

The method starts by fixing a coordinate system
$x^\mu = (t,x^j)$, and by writing Einstein's equations in that
coordinate system
in a form involving pseudotensors and partial derivatives as
\begin{equation}
\label{Einsteinsequations}
{\cal H}^{\mu\alpha\nu\beta}_{\,\,\,\,\,\,\,\,\,\,\,\;,\alpha\beta} = 16\pi
\left[(-g)T^{\mu\nu} + {\cal T}^{\mu\nu}\right].
\end{equation}
Here $g = {\rm det} \, g_{\mu\nu}$ and the tensor density ${\cal
  H}^{\mu\alpha\nu\beta}$ is given by
\begin{equation}\label{LL}
{\cal H}^{\mu\alpha\nu\beta} =
\mathfrak{g}^{\mu\nu}\mathfrak{g}^{\alpha\beta} -
\mathfrak{g}^{\alpha\nu}\mathfrak{g}^{\beta\mu}
\end{equation}
where
\begin{equation}
\mathfrak{g}^{\alpha\beta} = \sqrt{-g}g^{\alpha\beta}.
\label{gothicmetricdef}
\end{equation}
In Eq.\ (\ref{Einsteinsequations}), $T^{\mu\nu}$ is the stress-energy
tensor and the pseudotensor ${\cal T}^{\mu\nu}$ is given by
\begin{eqnarray}\nn
&&{\cal T}^{\alpha\beta} = \frac{1}{16\pi}\bigg[
\mathfrak{g}^{\alpha\beta}_{\;\;\,,\lambda}
\mathfrak{g}^{\lambda\mu}_{\;\;\,,\mu}
-\mathfrak{g}^{\alpha\lambda}_{\;\;\,,\lambda}
\mathfrak{g}^{\beta\mu}_{\;\;\,,\mu}
+\frac{1}{2}g^{\alpha\beta}g_{\lambda\mu}
\mathfrak{g}^{\lambda\nu}_{\;\;\,,\rho}\mathfrak{g}^{\rho\mu}_{\;\;\,,\nu}
\nn \\ &&
 +g_{\lambda\mu}g^{\nu\rho}\mathfrak{g}^{\alpha\lambda}_{\;\;\,,\nu}
\mathfrak{g}^{\beta\mu}_{\;\;\,,\rho}- g_{\mu\nu}\left(g^{\alpha\lambda}
\mathfrak{g}^{\beta\nu}_{\;\;\,,\rho}\mathfrak{g}^{\mu\rho}_{\;\;\,,\lambda}
+ g^{\beta\lambda}
\mathfrak{g}^{\alpha\nu}_{\;\;\,,\rho}
\mathfrak{g}^{\mu\rho}_{\;\;\,,\lambda}\right)
\nn \\
& & +  \frac{1}{8}\left(2g^{\alpha\lambda}g^{\beta\mu} -
g^{\alpha\beta}g^{\lambda\mu}\right)\left(2g_{\nu\rho}g_{\sigma\tau} -
g_{\rho\sigma}g_{\nu\tau}\right)\mathfrak{g}^{\nu\tau}_{\;\;\,,\lambda}
\mathfrak{g}^{\rho\sigma}_{\;\;\,,\mu}
\bigg] .\nn \\
 \label{pseudotensor}
\end{eqnarray}
This formulation of Einstein's equations is due to Landau and Lifshitz
\cite{landaulifshitz}.

Next,
in a surface of constant $t$ within the domain of the coordinates,
we consider a three dimensional region $V$
whose topology is that of the interior of a sphere, and
whose boundary $\Sigma = \partial V$ has spherical topology.
We assume that the stress energy tensor $T^{\mu\nu}$ vanishes in a
neighborhood of $\Sigma$.  We define a quantity $P^i_\Sigma$ associated with
$\Sigma$ and with the choice of coordinate system
by
\begin{equation}\label{momentum}
P^i_{\Sigma}(t) \equiv \frac{1}{16\pi}\oint_{\Sigma} {\cal H}^{i\alpha
  0j}_{\,\,\,\,\,\,\,\,\,\,\,,\alpha} d^2\Sigma_j.
\end{equation}
Here $d^2\Sigma_j$ is the natural surface element determined by the
flat metric $(dx^1)^2 + (dx^2)^2 + (dx^3)^2$.
Using the flat-space Gauss's theorem, the field equation
(\ref{Einsteinsequations}), the symmetry properties of ${\cal
  H}^{\mu\alpha\nu\beta}$, and the assumption (\ref{eq:domain00}),
the definition (\ref{momentum}) can also be written as
\begin{equation}
\label{momentum1}
P^i_{\Sigma}(t) \equiv \int_V d^3x \left[ (-g) T^{0i} + {\cal
    T}^{0i} \right].
\end{equation}
Here $d^3x$ is the volume element associated with the
flat metric $(dx^1)^2 + (dx^2)^2 + (dx^3)^2$.
As is well known, for an asymptotically flat spacetime the quantity
$P^i_\Sigma$ yields the ADM 3-momentum in the limit where $\Sigma$
tends to infinity.  For finite $\Sigma$, however, $P^i_\Sigma$ does
not have any invariant physical or geometric meaning.
Nevertheless, we can still use this quantity as an intermediate tool
in mathematical calculations in deriving relations between quantities
whose transformation properties and meaning are well understood, such
as the multipole and tidal moments discussed in the previous section.
For convenience, we will refer to $P^i_\Sigma$ as the ``enclosed
3-momentum'' in $\Sigma$, even though it is not an invariant quantity.

It follows from the formula (\ref{momentum1}) and from the form $\left[(-g)
T^{\mu\nu} + {\cal T}^{\mu\nu}\right]_{,\nu}=0$ of stress-energy
conservation that the time derivative of the enclosed 3-momentum is
\footnote{Note that the derivation of this conservation law does not
  require the
assumption that Einstein's equations are valid for $r<r_-$, since we
can take the point of view that the stress-tensor $T^{\mu\nu}$ is
defined by Eq.\ (\protect{\ref{Einsteinsequations}}). In other words,
a different theory of gravity could be applicable in the strong field
region $r<r_-$, with the correction to the field equations being
incorporated into
the definition of $T^{\mu\nu}$.  Our application of the conservation
law (\protect{\ref{conservationlaw}}) to derive the equation of motion
(\protect{\ref{mainresult}}) will therefore apply to any
theory of gravity for which the vacuum field equations coincide with
  those of general relativity.}
\begin{equation}\label{conservationlaw}
\dot{P}^i_\Sigma = -\oint_{\Sigma} {\cal T}^{ij}d^2\Sigma_j.
\end{equation}
The core of the surface integral method is to compute both sides of
Eq.\ (\ref{conservationlaw}) explicitly.   Namely, we compute the
left-hand side by performing the surface integral (\ref{momentum}) and
taking a time derivative of the result, and we compute the right-hand
side using Eq.\ (\ref{pseudotensor}). These surface integrals are
evaluated using the general solutions (\ref{basicmodelA}) --
(\ref{reducedzeta}) to the vacuum post-1-Newtonian field equations.
The dependence on the surface $\Sigma$ drops out and one obtains in this way
relations between the moments and their time derivatives that reduce
to Eqs.\ (\ref{newtonianlom}) and (\ref{mainresult}) at Newtonian
and post-Newtonian order, respectively.

\subsection{Post-2-Newtonian field equations}
\label{sec:post2}

In applying the surface-integral method to derive the laws of motion to some
post-Newtonian order $m$, one needs to compute some pieces of the
metric to post-Newtonian order $m+1$.  This was emphasized in the
original work of Einstein, Infeld and Hoffmann \cite{eih}.
For example, the Newtonian mass
currents, whose conservation law can be used to derive laws of motion at Newtonian order,
are the source for the gravitomagnetic potential $\bm{\zeta}$
[cf. Eq.\ (\ref{fieldeqC})]. Therefore, if one wants to avoid dealing
with the matter distribution itself and to use instead only the far
field metric to derive Newtonian laws of motion, one needs
to use the post-1-Newtonian gravitomagnetic potential.
Note however that knowledge of the post-1-Newtonian scalar potential
$\psi$ is not required for this purpose.

Therefore, for our goal of computing the post-1-Newtonian laws of
motion, we need to compute some pieces of the metric to
post-2-Newtonian order.  In this subsection we derive the harmonic gauge vacuum field equations satisfied by these post-2-Newtonian fields.

It turns out that we need the post-2-Newtonian corrections only
to $g_{ij}$ and to $g_{0i}$.  We can
therefore parameterize the metric as the post-1-Newtonian metric
(\ref{metric}) together with the appropriate correction terms:
\begin{eqnarray}\nn
ds^2 &=& - \frac{1}{\varepsilon^2}\left[1 + 2\varepsilon^2\Phi +
2\varepsilon^4(\Phi^2 + \psi) + O(\varepsilon^6)\right]dt^2 \\\nn
& & + \left[2\varepsilon^2\zeta_i + 2 \varepsilon^4 ( 2 \Phi \zeta_i +
\xi_i) + O(\varepsilon^5)\right]dx^i dt
\\\label{eq:metricP2N}
& & + \big[\delta_{ij} -2\varepsilon^2\Phi \delta_{ij} + \varepsilon^4 [ (2
\Phi^2 - 2 \psi  + \chi_{kk}) \delta_{ij} -
\chi_{ij}]
\nn \\ && + O(\varepsilon^5)\big]dx^idx^j.
\end{eqnarray}
Here the post-2-Newtonian fields are the vector field $\xi_i$ and the
spatial symmetric tensor $\chi_{ij}$.  We have chosen the particular
parameterization of the post-2-Newtonian pieces of the metric
(\ref{eq:metricP2N}) to simplify the gothic metric and the field equations.
From the definition (\ref{gothicmetricdef}) of the gothic metric we obtain
\begin{eqnarray}
\mathfrak{g}^{00} &=&
-\varepsilon + 4\varepsilon^3\Phi - \varepsilon^5\left(8\Phi^2 - 4\psi +
  \chi_{kk}\right) + O(\varepsilon^6), \nn \\
\mathfrak{g}^{0i} &=&
\varepsilon^3\zeta^i +  \varepsilon^5 \xi^i  + O(\varepsilon^6), \nn\\
\mathfrak{g}^{ij} &=&
\frac{1}{\varepsilon}\delta^{ij}
  +\varepsilon^3 \chi_{ij} + O(\varepsilon^4).
\label{gothicmetric}
\end{eqnarray}
Inserting these expressions into the harmonic gauge condition
(\ref{Harmonicgauge}) yields the equations
\begin{equation}\label{gaugeA}
\dot{\zeta}^i + \chi^{ij}_{\ \ \, ,j}=0
\end{equation}
and
\begin{equation}\label{gaugeB}
\xi^i_{\ \,,i} - \dot{\chi}_{kk} + 4\dot{\psi} -16\Phi\dot{\Phi} =0.
\end{equation}
The harmonic-gauge vacuum field equations for the fields $\xi^i$ and
$\chi^{ij}$ can be derived by substituting the expansion
(\ref{gothicmetric}) of the gothic metric into Eq.\
(\ref{Einsteinsequations}). The results are
\begin{equation}
\nabla^2 \xi^i = \ddot{\zeta}^i + 12\dot{\Phi} \partial_i\Phi +
8 \partial_{[i}\zeta_{k]} \partial_k\Phi
\label{2pnfieldeqA}
\end{equation}
and
\begin{equation}
\nabla^2 \chi_{ij} = 4 \partial_i\Phi\partial_j\Phi -
2 \delta_{ij} \partial_k\Phi\partial_k\Phi.
\label{2pnfieldeqB}
\end{equation}

We next compute the expansions of the enclosed momentum $P_\Sigma^i$
and the spatial components ${\cal T}^{ij}$ of the Landau-Lifshitz
pseudotensor that appear in conservation law (\ref{conservationlaw}).
Substituting the gothic metric components (\ref{gothicmetric}) into
the definition (\ref{momentum}) of enclosed momentum yields
\begin{equation}
P^i_\Sigma = \varepsilon^2 \nP_\Sigma^i + \varepsilon^4 \pnP^i_\Sigma +
O(\varepsilon^5),
\label{expandedmomentum}
\end{equation}
where the Newtonian piece is
\begin{equation}
\nP^i_\Sigma = \frac{1}{16\pi}\oint_{\Sigma} \partial_j \zeta^i
d^2\Sigma_j
\label{eq:momentumN}
\end{equation}
and the post-Newtonian piece is
\begin{equation}
{\ \ \ \ \ \ \ \ \ } \pnP^i_\Sigma =
\frac{1}{16\pi}\oint_{\Sigma} \left[ \partial_j \xi^i + {\dot
    \chi}_{ij} \right] d^2\Sigma_j.
\label{PiIV}
\end{equation}
The corresponding expansion of ${\cal T}^{ij}$ is,
from Eqs.\ (\ref{pseudotensor}),
(\ref{eq:metricP2N}) and (\ref{gothicmetric}),
\begin{equation}
{\cal T}^{ij} = \varepsilon^2 \ncalT^{ij} + \varepsilon^4 \pncalT^{ij}
+ O(\varepsilon^5),
\label{eq:calTexpand}
\end{equation}
where the Newtonian piece
is
\begin{equation}
{\ \ \ \ \ \ \ \ } \ncalT^{ij} = \frac{1}{4\pi}\left(\partial_i\Phi\partial_j\Phi -
\frac{1}{2}\delta_{ij}\partial_k\Phi\partial_k\Phi \right)
\label{eq:LLN}
\end{equation}
and the post-Newtonian piece is
\begin{eqnarray}
\nn
\pncalT^{ij} &=&
\frac{1}{4\pi} \bigg[ \partial_{[i}\zeta_{k]}\partial_{[k}\zeta_{j]} +
  2\partial_{(i}\Phi\dot{\zeta}_{j)} +
  2\partial_{(i}\Phi\partial_{j)}\psi \nn \\
&&  - \frac{1}{2}\delta_{ij}
\bigg(\frac{1}{2}\partial_{[l}\zeta_{k]}\partial_{[k}\zeta_{l]}
+ 2\partial_k\Phi\dot{\zeta}_k
\nn \\ &&
 + 2\partial_k\Phi\partial_k\psi
+ 3\dot{\Phi}^2\bigg)\bigg].
\label{eq:LLP2N}
\end{eqnarray}

\subsection{Newtonian order derivation}
\label{sec:nod}

To illustrate the method of computation, we first
derive the Newtonian law of motion (\ref{newtonianlom}) from the
conservation law (\ref{conservationlaw}).  A similar derivation has
been given by Futamase \cite{Futamasenew}.
We choose the 2-surface
$\Sigma$ to be the coordinate sphere $r = R$, for some $R$ with $r_- <
R < r_+$, and we henceforth drop the
subscript $\Sigma$ in $P^i_\Sigma$ for simplicity.  The formula
(\ref{eq:momentumN}) then reduces to
\begin{equation}
\nP^i = \frac{R^2}{16\pi}\oint  n^j\partial_j \zeta^i(R {\bf n}) \,\,d\Omega,
\label{eq:momentumN1}
\end{equation}
where the integral is over the unit sphere and $n^j = x^j/|{\bf
x}|$.
Plugging in the general solution (\ref{basicmodelC}) for the
gravitomagnetic potential and using the identities
(\ref{eq:identity1}) and (\ref{eq:identity2}) yields
\begin{widetext}
\begin{eqnarray}\nn
\nP^i &=& \frac{R^2}{16\pi}\oint \,n^j\left[\sum_{l=0}^\infty
  \frac{(-1)^{l+1}}{l!}Z_{iL}\partial_{jL}\frac{1}{|\bm{x}|} -
  \frac{l}{l!}Y_{ijL-1} x^{L-1}\right]_{r=R}\,\,d\Omega \\\nn
&=& \frac{R^2}{16\pi}\oint\,\left[\sum_{l=0}^\infty
  \frac{(-1)^{l+1}}{l!}\frac{1}{R^{l+2}}\left(a_lZ_{ijL-1}n^jn^{L-1}
  - b_lZ_{iL}n_jn^jn^L\right) -
  \frac{l}{l!}R^{l-1}Y_{ijL-1}n^jn^{L-1}\right] \,\,d\Omega\\
&=& \frac{R^2}{16\pi}\oint\,\left[\sum_{l=0}^\infty
  \frac{(-1)^{l+1}}{l!}\frac{1}{R^{l+2}}(a_l-b_l)Z_{iL}n^L -
  \frac{l}{l!}R^{l-1}Y_{iL}n^L\right]\,\,d\Omega.
\end{eqnarray}
\end{widetext}
Since $Z^A_{iL}$ and $Y^A_{iL}$ are STF on $L$, only the $l=0$ terms
can contribute [cf. Eqs.\ (\ref{unitvectorintegralA}) and
  (\ref{unitvectorintegralB})]. This implies that
\begin{eqnarray}\nn
\nP^i &=& \frac{R^2}{16\pi}\frac{4\pi b_0}{R^2}Z_i \\\label{newtonianP}
&=& {\,}^{\scriptscriptstyle \text{n}}\!\dot{M}_i,
\end{eqnarray}
where we have used the relation (\ref{hgB}).  Note
that the contribution from the gravitomagnetic tidal terms
vanishes identically.  Thus, the Newtonian enclosed momentum is the
time derivative of the Newtonian mass dipole.

We turn next to the surface integral of the
Landau-Lifshitz pseudotensor on the right hand side of Eq.\
(\ref{conservationlaw}).  To Newtonian order it is given by
\begin{eqnarray}\nn
\oint \, \ncalT^{ij}\,d^2\Sigma_j &=& \oint \,
\frac{1}{4\pi}\left(\partial_i\Phi\partial_j\Phi -
\frac{1}{2}\delta_{ij}\partial_k\Phi\partial_k\Phi \right)
d^2\Sigma_j,\\\label{Newtoniantij}
\end{eqnarray}
from Eq.\ (\ref{eq:LLN}).
This surface integral can be computed using the expansion
(\ref{basicmodelA}) of the Newtonian potential and the integrals
(\ref{unitvectorintegralA}) --- (\ref{STFintegralB}). First we write the derivative of the
expansion (\ref{basicmodelA}) of the Newtonian potential as
\begin{equation}
\partial_i \Phi = \sum_{l=0}^\infty \frac{1}{l!}\left(V^{\Phi}_Ln^i -
\bar{V}^{\Phi}_{iL}\right)n^L,
\end{equation}
where
\begin{eqnarray}
V^{\Phi}_L &=& \frac{(2l+1)!!}{r^{l+2}}\nM_L,
\label{eq:V1}
\end{eqnarray}
and
\begin{eqnarray}
\bar{V}^{\Phi}_{iL} &=& \frac{(2l+1)!!}{r^{l+3}}\nM_{iL} +
r^l\nG_{iL}.
\label{eq:V2}
\end{eqnarray}
Substituting this into the right-hand side of
Eq.\ (\ref{Newtoniantij}) and using the integrals
(\ref{STFintegralA}) and (\ref{STFintegralB}) gives
\begin{widetext}
\begin{eqnarray}\nn
\oint \ncalT^{ij} d^2\Sigma_j &=&
\frac{R^2}{4 \pi}\sum_{p=0}^\infty\sum_{q=0}^\infty\frac{1}{p!q!}
\oint n^P n^Q
\left[
  \left(V^{\Phi}_Pn^i - \bar{V}^{\Phi}_{iP}\right)\left(V^{\Phi}_Q
  - q\bar{V}^{\Phi}_{Q}\right) - \frac{n^i}{2}\left(V^{\Phi}_Pn^k -
  \bar{V}^{\Phi}_{kP}\right) \left(V^{\Phi}_Qn^k -
  \bar{V}^{\Phi}_{kQ}\right)\right]d\Omega \\\nn
&=&
\frac{R^2}{4 \pi}
\sum_{p=0}^\infty\sum_{q=0}^\infty
\frac{1}{p!q!}\oint\left[\frac{1}{2}\left(V^{\Phi}_PV^{\Phi}_Q
- \bar{V}^{\Phi}_{kP}\bar{V}^{\Phi}_{kQ}\right)n^in^Pn^Q +
\left(q\bar{V}^{\Phi}_{iP}\bar{V}^{\Phi}_Q -
\bar{V}^{\Phi}_{iP}V^{\Phi}_Q\right)n^Pn^Q \right]d\Omega \\\nn
&=& R^2 \sum_{l=0}^\infty
\frac{1}{l!}\left[\frac{1}{(2l+3)!!}\left(V^{\Phi}_{iL}V^{\Phi}_L -
\bar{V}^{\Phi}_{ikL}\bar{V}^{\Phi}_{kL}\right) +
\frac{1}{(2l+1)!!}\left(l\bar{V}^{\Phi}_{iL}\bar{V}^{\Phi}_L -
\bar{V}^{\Phi}_{iL}V^{\Phi}_L\right)\right] \\\nn
&=&
R^2\sum_{l=0}^\infty\frac{1}{l!(2l+1)!!}
V^{\Phi}_L\left(\frac{V^{\Phi}_{iL}}{2l+3}
- \bar{V}^{\Phi}_{iL}\right) \\
&=& - \sum_{l=0}^\infty \frac{1}{l!}\nM^A_L\nG^A_{iL}.
\end{eqnarray}
\end{widetext}
Here the last line has been obtained using the expressions (\ref{eq:V1})
and (\ref{eq:V2})
for $V^{\Phi}_L$ and $\bar{V}^{\Phi}_{iL}$ in terms of $\nM_L$,
$\nG_L$ and $r$, and performing some easy algebra. We thus have the key result:

\begin{equation}\label{integratednewtoniantij}
-\oint \, \ncalT^{ij}\,d^2\Sigma_j = \sum_{l=0}^\infty
 \frac{1}{l!}\nM^A_L\nG^A_{iL}.
\end{equation}
Substituting Eqs.\ (\ref{newtonianP}) and (\ref{integratednewtoniantij})
into the expansions (\ref{expandedmomentum}) and (\ref{eq:calTexpand}) and
then into the conservation law (\ref{conservationlaw}) finally gives
the Newtonian law of motion (\ref{newtonianlom}).

\subsection{Post-Newtonian order derivation}
\label{sec:pnderiv}

We now proceed with the derivation of the post-1-Newtonian law of motion
(\ref{mainresult}) from the conservation law (\ref{conservationlaw}).
Our method of derivation will be somewhat different from that used
above in the Newtonian case.  We start by describing the differences.

The differences are related to
the pieces of the computation that go
to one higher post-Newtonian order (post-1-Newtonian
order in the last subsection, and post-2-Newtonian order here).
In the Newtonian case, we had available the explicit parameterization
(\ref{basicmodelB}) -- (\ref{reducedzeta}) of the post-1-Newtonian
potentials in terms of
(i) the Newtonian order moments $\nM_L$, $\nG_L$; (ii) the
post-1-Newtonian order moments $\pnM_L$, $\pnG_L$, $H_L$ and $S_L$,
and (iii) the gauge moments $\mu_L$ and $\nu_L$.  The computation of
the Newtonian enclosed momentum (\ref{eq:momentumN}) did involve the
post-1-Newtonian potentials, but using the parameterization
(\ref{reducedzeta}) we found that the dependencies on the
post-Newtonian multipole, tidal and gauge moments dropped out.  Thus,
we obtained equations of motion in terms of the purely Newtonian
variables.
Similarly, in the post-1-Newtonian case, the computation of the
post-1-Newtonian momentum
(\ref{PiIV}) involves
the post-2-Newtonian potentials $\xi^i$
and $\chi_{ij}$.  Those potentials can presumably be parameterized
in terms of Newtonian and post-1-Newtonian moments, a set of
post-2-Newtonian moments, and
gauge degrees of freedom, via expansions
analogous to
(\ref{basicmodelB}) -- (\ref{reducedzeta}).
Using such expansions we could in principle proceed as in
the Newtonian computation.
However, this approach turns out to be extremely tedious\footnote{
The surface integrals encountered in the derivation of the
evolution laws (\protect{\ref{eq:mainresult0}})
and (\protect{\ref{eq:spinresult}})
for the mass monopole and spin are significantly simpler than those
for the dipole evolution law (\protect{\ref{mainresult}}).
We have derived explicit expressions for one particular solution of the
post-2-Newtonian field equations for $\xi^i$ and $\chi^{ij}$, and
in a later paper \protect{\cite{Racine}} those expressions will be used
to derive the mass monopole and spin evolution laws
(\protect{\ref{eq:mainresult0}}) and (\protect{\ref{eq:spinresult}})
using the same explicit method as used here in the Newtonian case.},
and we will proceed instead as follows.

Our strategy will be to argue indirectly that
there is no dependence on post-2-Newtonian degrees of freedom in the
post-1-Newtonian momentum (\ref{PiIV}).
The space of post-2-Newtonian potentials $(\xi^i,\chi_{ij})$ that
satisfy in ${\cal D}$ the field equations
(\ref{2pnfieldeqA}) and (\ref{2pnfieldeqB}) and the gauge conditions
(\ref{gaugeA}) and (\ref{gaugeB}) has the structure of an affine
space.  Given a particular solution
$(\xi^i_\text{p},\chi^{ij}_\text{p})$, any other solution can be
written as the sum
\begin{subequations}
\begin{eqnarray}
\label{eq:xisplit}
\xi^i &=& \xi^i_\text{p} + \xi^i_\text{h}  \\
\chi^{ij} &=& \chi^{ij}_\text{p} + \chi^{ij}_\text{h}
\label{eq:chisplit}
\end{eqnarray}
\end{subequations}
of the particular solution and a homogeneous solution
$(\xi^i_\text{h},\chi^{ij}_\text{h})$, where the homogeneous solution
satisfies homogeneous versions of the field equations
(\ref{2pnfieldeqA}) and (\ref{2pnfieldeqB}) and gauge conditions
(\ref{gaugeA}) and (\ref{gaugeB}):
\begin{subequations}
\begin{eqnarray}
\label{eq:homogeneousfieldeq}
\nabla^2 \xi_\text{h}^i &=& 0, \ \ \ \ \ \nabla^2 \chi^{ij}_\text{h}=0,
\\
\partial_j \chi^{ij}_\text{h} &=& 0, \ \ \ \ \ \partial_i \xi^i_\text{h} -
    {\dot \chi}^{kk}_\text{h}=0.
\label{eq:homogeneousgauge}
\end{eqnarray}
\end{subequations}
The general solutions to Eqs.\ (\ref{eq:homogeneousfieldeq}) can be
expanded as
\begin{subequations}
\begin{eqnarray}
\label{2pnsolA}
\xi^i_\text{h} &=& \sum_{l=0}^\infty
\frac{(-1)^{l+1}}{l!}X_{iL}\partial_L\frac{1}{|\bm{x}|} -
\frac{1}{l!}W_{iL}x^L, \\
\chi^{ij}_\text{h} &=& \sum_{l=0}^\infty
\frac{(-1)^{l+1}}{l!}C_{ijL}\partial_L\frac{1}{|\bm{x}|} -
\frac{1}{l!}B_{ijL}x^L.
\label{2pnsolB}
\end{eqnarray}
\end{subequations}
These multipole expansions define the multipole and tidal moments $X_{iL}$,
$W_{iL}$, $C_{ijL}$ and $B_{ijL}$, all of which are STF on the indices
$L$.  Inserting these expansions into the gauge conditions
(\ref{eq:homogeneousgauge}) yields the constraints
\begin{subequations}
\begin{eqnarray}
\label{eq:constraint1}
C_{i<L>}&=&0, \ \ \ \ \ B_{ijjL}=0, \\
l X_{<L>} + {\dot C}_{jjL}&=&0 \ \ \ \ \ W_{jjL} - {\dot B}_{jjL} =0.
\label{eq:constraint2}
\end{eqnarray}
\end{subequations}

Next, we insert the decompositions (\ref{eq:xisplit}) --
(\ref{eq:chisplit}) and the expansions (\ref{2pnsolA}) --
(\ref{2pnsolB}) into the formula (\ref{PiIV}) for the
post-1-Newtonian momentum.
Using the integrals (\ref{unitvectorintegralA}) -- (\ref{STFintegralB})
then gives
\begin{eqnarray}\nn
\pnP^i &=& \frac{1}{16\pi} \oint_\Sigma  \,
\left[ \partial_j \xi^i_{\text{p}} +  {\dot \chi}_{\text{p}}^{\,ij}  \right]
d^2\Sigma_j \\ \label{1pnmomentum}
& & + \frac{1}{4}X_i - \frac{1}{12}\dot{C}_{ijj} - \frac{R^3}{12}\dot{B}_{ijj}.
\label{eq:pnPans1}
\end{eqnarray}
Here as before we have chosen the 2-surface $\Sigma$ to be the sphere
$|{\bm{x}}|=R$.  The last three terms in this expression give the
dependence of $\pnP^i$ on the homogeneous solution.  From the
constraints (\ref{eq:constraint1}) -- (\ref{eq:constraint2}) it follows
that the sum of these three terms vanishes, so that
\begin{eqnarray}
\pnP^i &=& \frac{1}{16\pi} \oint  \,
\left[ \partial_j \xi^i_{\text{p}} +  {\dot \chi}_{\text{p}}^{\,ij}  \right]
d^2\Sigma_j.
\label{eq:pnPans2}
\end{eqnarray}
Thus, the enclosed momentum $\pnP_i$ is independent of which solution
of the post-2-Newtonian field equations is chosen.  Therefore, it must
be a function only of the Newtonian and post-1-Newtonian fields, or
equivalently of the Newtonian and post-1-Newtonian multipole and tidal
moments, as well as of the radius $R$.

Next, we show that
the enclosed momentum can be written as the sum of
the time derivative of the post-Newtonian mass dipole $\pndotM_i$ and
terms that are independent of $\pnM_i$.  To see this, note that
any solution $(\Phi,\zeta^i,\psi)$ of the post-1-Newtonian field equations
(\ref{fieldeqA}) -- (\ref{fieldeqC}) and gauge condition
(\ref{harmonicgauge}) can be decomposed as
\begin{eqnarray}
\Phi &=& \Phi_0, \ \ \ \ \ \zeta^i = \zeta^i_0, \\
\psi &=& \psi_0 + \pnM_i \partial_i \frac{1}{|{\bm{x}}|}.
\end{eqnarray}
Here $(\Phi_0,\zeta^i_0,\psi_0)$ is another
solution with vanishing post-Newtonian mass dipole.  Inserting this
decomposition into the post-2-Newtonian field equations and gauge
conditions (\ref{gaugeA}) -- (\ref{2pnfieldeqB}) yields a
corresponding decomposition of the post-2-Newtonian potentials:
\begin{eqnarray}
\label{eq:decompos4}
{\chi}^{ij} &=& \chi^{ij}_0, \\
\xi^i &=& \xi^i_0 - 4 \pndotM_i \frac{1}{|{\bm{x}}|}.
\label{eq:decompos5}
\end{eqnarray}
Here the potentials $(\Phi_0,\zeta_0^i,\psi_0,\chi^{ij}_0, \xi^i_0)$
are a solution of the
field equations and gauge conditions and are independent of $\pnM_i$.
Inserting the decompositions (\ref{eq:decompos4}) and
(\ref{eq:decompos5}) into Eq.\ (\ref{PiIV}) gives
\begin{eqnarray}
\pnP^i &=& \pndotM_i + \frac{1}{16\pi} \oint  \,
\left[ \partial_j \xi^i_0 +  {\dot \chi}^{ij}_0  \right]
d^2\Sigma_j.
\label{eq:pnPans3}
\end{eqnarray}
Inserting this into
the expansions (\ref{expandedmomentum}) and (\ref{eq:calTexpand}) and
then into the conservation law (\ref{conservationlaw}) now yields
\begin{eqnarray}\nn
{\,}^{\scriptscriptstyle \text{pn}}\!\ddot{M}_i &=& - \oint_\Sigma \left\{
\frac{1}{16\pi} \partial_j { {\dot \xi}}^i_0
+  \frac{1}{16 \pi}{\ddot {\chi}}^{ij}_0 +
\pncalT^{ij}
\right\}d^2\Sigma^j.\\\label{1pneom1}
\end{eqnarray}

Equation (\ref{1pneom1}) is the key result of this subsection.
It shows that the second time derivative of the post-Newtonian mass
dipole can be expressed purely in terms of the Newtonian and
post-1-Newtonian fields, and is independent of the post-2-Newtonian
degrees of freedom.  This independence was derived above for the first
two terms on the right hand side of Eq.\ (\ref{1pneom1}), and for the
third term it follows from the fact that the expression (\ref{eq:LLP2N}) for
$\pncalT^{ij}$
depends only on the Newtonian and the post-1-Newtonian fields.
Therefore the right hand side of Eq.\ (\ref{1pneom1}) is some function
of the surface $\Sigma$ as well as of the multipole and tidal moments
of the solution $(
\Phi,\zeta^i,\psi)$ [since the multipole and tidal moments of the
solution $(\Phi_0,\zeta^i_0,\psi_0)$ coincide with those of the
original solution $(\Phi,\zeta^i,\psi)$ except for the
post-1-Newtonian mass dipole].

We can deduce some properties of the functional dependence of
${\,}^{\scriptscriptstyle \text{pn}}\!\ddot{M}_i$ on the moments as
follows.  From the post-2-Newtonian field equations and gauge
conditions (\ref{gaugeA}) -- (\ref{2pnfieldeqB}) it follows that
the potentials $\xi^i_0$ and ${\dot \chi}_0^{ij}$ depend
linearly on ${\ddot \zeta}^i_0$ and ${\dot \psi}_0$, and
quadratically on $\Phi_0$, ${\dot \Phi}_0$, and
$\partial^{[i} \zeta^{j]}_0$.   Also from Eq.\ (\ref{eq:LLP2N}) it
follows that $\pncalT^{ij}$
depends quadratically on $\Phi$, ${\dot \Phi}$, $\partial^{[i}
\zeta^{j]}$, and ${\dot \zeta}^i + \partial^i \psi$, all of
which are independent of the gauge moments $\mu_L$ and $\nu_L$.
From Eq.\ (\ref{1pneom1}) and the expansions (\ref{basicmodelA}) --
(\ref{reducedzeta}) it now follows that
\begin{eqnarray}
{\,}^{\scriptscriptstyle \text{pn}}\!\ddot{M}_i =
&& \mathcal{F}_i\bigg(\nM_L,\ndotM_L,\nddotM_L,\nG_L,\ndotG_L,\nddotG_L,
H_L,{\dot H}_L, \nn \\
&& S_L,{\dot S}_L,\pnM_L,\pnG_L ;R \bigg) \nn \\
&& + {\cal G}_i\bigg(
{\,}^{\scriptscriptstyle \text{n}}\!\ddddot{M}_L,
{\,}^{\scriptscriptstyle \text{n}}\!\ddddot{G}_L,
{\dddot H}_L,{\dddot S}_L,{\dddot \mu}_L,{\dddot \nu}_L;R
\bigg).
\nn \\
\label{eq:schematic}
\end{eqnarray}
Here ${\cal G}_i$ is a linear function of all of the moments that
appear as its arguments, and can be an arbitrary function of the
radius $R$ that defines the 2-surface $\Sigma$.  Similarly the function
${\cal F}_i$ is a quadratic function of all the moments that appear as
its arguments, and can be an arbitrary function of $R$ \footnote{It is easy to
see that the right hand side of Eq.\ (\protect{\ref{eq:schematic}})
must be independent of the radii $r_-$ and $r_+$ that define the
domain ${\cal D}$.}.

In Appendix \ref{checks} we compute explicitly the linear term and
show that it vanishes identically:\footnote{The fact that the
right hand side of Eq.\ (\protect{\ref{eq:schematic}})
is independent of the gauge
moments $\mu_L$ and $\nu_L$ can alternatively be derived as follows.
As discussed in Sec. \ref{sec:bodyadapted} above, by
making a gauge transformation of the type (\ref{1pntimegauge}) we
can alter the values of the gauge moments $\mu_L$ and $\nu_L$
without altering any of the tidal and multipole
moments $\nM_L$, $\nG_L$, $\pnM_L$, $\pnG_L$, $S_L$, $H_L$ or their time
derivatives.  Under such a transformation, the left hand side of Eq.\
(\ref{eq:schematic}) is
invariant.  It follows that the right hand side does not depend on the gauge
moments or their time derivatives.}
\begin{equation}
{\cal G}_i = 0.
\end{equation}
Next, the left hand side of Eq.\
(\ref{eq:schematic}) is independent of the radius $R$,
as are the definitions of all the moments that appear as the arguments
of the function ${\cal F}_i$.  It follows that ${\cal F}_i$ is
independent of $R$, as one would expect.
Using these simplifications we can rewrite Eq.\ (\ref{eq:schematic})
as
\begin{eqnarray}
{\,}^{\scriptscriptstyle \text{pn}}\!\ddot{M}_i =
&& \mathcal{F}_i\bigg(\nM_L,\ndotM_L,\nddotM_L,\nG_L,\ndotG_L,\nddotG_L,
H_L,{\dot H}_L, \nn \\
&& S_L,{\dot S}_L,\pnM_L,\pnG_L\bigg),
\label{eq:schematic1}
\end{eqnarray}
where the function ${\cal F}_i$ is now a quadratic function of all of
its arguments.  This quadratic function could in principle be computed
from the expression (\ref{1pneom1}).  However, it is simpler to
appeal to a special case from which the functional form of ${\cal
F}_i$ can be deduced, as suggested in a different context
by Thorne and Hartle \cite{thornehartle}.

Specifically, we now specialize to the case
considered by DSX where the post-1-Newtonian field equations are
assumed to hold throughout $r<r_-$.  Our analysis applies to that
special case, and therefore the functional ${\cal F}_i$ coincides
with that computed by DSX, given in Eq. (4.21b) of Ref.\ \cite{dsxII}
and in Eq.\ (\ref{mainresult}) above.
The derivation of the form of ${\cal F}_i$ in this case is reviewed in
Appendix \ref{functionalform}.  The key point here is that the general
argument of this subsection
establishes the result (\ref{mainresult}) up to the values of
the coefficients of the terms on the right hand side, and our argument
shows that those coefficients are
universal, applying both to the case of weakly self-gravitating bodies
analyzed by DSX and to the case of strongly self-gravitating bodies
considered here.

This argument, which enables us to avoid doing
the surface integral (\ref{1pneom1}) explicitly, could of course also
be used to avoid doing the surface integrals in Eq.\
(\ref{Newtoniantij}) when computing the Newtonian laws of motion.  In
that case, however, we were able to check explicitly that the surface
integral method gives the correct answer.

Similarly, we could deduce that the linear term ${\cal G}_i$ must
vanish by comparison with the case of weakly self-gravitating bodies.
Therefore the explicit verification of this result in Appendix
\ref{checks} is not really necessary.  That verification
is useful, however, as a consistency check of our argument
and formalism.

Thus, we have established that the law of motion
(\ref{mainresult}) is valid not just for the class of
weakly self-gravitating bodies considered by DSX, but also for the more
general class of strongly self-gravitating bodies considered here,
subject to the assumptions outlined in Sec.\ \ref{sec:eomoverview}.

%%%%%%%%%%%%%%%%%%%%%%%%%%%%%%%%%%%%%%%%%%%%%%%%%%%%%%%%%%%%%%%%%%%%%%%%%%%%%%%%%%%%%%%%%%%%%%%%%%%%%%%%%%%%%%%%%%%%%%%%%%%%%%%%%%%%%%%%%%%%%%%%

%%%%%%%%%%%%%%%%%%%%%%%%%%%%%%%%%%%%%%%%%%%%%%%%%%%%%%%%%%%%%%%%%%%%%%%%%%%%%%%%%%%%%%%%%%%%%%%%%%%%%%%%%%%%%%%%%%%%%%%%%%%%%%%%%%%%%%%%%%%%%%%%%%%%%%%%%%%%%%

\section{An N-body system: foundations}
\label{manybody}

We now turn to an analysis of a system consisting of $N$ bodies with
arbitrary internal structure.  The bodies' masses can be comparable,
but their typical separations must be large compared to their masses
in order that their gravitational interactions are well described by
the post-1-Newtonian approximation\footnote{Our formalism and
derivation does not require that the bodies' separations be large
compared to their typical sizes.  However, that requirement is in
practice necessary if one wants to achieve good accuracy using a truncated
version of the equation of motion (\protect{\ref{fulleom}})
containing only a small number of
multipoles.}.  In this section we lay the foundations for our analysis
by defining
local coordinate systems associated with each body, and an overall
global coordinate system.  We also derive relations between the
moments that characterize the potentials in each of these coordinate
systems.  In Sec.\ \ref{explicit} below we will combine the results derived
here with the single-body equation of motion (\ref{mainresult}) derived in the
previous section to obtain the explicit form of the $N$-body equation
of motion.

\subsection{Assumptions}
\label{sec:assumptions}

We start by describing our assumptions.  We consider a system of $N$
bodies, labeled by the index $A$ with $1 \le A \le N$.
We associate with each body a world tube ${\mathcal W}_A$ containing the
region where the stress-energy tensor is nonzero, and also containing
the strong-field region associated with the body.  Our key assumption
is that the post-Newtonian equations are satisfied everywhere outside
all of the worldtubes ${\mathcal W}_A$.

More precisely, we make the following assumptions:
\begin{itemize}
\item For each $A$ there exists a coordinate system
$(s_A, y^j_A)$ of the type discussed in Sec.\
\ref{sec:bodyadapted} above which covers the $A$th body.
Thus, there exist radii $r_{-,A}$ and $r_{+,A}$ such that
the range of the coordinates includes the product of the ball
$|\bm{y}_A| < r_{+,A}$ with an open interval of time, and that
the coordinates are harmonic, conformally Cartesian and body-adapted
in the buffer region
\begin{equation}
{\mathcal B}_A \equiv \left\{ (s_A, y^j_A) \right| \left.
r_{-,A} < | \bm{y}_A | < r_{+,A} \right\}.
\label{eq:bufferregion}
\end{equation}
\item The various buffer regions ${\mathcal B}_A$ are non-intersecting
  (see Fig.\ \ref{fig1} above).
\item We define the world tube associated with the $A$th body to be
${\mathcal W}_A \equiv \left\{ (s_A, y^j_A) \right| \left.
 | \bm{y}_A | < r_{-,A} \right\}$, and we define the spacetime region
 ${\cal D}$
 to be the complement of the union of all the worldtubes,
$$
{\cal D} = {\cal M} \setminus \bigcup_{A=1}^N {\mathcal W}_A,
$$
where ${\cal M}$ is the entire manifold.  The vacuum Einstein equations are
satisfied for the one-parameter family of metrics
(\ref{eq:metricP2N}) on the
spacetime region ${\cal D}$.
\item There exists a conformally Cartesian and harmonic
coordinate system $(t,x^i)$ which covers all of ${\cal D}$.
\end{itemize}
We will call the coordinate system $(t,x^i)$ the {\it global frame},
even though it does not cover the entire manifold.  We will
call the coordinate systems $(s_A, y^j_A)$ the {\it body
frames}.

Note that in the context of the spatially non-compact domain
${\cal D}$, the meaning of the $O(\varepsilon^n)$ symbols that appear
in Eqs.\ (\ref{eq:st}) and (\ref{eq:metricP2N})
corresponds to pointwise convergence, not uniform convergence.
As is well known, solutions of the post-1-Newtonian field equations
are not good approximations to exact solutions at distances $
\gtrsim 1/\varepsilon$.  That is, although they work well in the near
zone they break down in the local wave zone \cite{thorne}.
In order to obtain solutions which are good
approximations everywhere one has to perform a matching of
post-Newtonian solutions onto radiation zone post-Minkowskian solutions; see
for example Blanchet \cite{blanchet} and references
therein.  However, the corresponding corrections to the near zone
gravitational fields and to the dynamics of the
bodies arises at post-2.5-Newtonian order, and can thus be neglected
for the post-1-Newtonian analysis of this paper. \\

\subsection{Body-frame multipole and tidal moments}
\label{sec:body_frame_moments}

In each local coordinate system $(s_A, y^j_A)$ we define multipole
and tidal moments $\nM_L^A(s_A)$, $\nG_L^A(s_A)$, $\pnM_L^A(s_A)$,
$\pnG_L^A(s_A)$, $S_L^A(s_A)$ and $H_L^A(s_A)$ according to the
prescription described in Sec.\ \ref{sec:pnsolns} above.  We have added
superscripts $A$ to these moments to denote the $A$th body.
The corresponding expansions of the potentials are
\begin{widetext}
\begin{subequations}
\begin{eqnarray}\label{adaptedA}
\Phi^A(s_A,y^j_A) &=& \sum_{l=0}^\infty
\frac{(-1)^{l+1}}{l!}\nM_L^A(s_A)\partial_L\frac{1}{|\bm{y}_A|} -
\frac{1}{l!}\nG_L^A(s_A)y_A^L ,\\\nn
\psi^A(s_A,y^j_A) &=& \sum_{l=0}^\infty
\left\{\frac{(-1)^{l+1}}{l!}
\pnM_L^A(s_A) \partial_L
\frac{1}{|\bm{y}_A|}
\right.
\left.
+ \frac{(-1)^{l+1}}{l!}{\,}^{\scriptscriptstyle
  \text{n}}\!\ddot{M}_L^A(s_A)\partial_L \frac{|\bm{y}_A|}{2}\right. \\
\label{adaptedB}
& & \left.- \frac{1}{l!}
\pnG_L^A(s_A) y^L_A -
\frac{1}{l!}\frac{|\bm{y}_A|^2}{2(2l + 3)}\, ^{\scriptscriptstyle
  \text{n}}\!\ddot{G}_L^A(s_A) y^L_A \right\}, \\
\label{adaptedC}
\zeta_i^A(s_A,y^j_A) &=& \sum_{l=0}^\infty
\left\{\frac{(-1)^{l+1}}{l!} \left[\frac{4}{l+1}{\,}^{\scriptscriptstyle
    \text{n}}\!\dot{M}_{iL}^A(s_A) -
  \frac{4l}{l+1}\epsilon_{ji<a_l}S^A_{L-1>j}(s_A) \right]
\partial_L\frac{1}{|\bm{y}_A|}\right. \nn \\
& & \left.- \frac{1}{l!}\left[
  \frac{l}{l+1}\epsilon_{ji<a_l}H^A_{L-1>j}(s_A)-
  \frac{4(2l-1)}{2l+1}{\,}^{\scriptscriptstyle
    \text{n}}\!\dot{G}^A_{<L-1}(s_A)\delta_{a_l>i}\right] y_A^L
\right\} \\
&\equiv & \sum_{l=0}^\infty
\frac{(-1)^{l+1}}{l!}Z_{iL}^A(s_A)\partial_L\frac{1}{|\bm{y}_A|}  -
\frac{1}{l!}Y_{iL}^A(s_A)y_A^L,
\label{adaptedC0}
\end{eqnarray}
\end{subequations}
\end{widetext}
where overdots mean derivatives with respect to the time argument.
These expansions are obtained from the expansions
(\ref{basicmodelD}) --- (\ref{basicmodelF})
by replacing $x^j$ with $y^j_A$ and $t$ with $s_A$, by adding
superscripts $A$ to the potentials and the various moments
to denote the $A$th body, and by
omitting the gauge moments $\mu_L$ and $\nu_L$ which vanish since we
have specialized to body-adapted gauge.
The specialization to body-adapted gauge also implies that
\begin{equation}
\nM_i^A = \pnM_i^A = \nG^A = \pnG^A = H_i^A =0,
\label{eq:bodyadapted1}
\end{equation}
cf. Sec.\ \ref{sec:bodyadapted} above.

\subsection{Configuration variables for the $A$th body.}
\label{sec:configvars}

For each body, there is a non-empty region of overlap between the
domain of the body-frame coordinates $(s_A, y^j_A)$ and the
domain of the global-frame coordinates $(t,x^i)$, namely the buffer region
${\mathcal B}_A$ defined by Eq.\ (\ref{eq:bufferregion}).
Both of these coordinate systems are harmonic and conformally
Cartesian, and therefore the mapping between the
two coordinate systems can be parameterized using
the general analysis of Sec.\
\ref{sec:gaugefreedom} above.  From Eqs.\
(\ref{eq:generalcoordtransform}) -- (\ref{conshgC}) it follows that
there exist functions
$z_i^A(s_A)$, $\alpha^A_{\text{c}}(s_A)$, $h^A_{\text{c} \, i}(s_A)$ and
$R^A_k(s_A)$ and a harmonic function $\beta^A_{\text{h}}(s_A,y_A^j)$
such that in $\mathcal{B}_A$
\begin{eqnarray}\nn
x^i &=& y_A^i + z^A_i(s_A) + \varepsilon^2h^A_i(s_A,y_A^j) + O(\varepsilon^4),
\\
t &=& s_A + \varepsilon^2\alpha^A(s_A,y_A^j) +
\varepsilon^4\beta^A(s_A,y^j_A) + O(\varepsilon^6), \nn \\
\label{coordinatetransformationII}
\end{eqnarray}
where
\begin{eqnarray}
\alpha^A &=& \alpha^A_{\text{c}}(s_A) +
y^j_A\dot{z}^A_j(s_A) ,\\\nn
h^A_i &=& h^A_{\text{c} \, i}(s_A) +
\epsilon_{ijk}y_A^jR^A_k(s_A) + \frac{1}{2}\ddot{z}^A_i(s_A)y_A^jy_A^j
\\\nn
& & - y_A^i\dot{\alpha}^A_\text{c}(s_A)
- y_A^i y_A^j {\ddot z}^A_j(s_A)
+ \frac{1}{2}y_A^i\dot{z}^A_j(s_A)\dot{z}_j^A(s_A)
\nn \\
&& + \frac{1}{2}\dot{z}^A_i(s_A)\dot{z}^A_j(s_A)y_A^j
\end{eqnarray}
and
\begin{eqnarray}
\beta^A &=&
y_A^jy_A^j\left[\frac{1}{10}\dddot{z}^A_k(s_A)y_A^k +
  \frac{1}{6}\ddot{\alpha}^A_{\text{c}}(s_A)\right] +
\beta^A_{\text{h}}(s_A,y_A^j).  \nn \\
\label{eq:ct2}
\end{eqnarray}

Because the body-frame coordinates are uniquely determined, the
functions $z_i^A(s_A)$, $\alpha^A_{\text{c}}(s_A)$, $h^A_{\text{c} \,
i}(s_A)$ and $R^A_k(s_A)$ acquire the role of configuration variables
that specify the location, orientation etc.\ in the global coordinates
$(t,x^i)$ of the local rest frame
attached to body $A$.
This role is in contrast to the role of the
corresponding variables in Sec.\ \ref{gaugefreedom} above,
which were freely specifiable functions.  The task of
determining the motion of the $N$ different bodies reduces to solving
for the time evolution of these configuration variables.  Below we
will show that some of these variables can be obtained by solving
differential equations, and the remainder are obtained from algebraic
relations.

The particular combination of these configuration variables that
enters into the
equation of motion which we derive below is the {\it center of mass
worldline}.  In the special case where the post-1-Newtonian equations
are assumed to hold inside each body,
this center of mass worldline is defined simply as the origin of
spatial coordinates of the body-adapted\footnote{More
  precisely, of the slightly modified body-adapted coordinate system
discussed in the last paragraph of Sec.\
\protect{\ref{sec:momentsgaugetransformation}} above, whose domain
of definition includes the interior of the body.  The difference
between this coordinate system and the body-adapted coordinate system
arises only at order $O(\varepsilon^4)$ in the time coordinate, which does
not affect the definition of center of mass worldline to
post-1-Newtonian accuracy.}  coordinate system $(s_A, y^j_A)$.
This worldline can be expressed in terms
of the global-frame coordinates in parametric form as
\begin{eqnarray}
\label{eq:com1}
x^i(s_A) &=& z_i^A(s_A) + \varepsilon^2 h_{\text{c}\,i}^A(s_A) +
O(\varepsilon^4),  \\
t(s_A) &=& s_A + \varepsilon^2 \alpha_\text{c}^A(s_A) +
O(\varepsilon^4),
\end{eqnarray}
from the coordinate transformation (\ref{coordinatetransformationII})
 -- (\ref{eq:ct2}).
Eliminating $s_A$ gives $x^i = \cmz_i^A(t)$, where
\begin{equation}
\cmz_i^A(t) = z_i^A(t) + \varepsilon^2 \left[ h_{\text{c}\,i}^A(t) -
  {\dot z}_i^A(t) \alpha_{\text{c}}^A(t) \right] + O(\varepsilon^4).
\label{eq:com3}
\end{equation}
Here the superscript ``cm'' means ``center of mass''.

Consider now the more general context where the post-1-Newtonian
equations are not assumed to hold inside each body.  Then, the
body-frame coordinates $(s_A, y^j_A)$ can
be arbitrary for $|\bm{y}_A| < r_{-,A}$, so the worldline in spacetime
of the origin $\bm{y}_A=0$ of these coordinates has no special
significance.
Nevertheless, we can still use Eqs.\ (\ref{eq:com1}) --
(\ref{eq:com3}) to define the function $\cmz_i^A(t)$.
That is, we define $\cmz_i^A(t)$ to be the image of the origin
$\bm{y}_A=0$ not under the true coordinate transformation, but under
the extension to $|{\bm y}_A| < r_{-,A}$ of the formulae
(\ref{coordinatetransformationII}) -- (\ref{eq:ct2}) which a priori
are only valid for $r_{-,A} < |\bm{y}_A| < r_{+,A}$.
The resulting function $\cmz_i^A(t)$
continues to characterize the location of the local rest frame
attached to body $A$, even though it no longer corresponds to a
worldline in spacetime\footnote{The function $\cmz_i^A(t)$ does however
transform like a worldline under the group
(\protect{\ref{coordinatetransformation}}) of post-1-Newtonian coordinate
transformations.}, and even though the location $x^i = \cmz_i^A(t)$
will in general be outside the domain of definition of the global
coordinates.  We will continue to call this function the center of
mass worldline, in a slight but conventional abuse of terminology.

\subsection{Global-frame multipole moments}
\label{sec:globalcoords}

In this section we define, for each body $A$, multipole moments associated
with the global coordinate system $(t,x^i)$.
We define the global-frame multipole moments $\nM_L^{\text{g},A}(t)$,
$\pnM_L^{\text{g},A}(t)$, $S_L^{\text{g},A}(t)$ and
$\mu_L^{\text{g},A}(t)$
to be the moments about the Newtonian-order center-of-mass worldline
${\bm x} = {\bm z}^A(t)$ of body $A$ [cf. Eq.\
(\ref{eq:com3}) above], using the
prescription discussed in Sec.\ \ref{sec:moments1}.
Using these multipole moments we can write
down multipole expansions of the global-frame potentials,
which we denote by $(\Phi^\text{g},\zeta^\text{g}_i,\psi^\text{g})$,
that are valid on the entire domain ${\cal D}$:
\begin{widetext}
\begin{subequations}
\begin{eqnarray}
\label{globalfieldsA}
\Phi^{\text{g}}(t,x^j) &=& \sum_{A=1}^N\sum_{l=0}^\infty \frac{(-1)^{l+1}}{l!}\nM^{\text{g},A}_L(t)\partial_L\frac{1}{|\bm{x} - \bm{z}^A(t)|}, \\
\label{globalfieldsB}
\psi^\text{g}(t,x^j) &=& \sum_{A=1}^N
\sum_{l=0}^\infty
\left(\frac{(-1)^{l+1}}{l!}\left\{ \pnM_L^{\text{g},A}(t)
\partial_L\frac{1}{|\bm{x} - \bm{z}^A(t)|}
+ \frac{(2l+1)}{(l+1)(2l+3)} \frac{\partial}{\partial t} \left[
  \mu_L^{\text{g},A}(t)
\partial_L\frac{1}{|\bm{x} - \bm{z}^A(t)|} \right]
\right\}
\right.
\nonumber \\  \mbox{} &&
\left.
+ \frac{(-1)^{l+1}}{l!} \frac{\partial^2}{\partial t^2} \left[
  \nM_L^{\text{g},A}(t)\partial_L \frac{|\bm{x} - \bm{z}^A(t)|}{2}
  \right]
\right), \\
\label{globalfieldsC}
\zeta_i^\text{g}(t,x^j) &=&
\sum_{A=1}^N\sum_{l=0}^\infty
\frac{(-1)^{l+1}}{l!}Z^{\text{g},A}_{iL}(t)\partial_L\frac{1}{|\bm{x}
  - \bm{z}^A(t)|} ,
\label{globalfieldsC1}
\end{eqnarray}
where
\begin{equation}
Z^{\text{g},A}_{iL} =
\frac{4}{l+1}\ndotM^{\text{g},A}_{iL} -
\frac{4l}{l+1}\epsilon_{ji<a_l}S^{\text{g},A}_{L-1>j}
  +  \frac{2l-1}{2l+1}\delta_{i<a_l}\mu^{\text{g},A}_{L-1>}
+ 4 {\dot z}^A_{<i} \nM^{\text{g},A}_{L>}.
\label{eq:Z1defg}
\end{equation}
\end{subequations}
\end{widetext}
Here the superscript ``g'' on the potentials and on the moments stands for
"global".

The form of the expansions (\ref{globalfieldsA}) --
(\ref{globalfieldsC}) is dictated by the following considerations:
\begin{itemize}
\item The expansions take the form of a linear superposition of
solutions, one for each body $A$.  This follows from the linearity of
the vacuum field equations (\ref{fieldeqA}) -- (\ref{fieldeqC}).
\item We choose to use a gauge for the global coordinates in which all
the potentials go to zero as $|\bm{x}| \rightarrow \infty$.  This
eliminates any tidal terms associated with acceleration of the
reference frame, cf. the discussion in Sec.\
\ref{sec:gaugefreedom} above.
\item There are no other tidal terms, since the terms in the sum over
$B$ with $B\ne A$ play the role of tidal terms for body $A$.
With this identification,
the expansions (\ref{globalfieldsA}) -- (\ref{globalfieldsC})
agree with the formulae (\ref{basicmodelA0}) -- (\ref{reducedzeta0})
of Sec.\ \ref{sec:moments1} above that define the multipole and gauge
moments, in the buffer region $\mathcal{B}_A$ about the $A$th body.
\end{itemize}

We will derive transformation laws relating
the global-frame
multipole moments $\nM^{\text{g},A}_L$, $\pnM^{\text{g},A}_L$ and
$S^{\text{g},A}_{L}$ to the
body-frame multipole moments $\nM^A_L$, $\pnM^A_L$ and $S^A_{L}$
in the next subsection.
Note that the moments that would be measured by observers residing in
the buffer region ${\cal B}_A$ about the $A$th body are the body-frame
moments and not the global-frame moments.

We next discuss the gauge freedom in the global coordinate system.
If we make a gauge transformation of the form (\ref{1pntimegauge})
with the harmonic function $\beta_\text{h}$ chosen to be
\begin{equation}
\beta_\text{h}({\bar t},{\bar x}^j) = \sum_{A=1}^N
  \sum_{l=0}^\infty\frac{(-1)^{l+1}}{l!}\lambda^A_L({\bar t}) \partial_L
  \frac{1}{|\bar{\bm{x}}   - \bm{z}^A(\bar{t})|},
\end{equation}
then the gauge moments $\mu_L^{\text{g},A}(t)$ transform according to
\begin{equation}
\label{eq:mutransform2}
{\bar \mu}_L^{\text{g},A} = \mu_L^{\text{g},A} +
\frac{(l+1)(2l+3)}{2l+1} \lambda_L^A,
\end{equation}
cf. Eq.\ (\ref{eq:mutransform1}) above.  Therefore there is
enough freedom to set
\begin{equation}
\mu_L^{\text{g},A} =0
\label{eq:globalgauge0}
\end{equation}
for all $A$ and for all $l \ge 0$.  This requirement,
together with the requirement that the potentials go to zero as
$|\bm{x}| \to \infty$, reduces the residual gauge freedom to the
post-Galilean transformation group discussed in Sec.\
\ref{sec:gaugefreedom}.

\subsection{Computation of body-frame tidal moments}
\label{sec:cbf}

Our goal is to deduce equations of motion for the $N$-body system from
the single-body equation of motion (\ref{mainresult}).  To this end,
we would like to compute the body-frame tidal moments $\nG^A_L$,
$\pnG^A_L$ and $H^A_{L}$ felt by body $A$ in terms of the body-frame
multipole moments
$\nM^B_L$, $\pnM^B_L$ and $S^B_L$ and also the configuration variables
of the other bodies $B$ with $B \neq A$.  We shall perform this computation in
stages, by relating both sets of quantities to
the global-frame moments.

We start by expanding the time derivatives that appear in the expansion
(\ref{globalfieldsB}) of the global-frame potential $\psi^{\text{g}}$
and by expressing the results in terms of STF tensors.  Using the gauge
specialization (\ref{eq:globalgauge0}) this computation gives

\begin{eqnarray}\nn
\psi^{\text{g}} &=& \sum_{A=1}^N \sum_{l=0}^\infty
\frac{(-1)^{l+1}}{l!}\Bigg[N^{\text{g},A}_L
\partial_L\frac{1}{|\bm{x} - \bm{z}^A|}  \\ \label{globalmodelD}
& & + P^{\text{g},A}_L
\partial_L\frac{|\bm{x} - \bm{z}^A|} {2} \Bigg],
\end{eqnarray}
where the STF tensors $N^{\text{g},A}_L$ and $P^{\text{g},A}_L$
are given by
[cf.\ Eqs.\ (\ref{eq:Ndef}) and (\ref{eq:Pdef}) above]
\begin{eqnarray}
N^{\text{g},A}_L &=& \pnM^{\text{g},A}_L +
\frac{1}{2l+3}\left[\ddot{z}_j^A \nM^{\text{g},A}_{jL} +
\dot{z}^A_j\dot{z}^A_j\,\nM^{\text{g},A}_L \right.
\nn \\ \mbox{} && \left.
+
2\dot{z}_j^A{\,}^{\scriptscriptstyle
\text{n}}\!\dot{M}^{\text{g},A}_{jL} + 2l\dot{z}^A_j \dot{z}^A_{<a_l}\nM^{\text{g},A}_{L-1>j}\right],\label{globalN}
\end{eqnarray}
and
\begin{eqnarray}
P^{\text{g},A}_L &=& {\,}^{\scriptscriptstyle
\text{n}}\!\ddot{M}^{\text{g},A}_L +
2l\dot{z}^A_{<a_l}{\,}^{\scriptscriptstyle
\text{n}}\!\dot{M}^{\text{g},A}_{L-1>} +
l\ddot{z}^A_{<a_l}\nM^{\text{g},A}_{L-1>}
\nn \\ \mbox{} &&
+ l(l-1)\dot{z}^A_{<a_l} \dot{z}^A_{a_{l-1}}\nM^{\text{g},A}_{L-2>}
.\label{globalP}
\end{eqnarray}

Next, we expand the global potentials in the buffer region
$\mathcal{B}_A$ of body A using the
Taylor series
\begin{eqnarray}
\nn
|\bm{x} - \bm{z}^B|^p &=& |(\bm{z}^B - \bm{z}^A) - (\bm{x} -
\bm{z^A})|^p \\\label{taylor}
&=& \sum_{k=0}^\infty \frac{1}{k!}(x - z^A)^K \mathcal{T}^p_K(\bm{z}^{BA}),
\end{eqnarray}
where $p$ is any integer, $\bm{z}^{BA} = \bm{z}^B - \bm{z}^A$ and
\begin{equation}
\mathcal{T}^p_K(\bm{z}) \equiv \left(\partial^{\,}_K|\bm{z} -
\bm{x}|^p\right)_{\bm{x} = 0}.
\label{eq:calTKdef}
\end{equation}
For $p=-1$ we have
\begin{equation}
\mathcal{T}^{-1}_K(\bm{z}) = (2k-1)!! \frac{z^{<K>}} {|\bm{z}|^{2k+1}}.
\label{eq:calTKdef0}
\end{equation}
Substituting Eq.\ (\ref{taylor}) into Eqs.\ (\ref{globalfieldsA}),
(\ref{globalfieldsC1}) and (\ref{globalmodelD})
and using the identity
\begin{equation}
{\cal T}_{KL}^{+1} = {\cal T}_{K<L>}^{+1} + \frac{ l(l-1)}{2l-1}
\delta_{(a_{l-1} a_l} {\cal T}_{L-2)K}^{-1}
\end{equation}
yields
\begin{widetext}
\begin{subequations}
\begin{eqnarray}\label{globaltidalA}
\Phi^{\text{g}} &=& \sum_{l=0}^\infty
\frac{(-1)^{l+1}}{l!}\nM^{\text{g},A}_L\partial_L\frac{1}{|\bm{x} -
\bm{z}^A|}  - \frac{1}{l!}\nG^{\text{g},A}_L(x - z^A)^L
,\\\nn
\psi^{\text{g}} &=& \sum_{l=0}^\infty
\left\{\frac{(-1)^{l+1}}{l!}\left[N^{\text{g},A}_L
\partial_L\frac{1}{|\bm{x} - \bm{z}^A|}  + P^{\text{g},A}_L
\partial_L\frac{|\bm{x} - \bm{z}^A|} {2} \right]  -
\frac{1}{l!}\left[F^{\text{g},A}_L (x - z^A)^L +
J^{\text{g},A}_L\frac{|\bm{x} - \bm{z}^A|^2 (x -
z^A)^L}{2(2l+3)}\right]\right\},\\\label{globaltidalB}
&\,& \, \\\label{globaltidalC}
\zeta^{\text{g}}_i &=& \sum_{l=0}^\infty
\frac{(-1)^{l+1}}{l!}Z^{\text{g},A}_{iL}\partial_L\frac{1}{|\bm{x} -
\bm{z}^A|} - \frac{1}{l!}Y^{\text{g},A}_{iL}(x - z^A)^L ,
\end{eqnarray}
\end{subequations}
\end{widetext}
where
\begin{subequations}
\begin{eqnarray}\label{globaltidalnG}
\nG^{\text{g},A}_L &=& \sum_{B\neq A}\sum_{k=0}^\infty
\frac{(-1)^k}{k!}\nM^{\text{g},B}_K \mathcal{T}^{-1}_{KL}(\bm{z}^{BA}),
\end{eqnarray}
\begin{equation}\label{globaltidalY}
Y^{\text{g},A}_{iL} = \sum_{B\neq A}\sum_{k=0}^\infty
\frac{(-1)^k}{k!}Z^{\text{g},B}_{iK} \mathcal{T}^{-1}_{KL}(\bm{z}^{BA}),
\end{equation}
\begin{equation}\label{globaltidalJ}
J^{\text{g},A}_L = \sum_{B\neq A}\sum_{k=0}^\infty
\frac{(-1)^k}{k!} P^{\text{g},B}_K \mathcal{T}^{-1}_{KL}(\bm{z}^{BA}),
\end{equation}
and
\begin{eqnarray}
F^{\text{g},A}_L &=& \sum_{B\neq A}\sum_{k=0}^\infty
\frac{(-1)^k}{k!}
\bigg[ N^{\text{g},B}_K \mathcal{T}^{-1}_{KL}(\bm{z}^{BA})
\nn \\ \mbox{} &&
\left. +
\frac{1}{2}P^{\text{g},B}_K \mathcal{T}^{+1}_{K<L>}(\bm{z}^{BA})\right].
\label{globaltidalF}
\end{eqnarray}
\end{subequations}
Here $\nG^{\text{g},A}_L$, $F^{\text{g},A}_L$,
$Y^{\text{g},A}_{iL}$, and $J^{\text{g},A}_L$ are global-frame
tidal moments.  The post-Newtonian moments $F^{\text{g},A}_L$,
$Y^{\text{g},A}_{iL}$ and $J^{\text{g},A}_L$
could be parameterized in terms of the irreducible global-frame tidal
moments $\pnG_L^{\text{g},A}$, $H_L^{\text{g},A}$, and
$\nu_L^{\text{g},A}$ if desired via equations analogous to
Eqs. (\ref{eq:Fdef}), (\ref{eq:Jdef}) and (\ref{eq:Y1def}).  Here,
however, it will be more convenient to work directly with the moments
$F^{\text{g},A}_L$, $Y^{\text{g},A}_{iL}$ and $J^{\text{g},A}_L$.
 The function ${\bm
z}^B(s_B)$ that appear on the right hand
sides of Eqs.\ (\ref{globaltidalnG}) -- (\ref{globaltidalF}) is
evaluated at $s_B = t$.

Next, we apply the coordinate transformation
(\ref{coordinatetransformationII}) to the global potentials
(\ref{globaltidalA}) --- (\ref{globaltidalC}) using the formulae (\ref{transfA})
-- (\ref{transfC}).  We parameterize the harmonic function
$\beta_\text{h}^A$ which appears in Eq.\ (\ref{eq:ct2}) as
\begin{eqnarray}
\beta_\text{h}^A(s_A,y^j_A) &=&
  \sum_{l=0}^\infty\frac{(-1)^{l+1}}{l!}\lambda^A_L(s_A) \partial_L
  \frac{1}{|\bm{y}_A|} \nn \\
&& -
  \sum_{l=0}^\infty\frac{1}{l!}\tau^A_L(s_A) y_A^L,
\end{eqnarray}
where the tensors $\lambda^A_L$ and $\tau^A_L$ are STF.
Comparing the result with the expansions
(\ref{adaptedA}) -- (\ref{adaptedC}) allows us to compute the
body-frame moments in terms of the global-frame moments, as in Sec.\
\ref{sec:momentsgaugetransformation} above.
At Newtonian order we obtain [cf.\ Eqs.\ (\ref{newM}) -- (\ref{lambdaphii}) above]
\begin{subequations}
\begin{equation}\label{physicalnM}
\nM^A_L = \nM^{\text{g},A}_L,
\end{equation}
\begin{equation}\label{tidalnG}
\nG^A_L = \nG^{\text{g},A}_L - l!\Lambda^{\Phi}_L,
\end{equation}
\end{subequations}
where the nonzero inertial moments $\Lambda^{\Phi}_L$ are given by
\begin{subequations}
\begin{equation}\label{inertialA}
\Lambda^{\Phi} = -\frac{1}{2}\dot{z}^A_j\dot{z}^A_j + \dot{\alpha}^A_c
\end{equation}
and
\begin{equation}\label{inertialB}
\Lambda^{\Phi}_i = \ddot{z}^A_i.
\end{equation}
\end{subequations}
Transforming the gravitomagnetic potential gives [cf.\ Eqs.\
(\ref{newZ}) -- (\ref{newY}) above]
\begin{subequations}
\begin{equation}\label{physicalZ}
Z^A_{iL} = Z^{\text{g},A}_{iL} - 4\dot{z}^A_i \nM^{\text{g},A}_L +
l\delta_{i<a_l}\lambda^A_{L-1>},
\end{equation}
\begin{equation}\label{tidalY}
Y^A_{iL} = Y^{\text{g},A}_{iL} - 4\dot{z}^A_i\nG^{\text{g},A}_L -
\tau^A_{iL} - l!\Lambda^{\bm{\zeta}}_{iL},
\end{equation}
\end{subequations}
where the nonzero inertial moments $\Lambda^{\bm{\zeta}}_{iL}$ are given by
\begin{subequations}
\begin{equation}\label{inertialC}
\Lambda^{\bm{\zeta}}_i = \dot{h}^A_{\text{c}\,\, i} + \dot{z}^A_i\dot{z}^A_j\dot{z}^A_j - \epsilon_{ijk}\dot{z}^A_j R^A_k  - 2\dot{\alpha}^A_c\dot{z}^A_i,
\end{equation}
\begin{equation}\label{inertialD}
\Lambda^{\bm{\zeta}}_{ij} = -\dot{z}^A_i\ddot{z}^A_j -\dot{z}^A_{(i}\ddot{z}^A_{j)} + \epsilon_{ijk}\dot{R}^A_k + 2\delta_{ij}\dot{z}^A_k\ddot{z}^A_k  - \frac{4}{3}\delta_{ij}\ddot{\alpha}^A_c,
\end{equation}
and
\begin{equation}\label{inertialE}
\Lambda^{\bm{\zeta}}_{ijk} = -\frac{6}{5}\delta_{i<j}\dddot{z}^A_{k>}.
\end{equation}
\end{subequations}
Next, by combining the transformation laws (\ref{physicalZ}) and
(\ref{tidalY}) with the decompositions of $Z^A_{iL}$ and $Y^A_{iL}$
given by Eqs.\ (\ref{adaptedC}) -- (\ref{adaptedC0}), the
decomposition (\ref{eq:Z1defg}) of $Z^{\text{g},A}_{iL}$, and the gauge
condition (\ref{eq:globalgauge0}), we can solve for the coordinate
transformation functions $\lambda_L^A$ for $l \ge 0$ and $\tau_L^A$ for $l
\ge 1$.  The result is
\begin{subequations}
\begin{equation}
\lambda_L^A = \frac{4 (2 l +1)}{(l+1) (2l+3)} \dot{z}^A_j
\nM_{jL}^{\text{g},A}
\label{eq:lambdaLans}
\end{equation}
and
\begin{equation}
\tau^A_{iL} = Y^{\text{g},A}_{<iL>} -
4\dot{z}^A_{<i}\nG^{\text{g},A}_{L>} - l!\Lambda^{\bm{\zeta}}_{<iL>}.
\label{eq:tauLans}
\end{equation}
\end{subequations}
Finally, matching the post-Newtonian potentials and using the
definition (\ref{globalN})
of $N^{\text{g},A}_L$, the formulae
(\ref{eq:lambdaLans}) and (\ref{eq:tauLans}) for $\lambda_L^A$ and $\tau_{iL}^A$, and the identities
(\ref{eq:ident00}) -- (\ref{eq:ident04}) gives
\begin{widetext}
\begin{subequations}
\begin{eqnarray}\nn
\pnM^A_L  &=& \pnM^{\text{g},A}_L - \frac{4l}{l+1}\dot{z}^A_j\epsilon_{jk<a_l}S^{\text{g},A}_{L-1>k} - lh^A_{\text{c}\,<a_l}\nM^{\text{g},A}_{L-1>} - l \nM^{\text{g},A}_{j<L-1}\epsilon_{a_l>jk}R^A_k \\\nn
& & + \alpha^A_c{\,}^{\scriptscriptstyle \text{n}}\!\dot{M}^{\text{g},A}_L + (l+1)\dot{\alpha}^A_c\nM^{\text{g},A}_L +  l\alpha^A_c\dot{z}^A_{<a_l}\nM^{\text{g},A}_{L-1>} \\\nn
& & - \frac{l^2 - 3 l + 4}{2(l+1)} \dot{z}^A_j\dot{z}^A_j\nM^{\text{g},A}_L
+ \frac{l( 2l^2 - 13 l + 9)}{ 2 (l+1)(2l+1)}
\dot{z}^A_j\dot{z}^A_{<a_l}\nM^{\text{g},A}_{L-1>j}
\\\label{physicalpnM}
& & + \left[(l+1) + \frac{4
    (2l+1)}{(l+1) (2l+3)}\right]\ddot{z}^A_j\nM^{\text{g},A}_{jL}
+ \left[1 -  \frac{8}{(l+1) (2l+3)}\right]\dot{z}^A_j{\,}^{\scriptscriptstyle
  \text{n}}\!\dot{M}^{\text{g},A}_{jL}
\end{eqnarray}
and
\begin{eqnarray}\nn
\pnG^A_L &=& F^{\text{g},A}_L  + \dot{Y}^{\text{g},A}_{<L>} - \dot{z}^A_jY^{\text{g},A}_{jL} + (h^A_{\text{c}\, j} - \alpha^A_{\text{c}}\dot{z}^A_j)\nG^{\text{g},A}_{jL} - l\nG^{\text{g},A}_{j<L-1}\epsilon_{a_l>jk}R^A_k \\\nn
& & + \alpha^A_{\text{c}}{\,}^{\text{n}}\!\dot{G}^{\text{g},A}_L -
l\dot{\alpha}^A_{\text{c}}\nG^{\text{g},A}_L +
\frac{l+4}{2}\dot{z}^A_j\dot{z}^A_j\nG^{\text{g},A}_L -
\frac{l}{2}\dot{z}^A_j\dot{z}^A_{<a_l}\nG^{\text{g},A}_{L-1>j}
\\\label{tidalpnG}
& & - (l^2 - l +4)\ddot{z}^A_{<a_l}\nG^{\text{g},A}_{L-1>} +
(l-4)\dot{z}^A_{<a_l}{\,}^{\text{n}}\!\dot{G}^{\text{g},A}_{L-1>} -
l! \Lambda^{\psi_{\text{h}}}_L - (l-1)!
\dot{\Lambda}^{\bm{\zeta}}_{<L>} + \delta_{l0} {\dot \tau}^A.
\end{eqnarray}
\end{subequations}
\end{widetext}
Here then nonzero inertial moments
$\Lambda^{\psi_{\text{h}}}_L$ are given by
\begin{subequations}
\begin{equation}\label{inertialF}
\Lambda^{\psi_{\text{h}}} = -\dot{z}^A_j\dot{h}^A_{\text{c}\,\, j} -
\frac{1}{4}(\dot{z}^A_j\dot{z}^A_j)^2 -
\frac{1}{2}(\dot{\alpha}^A_c)^2 +
\dot{\alpha}^A_c\dot{z}^A_j\dot{z}^A_j,
\end{equation}
\begin{eqnarray}\nn
\Lambda^{\psi_{\text{h}}}_i &=& \epsilon_{ijk}\dot{z}^A_j\dot{R}^A_k +
\frac{1}{2}\ddot{z}^A_i\dot{z}^A_j\dot{z}^A_j -
\frac{3}{2}\dot{z}^A_i\ddot{z}^A_j\dot{z}^A_j \\\label{inertialG}
& & - \dot{\alpha}^A_c\ddot{z}^A_i + \ddot{\alpha}^A_c\dot{z}^A_i,
\end{eqnarray}
and
\begin{equation}\label{inertialH}
\Lambda^{\psi_{\text{h}}}_{jk} =
-\frac{1}{2}\ddot{z}^A_{<j}\ddot{z}^A_{k>} +
\dot{z}^A_{<j}\dddot{z}^A_{k>}.
\end{equation}
\end{subequations}
In Eq.\ (\ref{tidalpnG}) it is understood that the moments $Y_L^{\text{g},A}$
and $\Lambda^{\bm{\zeta}}_L$ are zero for $l=0$.

The left-hand sides of Eqs.\
(\ref{physicalnM}), (\ref{tidalnG}), (\ref{physicalZ}),
(\ref{tidalY}), (\ref{physicalpnM}) and (\ref{tidalpnG})
are functions of the time
coordinate $s_A$ of the body-adapted coordinate system for the $A$th body,
cf. Eqs.\ (\ref{adaptedA}) --- (\ref{adaptedC}) above.
The right-hand sides are expressed as functions of $s_A$ by evaluating
the global moments, which are functions of the global time coordinate
$t$, at $t=s_A$, cf. the discussion in the last paragraph of
Sec.\ \ref{sec:momentsgaugetransformation} above.

Finally, by combining the transformation laws
(\ref{tidalnG}), (\ref{tidalY}) and (\ref{tidalpnG})
with the gauge specializations (\ref{eq:bodyadapted1}) of the
body-adapted coordinates we can deduce the values of some of the
configuration variables of the $A$th body.  We obtain
\begin{equation}
\alpha^A_{\text{c}} = \int ds_A \left[\nG^{\text{g},A} +
  \frac{1}{2}\dot{z}^A_j\dot{z}^A_j \right],
\label{eq:alphacans}
\end{equation}
\begin{eqnarray}\nn
\tau^A &=& \int ds_A \bigg[ - F^{\text{g},A} + \dot{z}^A_jY^{\text{g},A}_j -
2\dot{z}^A_j\dot{z}^A_j\nG^{\text{g},A} \\\nn
& & - (h^A_{\text{c}\, j} - \alpha^A_c\dot{z}^A_j)\nG^{\text{g},A}_j -
\alpha^A_{\text{c}} \ndotG^{\text{g},A} + \Lambda^{\psi_{\text{h}}} \bigg],\\
\end{eqnarray}
and
\begin{eqnarray}\nn
R^A_k &=& \frac{1}{2}\epsilon_{ijk} \int ds_A \left[\dot{z}^A_i\ddot{z}^A_j + Y^{\text{g},A}_{ij} - 4\dot{z}^A_i\nG^{\text{g},A}_j\right].\\
\label{eq:Rkans}
\end{eqnarray}
The only remaining configuration variables that are undetermined are
the variables $h_{\text{c}\,i}^A$ and $z^A_i$ that determine the center of
mass worldline (\ref{eq:com3}).

To summarize, the main results of this subsection are the explicit
expressions (\ref{tidalnG}), (\ref{tidalY}) and (\ref{tidalpnG}) for
the body-frame tidal moments $\nG^A_L$, $\pnG^A_L$ and $Y^A_{iL}$ which
act on body $A$ in terms of the configuration variables of all the bodies,
as well as the global-frame mass and current moments
$\nM^{\text{g},B}_L$, $\pnM^{\text{g},B}_L$ and $S^{\text{g},B}_L$ of
the other bodies. These expressions are given by combining
(\ref{tidalnG}), (\ref{tidalY}) and (\ref{tidalpnG}) with
Eqs.\ (\ref{globaltidalnG}) --- (\ref{globaltidalF}).  Also, the
global-frame multipole moments $\nM_L^{\text{g},B}$, $\pnM_L^{\text{g},B}$ and $S_L^{\text{g},B}$
of body $B$ can be reexpressed in terms of the body-frame multipole
moments $\nM^B_L$, $\pnM^B_L$ and $S^B_L$ of that body using the relations (\ref{physicalnM}), (\ref{physicalZ}) and
(\ref{physicalpnM}).

\subsection{Definition of body frame multipole moments $\bodyM^A_L$
  and $\bodyS^A_L$.}
\label{sec:mdefs1}

As discussed in the introduction, it is useful to use instead of
the body-frame multipole moments $M^A_L$ and $S^A_L$ a modified set of
body-frame moments defined as follows.  We define for each body $A$ a
coordinate system $({\tilde s}^A,{\tilde y}^A_i)$ which is identical
to the body-frame coordinate system $(s^A,y^A_i)$ except that it is
non-rotating with respect to the global frame coordinates $(t,x^i)$
(i.e., non-rotating with respect to fixed stars).
We define the moments $\bodyM^A_L(t)$ and $\bodyS_L^A(t)$
to be the multipole moments of body $A$ in this non-rotating
coordinate system, expressed as functions of the global time coordinate $t$.
These are given by the equations
\begin{eqnarray}\nn
\bodyM^A_{a_1 \ldots a_l}(t) &=& U_{a_1}^{A\,a'_1}(t) \ldots U_{a_l}^{A\,a'_l}(t)
M^A_{a'_1 \ldots a'_l}[s_A(t)] \,\,\,\,\,\,\, \\
\label{eq:calMdef}
\end{eqnarray}
and
\begin{eqnarray}
\bodyS^A_{a_1 \ldots a_l}(t) &=& U_{a_1}^{A\,a'_1}(t) \ldots U_{a_l}^{A\,a'_l}(t)
S^A_{a'_1 \ldots a'_l}[s_A(t)],\,\,\,\,\,\,\,
\label{eq:calSdef}
\end{eqnarray}
where $s_A(t)$ is the value of the body-frame time coordinate $s_A$
evaluated at what would be the intersection of the worldline of body
$A$ with the spacelike hypersurface of constant $t$.  From Eq.\
(\ref{coordinatetransformationII}) this function is given by
\begin{equation}
s_A(t) = t - \varepsilon^2 \alpha_{\text{c}}^A(t) + O(\varepsilon^4).
\end{equation}
Also the rotation matrices $U^{A\,a'}_a$ are defined by the formula
\begin{equation}
U^{A\,a'}_a = \delta_{a'a} + \varepsilon^2 \epsilon_{aa'j} R^A_j.
\label{eq:rotUdef}
\end{equation}
From Eqs.\ (\ref{eq:totalMLdef}), (\ref{eq:totalGLdef}) and
(\ref{eq:calMdef}) -- (\ref{eq:rotUdef}) we can write these moments as
\begin{eqnarray}\nn
\bodyM^A_L &=& \nM^A_L+ \varepsilon^2 \bigg[ \pnM^A_L - \alpha^A_c
  \ndotM^A_L
\label{eq:bodyMformula}
 \\
\mbox{} && +
l \epsilon_{jk<a_l}\nM^A_{L-1>j} R^A_k \bigg] + O(\varepsilon^4).
\\ \mbox{}
\bodyS^A_L &=& S^A_L+ O(\varepsilon^2).
\label{eq:bodySformula}
\label{nosupmass}
\end{eqnarray}
As indicated by Eq.\ (\ref{eq:bodySformula}), when working to
post-1-Newtonian order we can identify the moments $\bodyS^A_L$ and
$S^A_L$.  Nevertheless it might be useful in some circumstances to use
the more accurate relation (\ref{eq:calSdef}), for example for
systems which evolve for sufficiently long times that the rotation
matrices $U_a^{A\,a'}$ become significantly different from unity.

All the tools are now set up
to compute explicit equations of motion for the center of mass
worldlines. \\

%%%%%%%%%%%%%%%%%%%%%%%%%%%%%%%%%%%%%%%%%%%%%%%%%%%%%%%%%%%%%%%%%%%%%%%%%%%%%%%%%%%%%%%%%%%%%%%%%%%%%%%%%%%%%%%%%%%%%%%%%%%%%%%%%%%%%%%%%%%%%%%%%%%%%%%%%%%%%%

\section{Explicit equations of motion for an N-body system}
\label{explicit}

In this section we derive explicit equations of motion for the center
of mass worldlines $\cmz_i^A(t)$ of each body as seen from the global
coordinate system, by combining the single-body equations of
motion (\ref{newtonianlom}) and (\ref{mainresult}) with the moment
transformation formula derived in Sec.\ \ref{manybody} above.

We start by deriving the well-known Newtonian equations of motion,
in order to illustrate the computational method.
The Newtonian single-body equation of motion
(\ref{newtonianlom}) applied to body $A$ implies that
\begin{equation}
\sum_{l=0}^\infty \frac{1}{l!}\nM^A_L(s_A)\nG^A_{iL}(s_A) = 0,
\label{eq:ss2}
\end{equation}
since the body-adapted coordinates are mass-centered, i.e. $\nM^A_i
= 0$ for all $A$.
Using the relation (\ref{tidalnG}) between the body frame tidal
moments $\nG^A_L$ and the global frame tidal moments
$\nG^{\text{g},A}_L$ we can rewrite this as
\begin{eqnarray}\nn
\ddot{z}^A_i &=& \nG^{\text{g},A}_i + \sum_{l=2}^\infty
\frac{1}{l!}\frac{\nM^A_L}{\nM^A}\nG^{\text{g},A}_{iL} \\
&=& \sum_{l=0}^\infty
\frac{1}{l!}\frac{\nM^A_L}{\nM^A}\nG^{\text{g},A}_{iL}.
\end{eqnarray}
Here the acceleration $\ddot{z}^A_i$ of the Newtonian-order center of
mass worldline has appeared via the transformation law for $\nG^A_i$.
Next, we substitute the expression (\ref{globaltidalnG}) for the global-frame
tidal moments $\nG^{\text{g},A}_{iL}$ in terms of mass multipole
moments $\nM^{\text{g},B}_L$ of the other bodies, and use Eq.\ (\ref{physicalnM}).
This gives
\begin{eqnarray}\nn
\ddot{z}^A_i &=& \sum_{B \neq
  A}\sum_{k=0}^\infty\sum_{l=0}^\infty
\frac{(-1)^k}{k!l!}\frac{\nM^A_L}{\nM^A}\nM^B_K\\\label{NeomA}
& & \times
\mathcal{T}^{-1}_{iKL}(\bm{z}^B - \bm{z}^A).
\label{eq:newteoma}
\end{eqnarray}
Here the quantities $\bm{z}^A$ and $\nM_L^A$ are functions of $s_A$, while
the quantities $\bm{z}^B$ and $\nM_K^B$ are functions of $s_B$,
evaluated at $s_B = s_A$.
Writing the dependent variable as $t$ instead of $s_A$ and
using the definition (\ref{eq:calTKdef}) of $\mathcal{T}^p_K$ we can
rewrite Eq.\ (\ref{eq:newteoma})
in the more explicit, well-known form \cite{dsxI}
\begin{eqnarray}\nn
\ddot{z}^A_i(t) &=& \sum_{B \neq A}\sum_{k=0}^\infty\sum_{l=0}^\infty
\frac{(-1)^k}{k!l!}\frac{\nM^A_L(t)}{\nM^A}\nM^B_K(t)\\\label{NeomB}
& & \times
\left[\partial^{(\bm{u})}_{iKL}\frac{1}{|\bm{u}|}\right]_{\bm{u} =
  \bm{z}^A(t) - \bm{z}^B(t)}.
\end{eqnarray}

The analogous computation carried to post-1-Newtonian order is
similar but much more involved.
We start by focusing on the first two terms on the right hand side of
the single body equation of motion (\ref{mainresult}), and
evaluating explicitly the $l=0$ pieces using Eqs.\ (\ref{tidalnG}) and
(\ref{tidalpnG}).  [As before the $l=1$ pieces vanish
since the body-frame coordinates are mass-centered, by Eq.\
(\ref{eq:bodyadapted1})].
Using the Newtonian equation of motion (\ref{eq:newteoma}), the definition
(\ref{eq:bodyMformula}) of $\bodyM^A_L$ and the definition
(\ref{eq:com3}) of $\cmz_i^A$, the result can be written in the form
\begin{eqnarray}
\bodyM^A {\cmddotz}^A_i &=& \bodyM^A (\nG_i^{\text{g},A} +
\varepsilon^2 f^A_i) +
\sum_{l=2}^\infty \bodyM^A_L {\mathcal G}^A_{iL} \nn \\
\mbox{} && +  \varepsilon^2 g^A_i + O(\varepsilon^4).
\label{eq:eomintermediate}
\end{eqnarray}
Here by analogy with Eq.\ (\ref{eq:bodyMformula}) we have defined
\begin{eqnarray}\nn
{\mathcal G}^A_L &=& \nG^A_L+ \varepsilon^2 \bigg[ \pnG^A_L - \alpha^A_c
  \ndotG^A_L
 \\\label{eq:bodyGformula}\mbox{} &&
+l \epsilon_{jk<a_l}\nG^A_{L-1>j} R^A_k \bigg],
\end{eqnarray}
and $g_i^A$ is defined to be all of the terms on the right hand side of
Eq.\ (\ref{mainresult}) except for the first two terms (with superscripts
$A$ added to all the moments).  Also we define
\begin{eqnarray}
f^A_i &=& F^{\text{g},A}_i + {\dot Y}^{\text{g},A}_i - {\dot z}^A_i
Y^{\text{g},A}_{ji} + h^A_{\text{c}\,j} \nG^{\text{g},A}_{ji}
- \alpha^A_{\text{c}} {\dot z}^A_j \nG^{\text{g},A}_{ji}
\nn \\ \mbox{} &&
+ {\dot z}^A_j {\dot z}^A_j \nG_i^{\text{g},A} - {\dot z}^A_i {\dot
  z}^A_j \nG^{\text{g},A}_j - 4 \nG^{\text{g},A}_i \nG^{\text{g},A}
\nn \\ \mbox{} &&
- 3 {\dot z}^A_i \ndotG^{\text{g},A} + \epsilon_{ijk} {\bar f}^A_j
R^A_k + {\dot \alpha}^A_\text{c} {\bar f}^A_i - \frac{3}{2} {\dot
  z}^A_j {\dot z}^A_j {\bar f}^A_i
\nn \\ \mbox{} &&
- \frac{1}{2} {\dot z}^A_i {\dot z}^A_j {\bar f}^A_j - 4 {\bar f}^A_i \nG^{\text{g},A},
\end{eqnarray}
where
\begin{eqnarray}
{\bar f}^A_i &=& {\ddot z}^A_i - \nG^{\text{g},A}_i \nn \\
&=& \sum_{l=2}^\infty \frac{1}{l!} \frac{\nM^A_L}{\nM^A}
\nG^{\text{g},A}_{iL}.
\end{eqnarray}

In order to explicitly evaluate the tidal moments that appear on the
right hand side of Eq.\ (\ref{eq:eomintermediate}),
we perform the following
sequence of moment transformations: (i) Start with the body frame
multipole moments $\nM_L^B$, $\pnM_L^B$ and $S_L^B$ of body $B$. (ii)
Compute from these the global frame multipole moments $\nM^{\text{g},B}$,
$N^{\text{g},B}_L$, $P^{\text{g},B}_L$ and $Z^{\text{g},B}_L$ of body $B$
using Eqs.\ (\ref{eq:SLdef}), (\ref{eq:Z1defg}), (\ref{eq:Ndef}),
(\ref{eq:Pdef}), (\ref{physicalZ}), (\ref{eq:lambdaLans})
and (\ref{physicalpnM}).  The results are
\begin{subequations}
\begin{eqnarray}
\nM_L^{\text{g},B} &=& \nM_L^B,
\\ \mbox{}
Z_{iL}^{\text{g},B} &=& \frac{4}{l+1} \ndotM_{iL}^B - \frac{4l}{l+1}
\epsilon_{ji<a_l} S^B_{L-1>j} + 4 {\dot z}^B_i \nM_L^B
\nn \\ \mbox{}
&& - \frac{4 (2l-1)}{2l+1} {\dot z}^B_j \nM^B_{j<L-1} \delta_{a_l>i},
\label{eq:GLgans}
\\ \mbox{}
P_L^{\text{g},B} &=& \nddotM_L^B + 2 l {\dot z}^B_{<a_l}
\ndotM^B_{L-1>}
+ l {\ddot z}^B_{<a_l} \nM^B_{L-1>}
\nn \\ \mbox{}
&& + l(l-1) {\dot z}^B_{<a_l} {\dot z}^B_{a_{l-1}} \nM^B_{L-2>},
\end{eqnarray}
and
\begin{eqnarray}
&&N_L^{\text{g},B} = \pnM_L^B - \alpha_{\text{c}}^B \ndotM_L^B + l
\epsilon_{jk<a_l} \nM^B_{L-1>j} R^B_k
\nn \\ \mbox{}
&& + l h_{\text{c}\,<a_l}^B \nM^B_{L-1>} -l \alpha^B_{\text{c}} {\dot
  z}^B_{<a_l} \nM^B_{L-1>} -(l+1) {\dot \alpha}^B_{\text{c}} \nM^B_L
\nn \\ \mbox{}
&& + \frac{4l}{l+1} {\dot z}^B_j \epsilon_{jk<a_l} S^B_{L-1>k}
+ \frac{(l+2)(2l+7)}{2(2l+3)} {\dot z}^B_j {\dot z}^B_j \nM_L^B
\nn \\ \mbox{}
&& - \left[ \frac{l}{2} + 3 - \frac{10l + 21}{(2l+1)(2l+3)} \right]{\dot z}^B_j
  {\dot z}^B_{<a_l} \nM_{L-1>j}^B
\nn \\ \mbox{}
&& - \left[ l + 1 + \frac{7l + 3}{(l+1)(2l+3)} \right]{\ddot z}^B_j
  \nM_{jL}^B
\nn \\ \mbox{}
&& - \left[ 1 - \frac{2(l+5)}{(l+1)(2l+3)} \right]{\dot z}^B_j
  \ndotM_{jL}^B.
\end{eqnarray}
\end{subequations}
(iii) Compute the global-frame tidal moments $\nG_L^{\text{g},A}$,
$Y_L^{\text{g},A}$, $J_L^{\text{g},A}$ and $F_L^{\text{g},A}$ of body
$A$ in terms of the global-frame multipole moments of body $B$ using
Eqs.\ (\ref{globaltidalnG}) -- (\ref{globaltidalF}).  (iv) Compute the
body-frame tidal moments $H^A_L$, $\nG_L^A$ and $\pnG^A_L$ of body A in
terms of its global frame tidal moments $\nG_L^{\text{g},A}$,
$Y_L^{\text{g},A}$, $J_L^{\text{g},A}$ and $F_L^{\text{g},A}$.
Here the results are given by Eq.\ (\ref{tidalnG}) for $\nG_L^A$,
and by Eq.\ (\ref{tidalpnG}) for $\pnG_L^A$; note that for the
required values of $l$ ($l\ge 3$) the last three terms in Eq.\
(\ref{tidalpnG}) do not contribute.  For $H^A_L$ we have $H^A_i=0$ by
the gauge condition (\ref{eq:bodyadapted1}), while for $l\ge 2$ we
obtain from Eq.\ (\ref{tidalY}) that
\begin{equation}
H^A_L = Y^{\text{g},A}_{jk<L-1} \epsilon_{a_l>jk} - 4 {\dot z}^A_j
\nG^{\text{g},A}_{k<L-1} \epsilon_{a_l>jk}.
\end{equation}
(v) By combining the preceding steps, all the moments can be expressed
in terms of the multipole moments $\nM^B_L$, $\pnM^B_L$ and $S^B_L$ of
body $B$.

In the resulting expression, we eliminate the variables $z^C_i$ in
favor of $\cmz^C_i$ for all $C$ using the definition (\ref{eq:com3}), and we
eliminate the moments $\nM^C_L$, $S^C_L$ in favor of $\bodyM^C_L$,
$\bodyS^A_L$ for all $C$ using the definitions (\ref{eq:bodyMformula}) and
(\ref{eq:bodySformula}).  These substitutions generate correction
terms only in the $O(\varepsilon^0)$, Newtonian terms in Eq.\
(\ref{eq:eomintermediate}), and not in the $O(\varepsilon^2)$,
post-Newtonian terms\footnote{In particular since the current moments
$S^A_L$ do not enter at Newtonian order, there are no correction
terms generated when one eliminates $S^A_L$ in favor of
$\bodyS^A_L$.  Thus, we are free to use either $S^A_L$ or
$\bodyS^A_L$ in the equations of motion, to post-1-Newtonian order,
as noted in Sec.\ \protect{\ref{sec:mdefs1}} above.}, since we drop
all terms of order $O(\varepsilon^4)$.
We also eliminate ${\dot \alpha}^C_\text{c}$ using
Eq.\ (\ref{eq:alphacans}).  The resulting expression then depends only
on the variables $\cmz^C_i$, $\bodyM^C_L$ and $\bodyS^C_L$; the
dependencies on the variables $R^C_k$, $\alpha^C_\text{c}$ and
$h^C_{\text{c},i}$ cancel out.  Lastly, we set to one the formal
expansion parameter $\varepsilon$.
The result of this tedious computation is
\begin{eqnarray}
\cmddotz^A_i(t) &=& \sum_{B\neq A}  a^{AB}_i(t) +
  \sum_{B\neq A}\sum_{C\neq A} a^{ABC}_i(t) \nn \\
&& + \sum_{B\neq A}\sum_{C\neq B}
  \tilde{a}^{ABC}_i(t) + O(\varepsilon^4),
\label{fulleom}
\end{eqnarray}
where
\begin{widetext}
\begin{subequations}
\begin{eqnarray}
a^{AB}_i &=& \sum_{k=0}^\infty\sum_{l=0}^\infty \bigg[
{}^{(1)}{\cal D}_{KL}^{AB} \,
\frac{n^{BA}_{<iKL>}}{r^{k+l+2}_{BA}}
+ {}^{(2)}{\cal D}_{iKL}^{AB} \,
\frac{n^{BA}_{<KL>}}{r_{BA}^{k+l+1}}
+ {}^{(3)}{\cal D}_{ijKL}^{AB} \,
\frac{n^{BA}_{<jKL>}}{r^{k+l+2}_{BA}}
+ {}^{(4)}{\cal D}_{jKL}^{AB} \,
\frac{n^{BA}_{<ijKL>}}{r^{k+l+3}_{BA}} \nn \\
\mbox{} &&
+ {}^{(5)}{\cal D}_{ijmKL}^{AB} \,
\frac{n^{BA}_{<jmKL>}}{r^{k+l+3}_{BA}}
+ {}^{(6)}{\cal D}_{injmKL}^{AB} \,
\frac{n^{BA}_{<njmKL>}}{r^{k+l+4}_{BA}}
+ {}^{(7)}{\cal D}_{KL}^{AB} \,
\frac{n^{BA}_{<iKL>}}{r^{k+l}_{BA}}
\bigg],
\label{ab}
\end{eqnarray}
\begin{eqnarray}
a^{ABC}_i &=& \sum_{k=0}^\infty \sum_{l=0}^\infty \sum_{p=0}^\infty
{}^{(1)}{\cal D}_{KLP}^{ABC} \,
\frac{n^{CA}_{<P>}}{r^{p+1}_{CA}} \,
\frac{n^{BA}_{<iKL>}}{r^{k+l+2}_{BA}}
+ \sum_{k=0}^\infty \sum_{l=0}^\infty \sum_{p=0}^\infty
\sum_{q=0}^\infty \bigg[
{}^{(2)}{\cal D}_{KLPQ}^{ABC} \,
\frac{n^{CA}_{<iPQ>}}{r^{p+q+2}_{CA}} \, \frac{n^{BA}_{<KL>}}{r^{k+l+1}_{BA}}
\nn \\ \mbox{} &&
+ {}^{(3)}{\cal D}_{jKLPQ}^{ABC} \,
\frac{n^{CA}_{<jPQ>}}{r^{p+q+2}_{CA}}
\frac{n^{BA}_{<iKL>}}{r^{k+l+2}_{BA}}
+ {}^{(4)}{\cal D}_{iKLPQ}^{ABC} \,
\frac{n^{CA}_{<jPQ>}}{r^{p+q+2}_{CA}} \,
\frac{n^{BA}_{<jKL>}}{r^{k+l+2}_{BA}}
+ {}^{(5)}{\cal D}_{ijKLPQ}^{ABC} \,
\frac{n^{CA}_{<mPQ>}}{r^{p+q+2}_{CA}} \, \frac{n^{BA}_{<jmKL>}}{r^{k+l+3}_{BA}}
\bigg], \nn \\
\label{abc}
\end{eqnarray}
and
\begin{eqnarray} \nn
{\tilde a}^{ABC}_i &=&
\sum_{k=0}^\infty \sum_{l=0}^\infty \sum_{p=0}^\infty
{}^{(1)}{\tilde {\cal D}}_{KLP}^{ABC} \,
\frac{n^{CB}_{<P>}}{r^{p+1}_{CB}} \,
\frac{n^{BA}_{<iKL>}}{r^{k+l+2}_{BA}}
+ \sum_{k=0}^\infty \sum_{l=0}^\infty \sum_{p=0}^\infty
\sum_{q=0}^\infty \bigg[
{}^{(2)}{\tilde {\cal D}}_{ijKLPQ}^{ABC} \,
\frac{n^{CB}_{<jPQ>}}{r^{p+q+2}_{CB}} \, \frac{n^{BA}_{<KL>}}{r^{k+l+1}_{BA}}
\nn \\ \mbox{} &&
+{}^{(3)}{\tilde {\cal D}}_{KLPQ}^{ABC} \,
\frac{n^{CB}_{<iPQ>}}{r^{p+q+2}_{CB}} \,
\frac{n^{BA}_{<KL>}}{r^{k+l+1}_{BA}}
+ {}^{(4)}{\tilde {\cal D}}_{jKLPQ}^{ABC} \,
\frac{n^{CB}_{<jPQ>}}{r^{p+q+2}_{CB}}
\frac{n^{BA}_{<iKL>}}{r^{k+l+2}_{BA}}
\nn \\ \mbox{} &&
+ {}^{(5)}{\tilde {\cal D}}_{iKLPQ}^{ABC} \,
\frac{n^{CB}_{<jPQ>}}{r^{p+q+2}_{CB}} \,
\frac{n^{BA}_{<jKL>}}{r^{k+l+2}_{BA}}
+ {}^{(6)}{\tilde {\cal D}}_{ijKLPQ}^{ABC} \,
\frac{n^{CB}_{<mPQ>}}{r^{p+q+2}_{CB}} \, \frac{n^{BA}_{<jmKL>}}{r^{k+l+3}_{BA}}
\bigg].
\label{tildeabc}
\end{eqnarray}
\end{subequations}

Here the coefficients are given by
\begin{subequations}
\begin{eqnarray}
{}^{(1)}{\cal D}_{KL}^{AB} &=&
\frac{(-1)^k (2k+2l+1)!!}{k!l! \, \bodyM^A} \bigg\{
\bodyM^A_L \bodyM^B_K \left[ 1 + v^A_j v^A_j +
  \frac{6k+11}{2(2k+3)} v^B_j v^B_j - 4v^A_j v^B_j \right]
+  v^A_j \bodyM^A_{jL} {\dot\bodyM}^B_K
\nn \\
&& + \bodyM^A_L {\dot\bodyM}^B_{jK} \left[
 \frac{4}{k+1}v^{BA}_j- \frac{2k+1}{2k+3}v^B_j \right]
- \frac{1}{2k+2l+5} \bodyM^A_{jL} {\hat P}^B_{jK} -
\frac{4}{l+1}{\dot\bodyM}^A_{jL} \left(v^{BA}_j\bodyM^B_K + \frac{1}{k+1}{\dot\bodyM}^B_{jK}\right)
\bigg\}, \nn \\
\label{eq:coeff1}
\end{eqnarray}
\begin{eqnarray}
{}^{(2)}{\cal D}_{iKL}^{AB} &=&
\frac{(-1)^k (2k+2l+1)!!}{k!l! (2k + 2l + 1) \, \bodyM^A} \bigg\{
\bodyM^A_L \left[ 4{\dot\bodyM}^B_K v^{BA}_i + {\dot\bodyM}^B_Kv^A_i -
  \frac{1}{2k+2l+3} {\hat P}^B_{iK} + \frac{4}{k+1}\ddot{\bodyM}^B_{iK} \right]
\nn\\
& &
+
4{\dot\bodyM}^A_L \left[\frac{1}{k+1}{\dot\bodyM}^B_{iK} + \bodyM^B_Kv^{BA}_i \right]
- \frac{(2l^2 + 3l + 5)}{(l+1)}{\dot\bodyM}^A_{iL}{\dot\bodyM}^B_K \nn \\
& &  + \frac{1}{2l+3}\bodyM^A_{iL}\left[-(l+2)(2l+1)\ddot{\bodyM}^B_K -
\frac{2k}{(2k+2l+3)} {\hat P}^B_K\right] -
\frac{(l^2+l+4)}{(l+1)}\ddot{\bodyM}^A_{iL}\bodyM^B_K
\bigg\},
\end{eqnarray}
\begin{eqnarray}
{}^{(3)}{\cal D}_{ijKL}^{AB} &=&
\frac{(-1)^k (2k+2l+1)!!}{k!l! \, \bodyM^A} \bigg\{
\bodyM^A_L\left[- \frac{4}{k+1}v^{BA}_j{\dot\bodyM}^B_{iK}
  -\bodyM^B_K(4v^{BA}_{i} v^{BA}_j + v^A_iv^B_j) \right. \nn \\
\mbox{} && +
 \left. \frac{3}{\bodyM^A}\left(\bodyM^A_{ij}\ddot{\bodyM}^B_K + 2{\dot\bodyM}^A_{ij}{\dot\bodyM}^B_K
  + \ddot{\bodyM}^A_{ij}\bodyM^B_K\right)
+ \epsilon_{ijm}\left(\frac{4}{k+2}\dot{\bodyS}^B_{mK} -
  2\dot{\bodyS}^A_m\frac{\bodyM^B_K}{\bodyM^A} - \bodyS^A_m\frac{{\dot\bodyM}^B_K}{\bodyM^A}\right)
  \right]
\nn\\
& & +2 \bodyM^A_{iL} {\dot\bodyM}^B_K \left[\frac{(l+2)(2l+1)}{(2l+3)} v^{B}_j
  -(l+1) v^A_j \right] \nn\\\label{calDAB3}
& & + {\dot\bodyM}^A_L\left[\epsilon_{ijm}\left(\frac{4}{k+2}\bodyS^B_{mK} -
  \bodyS^A_m\frac{\bodyM^B_K}{\bodyM^A}\right)  +
  \frac{6}{\bodyM^A}\left(\bodyM^A_{ij}{\dot\bodyM}^B_K +
  {\dot\bodyM}^A_{ij}\bodyM^B_K\right)\right]  +
\frac{(2l^2+3l+5)}{(l+1)}v^{BA}_j{\dot\bodyM}^A_{iL}\bodyM^B_K  \nn\\
& &  + \frac{3}{\bodyM^A} \bodyM^A_{ij} \ddot{\bodyM}^A_L\bodyM^B_K +
\frac{4}{l+2}\epsilon_{ijm}\left(\bodyS^A_{mL}{\dot\bodyM}^B_K +
\dot{\bodyS}^A_{mL}\bodyM^B_K\right)
\bigg\},
\end{eqnarray}
\begin{eqnarray}
{}^{(4)}{\cal D}_{jKL}^{AB} &=&
\frac{(-1)^k (2k+2l+3)!!}{k!l! \, \bodyM^A} \bigg\{
\bodyM^A_L \left[\frac{2k+1}{2(2k+5)}v^B_{j} v^B_m \bodyM^B_{mK}
- \frac{4}{k+2}v^{BA}_m\epsilon_{jmn}\bodyS^B_{nK} \right] \nn\\
& & - \bodyM^A_{nL}\bodyM^B_K\left(v^A_nv^{BA}_j + \frac{1}{2}v^A_{n} v^A_j\right) +
\frac{4}{(l+1)(k+2)}\epsilon_{jmn}{\dot\bodyM}^A_{mL}\bodyS^B_{nK} \nn\\
& &  -
\frac{4}{l+2}\epsilon_{jmn}\bodyS^A_{nL}\left(v^{BA}_m\bodyM^B_K +
\frac{1}{k+1}{\dot\bodyM}^B_{mK}\right)
\bigg\},
\end{eqnarray}
\begin{eqnarray}
{}^{(5)}{\cal D}_{ijmKL}^{AB} &=&
\frac{(-1)^k (2k+2l+3)!!}{k!l! \, \bodyM^A} \bigg\{
\bodyM^A_L\left[\epsilon_{ijn}\bodyS^A_nv^{BA}_m\frac{\bodyM^B_K}{\bodyM^A} -
  \frac{4}{k+2}\epsilon_{ijn}\bodyS^B_{nK}v^{BA}_m -
  \frac{6}{\bodyM^A}v^{BA}_m\left(\bodyM^A_{ij}{\dot\bodyM}^B_K +
  {\dot\bodyM}^A_{ij}\bodyM^B_K\right)
  \right]
\nn\\
& & + \frac{2}{2l+3} \bodyM^A_{iL} \bodyM^B_K(v^B_jv^A_m)
- \bodyM^B_K \left[\frac{\bodyM^A_{iL}}{2l+3}\left(v^A_{j} v^A_m+
  (l+2)(2l+1)v^{BA}_{j} v^{BA}_m\right) \right] \nn \\
\mbox{} && - \bodyM^B_K \left[
  \frac{6}{\bodyM^A}v^{BA}_m{\dot\bodyM}^A_L\bodyM^A_{ij} +
  \frac{4}{l+2}\epsilon_{imn}\bodyS^A_{nL}v^{BA}_j\right]
\bigg\},
\end{eqnarray}
\begin{eqnarray}
{}^{(6)}{\cal D}_{injmKL}^{AB} &=&
\frac{(-1)^k (2k+2l+5)!!}{k!l! \, \bodyM^A} \bigg\{
\frac{3}{\bodyM^A} v^{BA}_{j} v^{BA}_{m} \bodyM^A_{in} \bodyM^A_L\bodyM^B_K -
\frac{4 \delta_{in}}{(k+2)(l+2)}\bodyS^A_{jL}\bodyS^B_{mK}
\bigg\},
\end{eqnarray}
\begin{eqnarray}
{}^{(7)}{\cal D}_{KL}^{AB} &=&
\frac{(-1)^k (2k+2l+1)!!}{k!l! (2k + 2l + 1) \, \bodyM^A} \bigg\{
- \frac{1}{2}\bodyM^A_L {\hat P}^B_K \bigg\},
\end{eqnarray}
\end{subequations}

\begin{subequations}
\begin{eqnarray}
{}^{(1)}{\cal D}_{KLP}^{ABC} &=&
\frac{(-1)^{k+p+1} l (2k+2l+1)!! (2p+1)!!}{k!l!p! (2p + 1) \, \bodyM^A} \,
\bodyM^A_L \bodyM^B_K \bodyM^C_P,
\end{eqnarray}
\begin{eqnarray}
{}^{(2)}{\cal D}_{KLPQ}^{ABC} &=&
\frac{(-1)^{k+p+1} (l+4) (2k+2l+1)!! (2p+2q+1)!!}{k!l!p!q! (2k + 2l + 1) \, \bodyM^A} \,
\,
\bodyM^A_L \bodyM^B_K \frac{\bodyM^A_Q}{\bodyM^A} \bodyM^C_P,
\end{eqnarray}
\begin{eqnarray}
{}^{(3)}{\cal D}_{jKLPQ}^{ABC} &=&
\frac{(-1)^{k+p+1} (l+1) (2k+2l+1)!! (2p+2q+1)!!}{k!l!p!q! \, \bodyM^A} \,
\, \bodyM^A_{jL} \bodyM^B_K \frac{\bodyM^A_Q}{\bodyM^A} \bodyM^C_P,
\end{eqnarray}
\begin{eqnarray}
{}^{(4)}{\cal D}_{iKLPQ}^{ABC} &=&
\frac{(-1)^{k+p+1} l (2k+2l+1)!! (2p+2q+1)!!}{k!l!p!q! \, \bodyM^A} \,
\, \bodyM^A_{iL} \bodyM^B_K \frac{\bodyM^A_Q}{\bodyM^A} \bodyM^C_P,
\end{eqnarray}
\begin{eqnarray}
{}^{(5)}{\cal D}_{ijKLPQ}^{ABC} &=&
\frac{3 (-1)^{k+p} (2k+2l+3)!! (2p+2q+1)!!}{k!l!p!q! \, \bodyM^A} \,
\, \frac{\bodyM^A_L}{\bodyM^A} \bodyM^A_{ij} \bodyM^B_K
\frac{\bodyM^A_Q}{\bodyM^A} \bodyM^C_P,
\end{eqnarray}
\end{subequations}

\begin{subequations}
\begin{eqnarray}
{}^{(1)}{\tilde {\cal D}}_{KLP}^{ABC} &=&
\frac{(-1)^{k+p+1} (k+1) (2k+2l+1)!! (2p+1)!!}{k!l!p! (2p + 1) \, \bodyM^A} \,
\bodyM^A_L \bodyM^B_K \bodyM^C_P,
\end{eqnarray}
\begin{eqnarray}
{}^{(2)}{\tilde {\cal D}}_{ijKLPQ}^{ABC} &=&
\frac{(-1)^{k+p+1} (2k+2l+1)!! (2p+2q+1)!!}{k!l!p!q!(2k+2l + 1)
  (2k+2l+3) \, \bodyM^A} \frac{\bodyM^B_Q}{\bodyM^B} \bodyM^C_P \nn \\
\mbox{} &&\times \left[ (k+1) \bodyM^A_L
  \bodyM^B_{<K} \delta_{i>j} + \frac{2 k^2}{2l+3} \bodyM^A_{iL} \bodyM^B_{<K-1}
  \delta_{b_k>j} \right],
\end{eqnarray}
\begin{eqnarray}
{}^{(3)}{\tilde {\cal D}}_{KLPQ}^{ABC} &=&
 \frac{4 (-1)^{k+p}  (2k+2l+1)!! (2p+2q+1)!!}{k!l!p!q! (2k + 2l + 1) \, \bodyM^A} \,
\,
\bodyM^A_L \bodyM^B_K \frac{\bodyM^B_Q}{\bodyM^B} \bodyM^C_P,
\end{eqnarray}
\begin{eqnarray}
{}^{(4)}{\tilde {\cal D}}_{jKLPQ}^{ABC} &=&
\frac{(-1)^{k+p+1} (2k+2l+1)!! (2p+2q+1)!!}{k!l!p!q! \, \bodyM^A} \,
\frac{\bodyM^B_Q}{\bodyM^B} \bodyM^C_P \nn \\
\mbox{} && \times \left[ \frac{(2k+1)(k+2)}{2k+3} \bodyM^A_{L} \bodyM^B_{jK} +
  \frac{k+1}{2k+2l+5} \bodyM^A_{Lr} \bodyM^B_{<K} \delta_{r>j} \right],
\end{eqnarray}
\begin{eqnarray}
{}^{(5)}{\tilde {\cal D}}_{iKLPQ}^{ABC} &=&
\frac{(-1)^{k+p} (l+2)(2l+1) (2k+2l+1)!! (2p+2q+1)!!}{k!l!p!q! (2l+3)\, \bodyM^A} \,
\, \bodyM^A_{iL} \bodyM^B_K \frac{\bodyM^B_Q}{\bodyM^B} \bodyM^C_P,
\end{eqnarray}
\begin{eqnarray}
{}^{(6)}{\tilde {\cal D}}_{ijKLPQ}^{ABC} &=&
\frac{3 (-1)^{k+p+1} (2k+2l+3)!! (2p+2q+1)!!}{k!l!p!q! \, \bodyM^A} \,
\, \frac{\bodyM^A_L}{\bodyM^A} \bodyM^A_{ij} \bodyM^B_K
\frac{\bodyM^B_Q}{\bodyM^B} \bodyM^C_P,
\label{eq:coeff10}
\end{eqnarray}
\end{subequations}
where
\begin{equation}
{\hat P}^B_K = \ddot{\bodyM}^B_K + 2kv^B_{<b_k}{\dot\bodyM}^B_{K-1>} +
k(k-1)v^B_{<b_k} v^B_{b_{k-1}}\bodyM^B_{K-2>}.
\end{equation}
\end{widetext}
Here we have denoted by $\bm{n}^{BA}$ the unit vector pointing
from the center of mass worldline of body $A$ to that of body $B$:
\begin{equation}
n^{BA}_i(t) \equiv \frac{\cmz^{B}_i(t) - \cmz^{A}_i(t)}{r_{BA}(t)},
\end{equation}
where
\begin{equation}
r_{BA}(t) \equiv |\cmbmz^B(t) - \cmbmz^A(t)|.
\label{eq:rbadef}
\end{equation}
We also have defined $v^A_i = {\cmdotz}^A_i$ and $v^{BA}_i =
{\cmdotz}^B_i - {\cmdotz}^A_i$.

The right hand sides of Eqs.\ (\ref{fulleom}) --
(\ref{tildeabc}) depend on the time derivatives ${\dot \bodyS}^A_i$,
${\dot \bodyM}^A$ and ${\ddot \bodyM}^A$ of the bodies' spins and mass
monopole moments.
These dependencies can be eliminated using the single-body laws
of motion (\ref{eq:masscons}), (\ref{eq:pnmasscons}), and
(\ref{eq:spinresult}), the last two of which are derived for strongly
self-gravitating bodies in paper II \cite{Racine}.
For the case of the mass monopole moments, this procedure
generates terms that are of post-2-Newtonian
order which can be neglected.  Thus, all time derivatives of mass
monopoles appearing in Eqs.\ (\ref{fulleom}) -- (\ref{tildeabc}) can
be neglected.  In other words, we can make the following substitutions
in Eqs.\ (\ref{eq:coeff1}) -- (\ref{eq:coeff10}):
\begin{equation}
{\dot \bodyM}^A_L \to (1 - \delta_{l0}) {\dot \bodyM}^A_L, \ \ \ \ \
{\ddot \bodyM}^A_L \to (1 - \delta_{l0}) {\ddot \bodyM}^A_L.
\label{eq:monopolesub}
\end{equation}
For the case of the spin time derivative terms, using
Eq.\ (\ref{eq:spinresult}) together with Eqs.\ (\ref{globaltidalnG})
and (\ref{tidalnG}) we obtain modified values of the coefficients
(\ref{eq:coeff1}) -- (\ref{eq:coeff10}) which are listed in Appendix
\ref{sec:newcoeffs}.

A simple special case of the above equations of motion is the
non-spinning point particle model, or monopole-truncated model.  This
is obtained by setting to zero all the mass multipole moments $M^A_L$
for $l \geq 1$,
and all the current multipoles $S_L^A$.  In this case
Eq.\ (\ref{fulleom}) reduces to the well-known
Lorentz-Droste-Einstein-Infeld-Hoffmann equations of
motion \cite{ld,eih}, which were also reported in the first DSX paper
[Eq.\ (7.20b) of Ref.\ \cite{dsxI}].

A second special case is the spinning point particle model or
monopole-spin truncated model, obtained
by setting to zero all the mass multipoles $\nM_L^A$ and $\pnM_L^A$
for $l \ge 1$, all the current multipoles $S_L^A$ for $l\ge2$, but
allowing non-zero spins $S_i^A$.
For this case our general equation of motion
(\ref{fulleom}) reduces to the equations of motion obtained
for this case by DSX [Eqs.\ (6.30) --- (6.34) of Ref.\ \cite{dsxII}].

Finally, we can obtain an explicit expression for the angular velocity
(\ref{eq:Rkans}) parameterizing the dragging of inertial frames by
using the Newtonian equation of motion (\ref{eq:newteoma}), the
formulae (\ref{globaltidalnG}) and (\ref{globaltidalY}) for
$\nG^{\text{g},A}_L$ and $Y^{\text{g},A}_L$, the formula
(\ref{eq:GLgans}) for $Z^{\text{g},B}_L$, the formula
(\ref{eq:calTKdef0}) for $\mathcal{T}^{-1}_K$, and the definitions
(\ref{eq:rotUdef}), (\ref{eq:bodyMformula}) and (\ref{eq:bodySformula}).
The result is
\begin{eqnarray}
&&\bigg[{\dot {\bf U}}^A \cdot \left({\bf U}^A\right)^{-1}\bigg]_{ij} =
\frac{1}{2} (\delta_{ir} \delta_{js} - \delta_{is} \delta_{jr})
\sum_{B\ne A} \sum_{k=0}^\infty \frac{(-1)^k}{k!} \nn \\
&& \times \bigg[ \sum_{l=0}^\infty \frac{(2k+2l+1)!!}{l!}
  \frac{\bodyM^A_L}{\bodyM^A} \bodyM^B_K v^A_r \frac{
    n^{BA}_{<sKL>}}{ r_{BA}^{k+l+2}} \nn \\
&& + \frac{4(2k+1)!!}{k+1} {\dot \bodyM}^B_{Kr} \frac{
    n^{BA}_{<sK>}}{r_{BA}^{k+2}} \nn \\
&& - \frac{4 k (2k+1)!!}{k+1} \bodyS^B_{mK-1} \epsilon_{b_kmr}
  \frac{n^{BA}_{<sK>}}{r_{BA}^{k+2}} \bigg] + O(\varepsilon^4).
\label{eq:Uevol1}
\end{eqnarray}

%%%%%%%%%%%%%%%%%%%%%%%%%%%%%%%%%%%%%%%%%%%%%%%%%%%%%%%%%%%%%%%%%%%%%%%%%%%%%%%%%%%%%%%%%%%%%%%%%%%%%%%%%%%%%%%%%%%%%%%%%%%%%%%%%%%%%%%%%%%%%%%%

\section{Conclusion}

In this paper, we have given a surface integral derivation of the
full post-1-Newtonian DSX laws of motion (\ref{mainresult}).
We have shown that these
laws of motion apply to a wide class of strongly self-gravitating
objects, provided that the mass and current moments are appropriately
defined in terms of the asymptotic weak field metric in the buffer
regions around each body.
We have given an explicit form for the coupled equations
of motion of the bodies' center of mass worldlines
including the effects of \textit{all} the post-Newtonian mass and
current multipole couplings.
To the best of our knowledge this is
the first time these equations of motion have been written out
explicitly.
The second paper in this series will include a surface integral derivations of the
evolution laws (\protect{\ref{eq:mainresult0}})
and (\protect{\ref{eq:spinresult}}) for the energy (mass monopole) and
the spin $S^A_i$ \cite{Racine}.

\begin{acknowledgments}
\'{E}.R. acknowledges support by Le Fonds Qu\'eb\'ecois de Recherche
sur la Nature et les Technologies (NATEQ, formerly FCAR).  We
thank Steve Drasco, Marc Favata, Nicholas Jones and James York for helpful discussions and comments. This research was supported in part by the Radcliffe Institute and by NSF grants PHY-9722189 and PHY-0140209.
\end{acknowledgments}

\appendix

\section{Useful identities involving STF tensors}
\label{stf}

In this appendix we give the general definition of the STF projection
of an arbitrary tensor.  We also review some identities that are
useful for manipulating expressions
involving STF tensors.

The STF projection of any tensor $T_L$ is obtained by taking the symmetric part of $T_L$, and then subtracting out all the
partial traces.  One obtains in this way a unique symmetric tensor
that is trace free on all pairs of indices.  The general formula for
this projection is
\begin{equation}
T_{<L>} \equiv \sum_{k=0}^{[l/2]}c_k^l
\delta_{(a_1a_2}...\delta_{a_{2k-1}a_{2k}}S_{a_{2k+1}...a_l)j_1j_1...j_kj_k},
\end{equation}
where
$[l/2]$ is the largest integer less than or equal to $l/2$, the
coefficients $c^l_k$ are given by
\begin{equation}
c^l_k = (-1)^k\frac{l!}{(l-2k)!}\frac{(2l-2k-1)!!}{(2l-1)!!(2k)!!},
\end{equation}
$l!!$ means $l (l-2) (l-4) \ldots (4)(2)$ or $l (l-2) (l-4)
\ldots (3)(1)$, and
\begin{equation}
S_L \equiv T_{(L)}
\end{equation}
is the symmetric part of $T$.  For example,
\begin{equation}
T_{<abc>} = S_{abc} - \frac{1}{5} \left[ \delta_{ab} S_{cdd} +
  \delta_{ac} S_{bdd} + \delta_{bc} S_{add} \right].
\end{equation}

From the definition of the STF projection one can derive the following
``peeling formula'' \cite{dsxI} for any STF tensor $T_L$ and vector $V_i$:
\begin{eqnarray} \nn
V_{<j}T_{L>} &=& \frac{1}{l+1} V_jT_L + \frac{l}{l+1}
T_{j<L-1}V_{a_l>}  \\
& &  - \frac{2l}{(l+1)(2l+1)}V_kT_{k<L-1}\delta_{a_l>j}.
\label{eq:peeling}
\end{eqnarray}
From this peeling formula one can obtain the identities
\begin{equation}
T_{i<L}\delta_{j>j} = \frac{2l+3}{2l+1}T_{iL}
\end{equation}
and
\begin{equation}
T_{j<L}\delta_{j>i} = \frac{1}{(l+1)(2l+1)}T_{iL},
\end{equation}
which are valid for any STF tensor $T_{L+1}$.

Next, some useful formulae involving derivatives
are
\begin{equation}
| {\bm x} |^2 \partial_L \frac{1}{|{\bm x}|} = - (2l -1)
  \partial_{<L>} |{\bm x}|,
\label{eq:ident00}
\end{equation}
and
\begin{equation}
\partial_{iL}|\bm{x}| = \partial_{<iL>}|\bm{x}| + \frac{l(l+1)}{2l+1}\delta_{(ia_l}\partial_{L-1)}\frac{1}{|\bm{x}|}.
\end{equation}
Also, given a sequence of STF tensors $T_L$, one for each $l$, we have
the identities
\begin{eqnarray}
\sum_{l=0}^\infty\frac{(-1)^{l}}{l!}{T}_Lx^j\partial_L\frac{1}{|\bm{x}|}
& =& \sum_{l=0}^\infty
 \frac{(-1)^{l}}{l!}\bigg[\left(\frac{2l+1}{2l+3}\right){T}_{jL}\partial_L\frac{1}{|\bm{x}|}
 \nonumber \\
\mbox{} && + {T}_L\partial_{<jL>}|\bm{x}|\bigg]
\end{eqnarray}
and
\begin{equation}
\sum_{l=0}^\infty\frac{1}{l!}{T}_Lx^{jL} = \sum_{l=0}^\infty\frac{1}{l!}\left({T}_Lx^{<jL>} + \frac{1}{2l+3}|\bm{x}|^2{T}_{jL}x^L\right).
\label{eq:ident04}
\end{equation}
These identities are used in Sec.\ \ref{sec:momentsgaugetransformation}
above in the computation of the transformation laws for the multipole
and tidal moments.  In Sec. \ref{eom} we use the identities
\begin{eqnarray}
T_L\,\partial_{iL}\frac{1}{|\bm{x}|} &=&
\frac{1}{|\bm{x}|^{l+2}}\left(a_l T_{iL-1}n^{L-1} - b_l
T_Ln^{iL}\right),\ \ \ \ \
\label{eq:identity1}
\end{eqnarray}
\begin{eqnarray}
T_L\,\partial_i \,x^L &=& l|\bm{x}|^{l-1} T_{iL-1}n^{L-1} ,
\label{eq:identity2}
\end{eqnarray}
where $a_l = (-1)^ll (2l-1)!!$, $b_l = (-1)^l (2l+1)!!$, and $n^i =
x^i / |\bm{x}|$.  We also use the following integrals over the
unit sphere given in Thorne \cite{thorne}
\begin{equation}\label{unitvectorintegralA}
\frac{1}{4\pi}\oint\, n^{2L+1} \,d\Omega = 0,
\end{equation}
\begin{equation}\label{unitvectorintegralB}
\frac{1}{4\pi}\oint\, n^{2L} \,d\Omega = \frac{1}{2l+1}\delta_{(i_1i_2}...\delta_{i_{2l-1}i_{2l})},
\end{equation}
\begin{eqnarray}\nn
\frac{1}{4\pi}\oint \, T_KS_L n^Kn^L\,\,d\Omega &=&
\frac{l!}{(2l+1)!!}T_LS_L \;\;\text{if}\;\; k=l
\\\label{STFintegralA}
&=& 0 \;\;\text{if}\;\; k\neq l ,
\end{eqnarray}
and
\begin{eqnarray}\nn
\frac{1}{4\pi}\oint \, T_KS_L n^Kn^Ln^i\,\,d\Omega &=& \frac{(l+1)!}{(2l+3)!!}T_{iL}S_L \;\;\text{if}\;\; k=l+1  \\\label{STFintegralB}
&=& 0 \;\;\text{if}\;\; |k- l|\neq 1.
\end{eqnarray}

Another useful peeling identity, valid for a $k+1$ index tensor which is STF
on its last $k$ indices, is
\begin{eqnarray}
&& T_{<i K} z_{L-K>} = \frac{k+1}{l+1} T_{i<K} z_{L-K>} \nonumber \\
\mbox{} && + \frac{l-k}{l+1}
T_{<K+1} z_{L-(K+1)>} z_i
\nonumber \\ \mbox{} &&
- \frac{2 (k+1)(l-k)}{(l+1)(2l+1)} z_j T_{j<K} z_{(L-1)-K} \delta_{a_l>i}
\nonumber \\ \mbox{} &&
- \frac{(l-k)(l-k-1)}{(l+1)(2l+1)} z_j z_j
T_{<K+1} z_{(L-1)-(K+1)}
\delta_{a_l>i}.\nn\\
\label{eq:ident14}
\end{eqnarray}
Finally, for any STF tensor $T$ and vectors $v_i$ and $z_i$ we have
the identity
\begin{eqnarray}
&& v_{<i} T_{LK>} z^K = \frac{l+1}{l+k+1} v_{<i} T_{L>K} z^K
\nn \\ \mbox{} &&
- \frac{k(l+1)}{(l+k+1)(2l+2k+1)} v_j T_{jK-1<iL-1} z_{a_l>} z^{K-1}
\nn \\ \mbox{} &&
- \frac{k(k-1)}{(l+k+1) (2l + 2k +1)} z_m z_m v_j T_{j<iL>K-2}
z^{K-2}
\nn \\ \mbox{} &&
+ \frac{k}{l+k+1} v_j z_j T_{<iL>K-1} z^{K-1}.
\label{eq:ident15}
\end{eqnarray}

\section{Derivation of gauge transformation parameterization}
\label{app:gaugefreedom}

In this appendix we consider harmonic, conformally Cartesian coordinate
systems on a spacetime region ${\cal D} \times (t_0,t_1)$, where ${\cal
D}$ is a simply connected spatial region and $(t_0,t_1)$ is an open
interval of time.  We show that the most general gauge transformation
between two such coordinate systems is of the form given by Eqs.\
(\ref{coordinatetransformation}) -- (\ref{conshgC}), up to constant
displacements in time and up to time-independent spatial
rotations.

We start by reviewing the well-known argument that gives this result
to Newtonian order.  Let the coordinate transformation to zeroth
order in $\varepsilon$ be
\begin{eqnarray}
x^i &=& x^i({\bar t},{\bar x}^j) + O(\varepsilon^2), \nonumber \\
t &=& t({\bar t},{\bar x}^j) + O(\varepsilon^2).
\end{eqnarray}
Substituting this into the metric expansion (\ref{metric}), we find
that the leading order expression for the spatial metric is
\begin{equation}
- \frac{1}{\varepsilon^2}
\frac{\partial t}{\partial {\bar x}^i}
\frac{\partial t}{\partial {\bar x}^j} d{\bar x}^i d{\bar x}^j + O(1).
\end{equation}
This is in conflict with the expansion (\ref{transformedmetric})
unless $\partial t / \partial {\bar x}^i=0$.
Similarly, the leading order expression for the time-time piece of the
line element is
\begin{equation}
- \frac{1}{\varepsilon^2}
\left( \frac{\partial t}{\partial {\bar t}} \right)^2
d{\bar t}^2+ O(1),
\end{equation}
which disagrees with the expansion (\ref{transformedmetric}) unless
$\partial t / \partial {\bar t} = \pm 1$.  Assuming that the
coordinate transformation preserves the time orientation and
neglecting constant displacements in time we obtain $t = {\bar t} +
O(\varepsilon^2)$.  Therefore we can write
\begin{equation}
t = {\bar t} + \varepsilon^2 \alpha({\bar t},{\bar x}^j) +
O(\varepsilon^4),
\label{eq:t0}
\end{equation}
where the function $\alpha({\bar t},{\bar x}^j)$ is as yet undetermined.

The leading order expression for the spatial metric is now
\begin{equation}
\delta_{kl} \frac{\partial x^k}{\partial {\bar x}^i}
\frac{\partial x^l}{\partial {\bar x}^j} d{\bar x}^i d{\bar x}^j +
O(\varepsilon^2) = \delta_{ij} d{\bar x}^i d {\bar x}^j + O(\varepsilon^2),
\end{equation}
where we have used the expansion (\ref{transformedmetric}).  Thus, for
each fixed ${\bar t}$, the function $x^i = x^i({\bar t},{\bar x}^j)$
is an isometry of 3-dimensional Euclidean space.  It follows that
\begin{equation}
x^i = R^i_{\ \,j}({\bar t}) {\bar x}^j + z^i({\bar t}) + O(\varepsilon^2)
\label{eq:x0}
\end{equation}
for some time-dependent rotation matrix $R^i_{\ \,j}({\bar t})$ and some
time-dependent displacement $z^i({\bar t})$.  Using Eqs.\
(\ref{metric}), (\ref{eq:t0}) and
(\ref{eq:x0}) the leading order expression for the space-time piece of
the line element is
\begin{equation}
\left\{ 2 \delta_{ik} R^k_{\ \,l}({\bar t}) \left[ {\dot R}^i_{\
    \,j}({\bar t}) {\bar x}^j + {\dot z}^i({\bar t}) \right] - 2
    \frac{\partial \alpha}{\partial {\bar x}^l} \right\} d{\bar t}
    d{\bar x}^l + O(\varepsilon^2).
\end{equation}
The first term here must vanish in order to be compatible with Eq.\
(\ref{transformedmetric}), which gives
\begin{equation}
\delta_{ik} R^k_{\ \,l}({\bar t}) \left[ {\dot R}^i_{\
    \,j}({\bar t}) {\bar x}^j + {\dot z}^i({\bar t}) \right] =
     \frac{\partial \alpha}{\partial {\bar x}^l}.
\label{eq:dds}
\end{equation}
If ${\dot R}^i_{\ \,j}({\bar t})$ is non-vanishing, it is impossible
to find any function $\alpha({\bar t},{\bar x}^j)$ which satisfies this
equation, since the left hand side is not a pure gradient.  Therefore
we conclude that the rotation matrix is time-independent, and we
choose the new coordinate system ${\bar x}^i$ so that $R^i_{\ \,j} =
\delta^i_j$.  We can now solve Eq.\ (\ref{eq:dds}) for the function
$\alpha$, which gives
\begin{equation}
\alpha({\bar t},{\bar x}^j) = \alpha_{\text{c}}({\bar t}) + {\dot
  z}^i({\bar t}) {\bar x}^i,
\label{eq:alphaans}
\end{equation}
where $\alpha_{\text{c}}({\bar t})$ is an arbitrary function of ${\bar
t}$, cf.\ Eq.\ (\ref{conshgA}) above.

To summarize, the coordinate transformation to Newtonian order is
given by
\begin{eqnarray}
x^i({\bar t},{\bar x}^j) &=& {\bar x}^i + z^i({\bar t}) +
O(\varepsilon^2) \nn \\
t({\bar t},{\bar x}^j) &=& {\bar t} + \varepsilon^2 \alpha({\bar t},{\bar
  x}^j) + O(\varepsilon^4),
\label{eq:tansN1}
\end{eqnarray}
where $\alpha$ is given by Eq.\ (\ref{eq:alphaans}).  The
transformation law for the Newtonian potential $\Phi$ can now be obtained
by substituting Eqs.\ (\ref{eq:tansN1}) into the
metric expansion (\ref{metric}) and comparing the time-time piece with
the metric expansion (\ref{transformedmetric}); the result is given by Eq.\
(\ref{transfA}).

We now turn to the post-Newtonian extension of this computation.
We assume that the coordinate transformation can be written as
the Newtonian-order
coordinate transformation (\ref{eq:tansN1})
plus arbitrary post-Newtonian correction terms:
\begin{eqnarray}
x^i &=& {\bar x}^i + z^i({\bar t}) + \ve^2 h^i({\bar t},{\bar x}^j)  +
O(\ve^4), \nonumber \\
\mbox{} t &=& {\bar t} + \ve^2 \alpha({\bar t},{\bar x}^j) + \ve^4
\beta({\bar t},{\bar x}^j) + O(\ve^6).
\label{eq:tansPN}
\end{eqnarray}
Here the functions $h^i({\bar t},{\bar x}^j)$ and $\beta({\bar t},{\bar
x}^j)$ are arbitrary.  As before we can compute the
transformed metric by combining the coordinate transformation
(\ref{eq:tansPN}) with the metric expansion (\ref{metric}).  The
resulting leading order expression for the spatial metric is
\begin{equation}
\left\{ \delta_{ij} + \varepsilon^2 \left[ -2 {\hat \Phi} \delta_{ij}
  - {\dot z}_i {\dot z}_j + h_{i,j} + h_{j,i} \right] \right\} d{\bar
  x}^i d{\bar x}^j,
\end{equation}
where we are using the notation (\ref{eq:hatnotation}).
Comparing this with the metric expansion (\ref{transformedmetric}) and
using Eqs.\ (\ref{transfA}) and (\ref{conshgA}) gives the following
differential equation for $h^i({\bar t},{\bar x}^j)$:
\be
h_{i,j} + h_{j,i} = -  2 \delta_{ij} \left[ {\ddot z}_k {\bar x}^k  +
{\dot \alpha}_{\text{c}} - \frac{1}{2} {\dot {\bf z}}^2 \right] + {\dot
z}_i {\dot z}_j.
\ee
The general solution to this equation consists of a homogeneous
solution plus an inhomogeneous solution.  The homogeneous solution is
just the general Killing vector of three dimensional Euclidean space,
which is
\be
h^i(\bar{t},\bar{x}^j) = h_{\text{c}}^i(\bar{t}) +
\epsilon^{ijk}\bar{x}_jR_k(\bar{t}),
\label{gggg2x0}
\ee
where the functions $h^i_{\text{c}}({\bar t})$ and $R_k({\bar t})$ are
arbitrary.  The full solution that we will use for $h^i$ is the sum of
(\ref{gggg2x0}) and the inhomogeneous
solution, which can be obtained by inspection.  The result is [cf.\
Eq.\ (\ref{conshgB}) above]
\begin{eqnarray}
h^i(\bar{t},\bar{x}^j) &=& h_{\text{c}}^i(\bar{t}) +
\epsilon^{ijk}\bar{x}_jR_k(\bar{t}) +
\frac{1}{2}\ddot{z}^i(\bar{t})\bar{x}_j\bar{x}^j
\nn \\
& & - \bar{x}^i\dot{\alpha}_{\text{c}}(\bar{t})
- \bar{x}^i {\bar x}_j \ddot{z}^j(\bar{t}) +
\frac{1}{2}\bar{x}^i\dot{z}_j(\bar{t})\dot{z}^j(\bar{t})
\nn \\
\mbox{} && +
\frac{1}{2}\dot{z}^i(\bar{t})\dot{z}^j(\bar{t})\bar{x}_j.
\label{conshgB1}
\end{eqnarray}

Next, we use Eqs.\ (\ref{metric}), (\ref{transformedmetric}),
(\ref{eq:tansPN}) and (\ref{conshgB1}) to compute the
transformed gravitomagnetic potential ${\bar \zeta}^i$.  The result is
\begin{eqnarray}
{\bar \zeta}_i({\bar t},{\bar x}^j) &=&  {\hat \zeta}_i({\bar t},{\bar
  x}^j) - \frac{1}{2} {\ddot z}^i({\bar t}) {\dot z}_j({\bar
  t}) {\bar x}^j
+ \frac{1}{2} {\dddot z}^i({\bar t}) {\bar x}_j {\bar x}^j
\nonumber  \\ \mbox{} &&
+ {\bar x}^i \left[  2 {\dot z}^j({\bar t}) {\ddot z}_j({\bar t})  -
{\bar x}^j {\dddot z}_j({\bar t}) - {\ddot \alpha}_{\text{c}}({\bar t}) \right]
\nonumber  \\ \mbox{} &&
- {\dot z}^i({\bar t}) \bigg[ 4 {\hat \Phi}({\bar t},{\bar x}^j) + 2
  {\dot \alpha}_{\text{c}}({\bar t}) + \frac{3}{2} x^j {\ddot
  z}_j({\bar t})
\nonumber  \\ \mbox{} &&
- {\dot z}^j({\bar t}) {\dot z}_j({\bar t})
\bigg]
- \frac{\partial \beta}{\partial {\bar x}^i}({\bar t},{\bar x}^j) +
  {\dot h}_{\text{c}}^i({\bar t})
\nonumber  \\ \mbox{} &&
+ \epsilon_{ijk} {\bar  x}^j {\dot
R}^k({\bar t}) + \epsilon_{ijk} R^j({\bar t})
{\dot z}^k({\bar t}).
\label{gggxx}
\end{eqnarray}
Combining this with the expression (\ref{transfA}) for the transformed
Newtonian potential, and using the harmonic gauge condition
(\ref{harmonicgauge})
applied to both the original and barred coordinate systems gives
the differential equation
\begin{equation}
{\bar \nabla}^2 \beta = {\dddot z}_j({\bar t}) {\bar x}^j + {\ddot
  \alpha}_{\text{c}}({\bar t}).
\end{equation}
The general solution to this equation is
\be
\beta({\bar t},{\bar x}^j) = \beta_{\text{h}}({\bar t},{\bar x}^j) +
\left[ \frac{1}{10} {\dddot z}_k({\bar
t}) {\bar x}^k  + \frac{1}{6} {\ddot \alpha}_{\text{c}}({\bar t})
  \right] {\bar x}_j {\bar x}^j,
\label{betaf}
\ee
where $\beta_{\text{h}}$ is an arbitrary harmonic function, cf.\ Eq.\
(\ref{conshgC}) above.  This completes the derivation.

\section{Piece of surface integral that depend linearly on moments}
\label{checks}

In this appendix we compute explicitly the piece of the surface
integral (\ref{1pneom1}) that depends linearly on the multipole, tidal
and gauge moments.  That linear piece appears on the right hand side of Eq.\
(\ref{eq:schematic}) as the function
${\cal G}_i(
{\,}^{\scriptscriptstyle \text{n}}\!\ddddot{M}_L,
{\,}^{\scriptscriptstyle \text{n}}\!\ddddot{G}_L,
{\dddot H}_L,{\dddot S}_L,{\dddot \mu}_L,{\dddot \nu}_L;R)$.  We will
show that this function vanishes.

We start by noting that the splitting of the surface integral
(\ref{1pneom1}) into pieces that are linear in the moments and pieces
that are quadratic in the moments is unambiguous for all the multipole
and tidal moments, except for the Newtonian mass dipole $\nM_i(t)$.
That mass dipole is constrained by the Newtonian equation of motion
(\ref{newtonianlom}), and therefore a term proportional to the fourth
time derivative of $\nM_i$ could be re-expressed as a quadratic
expression in the moments $\nM_L$ and $\nG_L$ and their time
derivatives up to second order.  We resolve this ambiguity by
demanding that there be no dependence on $\nM_i$ in ${\cal G}_i$;
the relevant term if present can be re-expressed as a quadratic
expression and moved into the function ${\cal F}_i$.

To compute the linear piece of the surface integral, we simply drop
all the quadratic source terms in the post-2-Newtonian field equations
and gauge conditions (\ref{gaugeA}) -- (\ref{2pnfieldeqB}).  We also
drop the term $\pncalT^{ij}$ in Eq. (\ref{1pneom1}), since the
expression (\ref{eq:LLP2N}) for $\pncalT^{ij}$ is explicitly
quadratic.  We also assume without loss of generality that $\nM_i=0$,
for the reason discussed above.  This yields
from Eqs.\ (\ref{1pneom1}) and (\ref{eq:schematic})
the set of equations
\begin{eqnarray}
{\cal G}_i
&=& - \frac{1}{16\pi} \oint \left[
\partial_j { {\dot \xi}}^i_0
+{\ddot {\chi}}^{ij}_0
\right] d^2\Sigma^j,\label{1pneom1a}
\end{eqnarray}
where
\begin{eqnarray}\label{gaugeA1}
\partial_j \chi^{ij}_0 &=& - {\dot \zeta}^i_0, \ \ \ \ \
\partial_i \xi^i_0 - \dot{\chi}^{kk}_0 =- 4\dot{\psi}_0, \\
\nabla^2 \xi^i_0 &=& \ddot{\zeta}^i_0, \ \ \ \ \ \ \
\nabla^2 \chi^{ij}_0 = 0.
\label{2pnfieldeqB1}
\end{eqnarray}
Here the subscripts $0$ indicate that the post-1-Newtonian mass dipole
associated with the potentials $(\Phi_0,\zeta^i_0,\psi_0)$ vanishes,
cf., the discussion in Sec.\ \ref{sec:pnderiv} above.

To compute the function ${\cal G}_i$\footnote{
One might think that the easiest way to evaluate the expression
(\protect{\ref{1pneom1a}}) for ${\cal G}_i$ is to use Gauss' theorem
to convert the surface integral to a volume integral.  However, this
strategy does not work: Because the fields are only defined on the
domain $r_0 < r < r_1$ one obtains a surface term at $r=r_0$ in
addition to the volume term.  It is impossible to extend the definitions
of the fields smoothly all the way to $r=0$, so one is always forced
to evaluate a surface term.}, we can pick
any solution of
the post-2-Newtonian equations (\ref{gaugeA1}) -- (\ref{2pnfieldeqB1}),
since we showed in Sec.\ \ref{sec:pnderiv} that the result is
independent of which solution is chosen.  A particular solution
$(\xi^i_1,\chi^{ij}_1)$ of
Eqs.\ (\ref{2pnfieldeqB1}) can be obtained using the expansion
(\ref{reducedzeta}) of the gravitomagnetic potential $\zeta^i$.  This gives
\begin{eqnarray}
\label{eq:e12}
\chi^{ij}_1 &=& 0, \\
\xi^i_1 &=& \sum_{l=0}^\infty \left[ \frac{(-1)^{l+1}}{2 l!} {\ddot
Z}_{iL} \partial_L |\bm{x}| - \frac{|\bm{x}|^2}{2 (2l+3) l!} {\ddot
Y}_{iL} x^L\right].\nn \\
\end{eqnarray}
These potentials do not satisfy the gauge conditions (\ref{gaugeA1}),
but we can fix this by adding appropriately chosen solutions of Laplace's
equation.  Thus, we define
\begin{eqnarray}
\label{2pnsolA1}
\xi^i_0 &=& \xi_1^i + \sum_{l=0}^\infty
\frac{(-1)^{l+1}}{l!}X_{iL}\partial_L\frac{1}{|\bm{x}|} -
\frac{1}{l!}W_{iL}x^L, \\
\chi^{ij}_0 &=& \chi^{ij}_1 + \sum_{l=0}^\infty
\frac{(-1)^{l+1}}{l!}C_{ijL}\partial_L\frac{1}{|\bm{x}|} -
\frac{1}{l!}B_{ijL}x^L.
\label{2pnsolB1}
\end{eqnarray}
Inserting Eqs.\ (\ref{eq:e12}) -- (\ref{2pnsolB1}) into the formula
(\ref{1pneom1a}) gives
\begin{equation}
{\cal G}_i = - \frac{1}{12} R^3 {\ddot \nu}_i +
\frac{1}{4}X_i - \frac{1}{12}\dot{C}_{ijj} - \frac{R^3}{12}\dot{B}_{ijj}.
\label{eq:calGans}
\end{equation}
To obtain the moments $X_i$, ${\dot C}_{ijj}$ and ${\dot B}_{ijj}$ we
substitute Eqs.\ (\ref{eq:e12}) -- (\ref{2pnsolB1}) into the gauge
conditions (\ref{gaugeA1}).  A useful intermediate result is
\begin{eqnarray}
\partial_i \xi^i_1 &=& - \sum_{l=0}^\infty \frac{(-1)^{l+1}}{l!}
\left[ 2 \ndddotM_L \partial_L |\bm{x}|
+ \frac{{\ddot \mu}_L}{2l+3} \partial_L \frac{1}{|\bm{x}|}
\right] \nn \\
&& + \sum_{l=0}^\infty \frac{x^L}{(2l+3) l!} \left[ 2 |\bm{x}|^2
\ndddotG_L - {\ddot \nu}_{iL} x^i\right].
\end{eqnarray}
This gives $B_{ijj} = - {\dot \nu}_i$ and $X_i - {\dot
C}_{ijj}/3 =0$, yielding from Eq.\ (\ref{eq:calGans}) that ${\cal
G}_i=0$.

\section{Law of motion for a single body with weak self-gravity}
\label{functionalform}

In this appendix we sketch briefly the DSX derivation \cite{dsxII} of
Eq.\ (\ref{mainresult}), translated into our
notation.  This equation is obtained by
direct computation of the second time derivative of the mass
dipole, making use of the stress-energy conservation law in the
interior of the body.  In our
notation, the total mass dipole is [cf.\ Eqs.\
(\ref{eq:integraldefine}) and (\ref{eq:pnMintegraldefine}) above]
\begin{widetext}
\begin{equation}\label{bdmassdipole}
\nM_i + \varepsilon^2\pnM_i \equiv \int_{r<r_-}
\left[x^i \,^{\scriptscriptstyle \text{n}}\!{T^{00}} +
  \varepsilon^2\left(x^i \,^{\scriptscriptstyle \text{pn}}\!{T^{00}} +
  x^i \,^{\scriptscriptstyle \text{n}}\!{T^{jj}} +
  \frac{x^i |\bm{x}|^2}{6}\frac{\partial^2\,^{\scriptscriptstyle \text{n}}\!{T^{00}}}
{\partial t^2} - \frac{6x^{<ij>}}{5}
\frac{\partial \,^{\scriptscriptstyle \text{n}}\!{T^{0j}}}{\partial
  t}\right)\right] \, d^3x.
\end{equation}
The conservation equations $\nabla_\mu T^{\mu\nu}$ can be written
using the expansions (\ref{metric}) and (\ref{eq:st})
in the form \cite{dsxII}
\begin{equation}\label{zerothcomponent}
\frac{\partial}{\partial t}\left( \,^{\scriptscriptstyle \text{n}}\!{T^{00}} +
\varepsilon^2 \,^{\scriptscriptstyle \text{pn}}\!{T^{00}} \right)+
\frac{\partial}{\partial x^j} \left(\,^{\scriptscriptstyle \text{n}}\!{T^{0j}}
+ \varepsilon^2 \,^{\scriptscriptstyle \text{pn}}\!{T^{0j}} \right)=
\,\,\varepsilon^2 \,^{\scriptscriptstyle \text{n}}\!{T^{00}} \frac{\partial
  \Phi}{\partial t} + O(\varepsilon^4),
\end{equation}
\begin{eqnarray}\nn
\frac{\partial}{\partial t} &\,& \!\!\!\!\!\!\!\left[\left(1 -
  4\varepsilon^2\Phi \right) \left(\,^{\scriptscriptstyle \text{n}}\!{T^{0i}} +
  \,\varepsilon^2 \,^{\scriptscriptstyle \text{pn}}\!{T^{0i}}\right)\right] +
\frac{\partial}{\partial x^j} \left[\left(1 - 4\varepsilon^2\Phi
  \right)\left(\,^{\scriptscriptstyle \text{n}}\!{T^{ij}} + \,\varepsilon^2
  \,^{\scriptscriptstyle \text{pn}}\!{T^{ij}}\right)\right] =
\\\label{ithcomponent}
& & -\left[ \,^{\scriptscriptstyle \text{n}}\!{T^{00}} + \varepsilon^2\left(
  \,^{\scriptscriptstyle \text{pn}}\!{T^{00}} +
  \,^{\scriptscriptstyle \text{n}}\!{T^{kk}}\right)\right]\left[\varepsilon^2\frac{\partial}{\partial
  t} \zeta_i + \frac{\partial}{\partial x^i} \left(\Phi +
  \varepsilon^2\psi \right) \right] -
  \varepsilon^2\,^{\scriptscriptstyle \text{n}}\!{T^{0j}}\left(\frac{\partial}{\partial
  x^j }\zeta_{i} - \frac{\partial}{\partial
  x^i}\zeta_{j}\right) + O(\varepsilon^4).
\end{eqnarray}
These conservation equations can be used to evaluate explicitly the time
derivatives appearing in Eq.\ (\ref{bdmassdipole}). Some
algebra leads to the following expression
\begin{eqnarray}\nn
\nM_i + \varepsilon^2\pnM_i &=& \int_{r < r_-} d^3x
\left[x^i \left(\,^{\scriptscriptstyle \text{n}}\!{T^{00}} + \varepsilon^2
  \,^{\scriptscriptstyle \text{pn}}\!{T^{00}} \right)
+ \varepsilon^2 \,^{\scriptscriptstyle \text{n}}\!{T^{00}}\left(
  x^i x^j \frac{\partial \Phi}{\partial x^j} -
  \frac{|\bm{x}|^2}{2}\frac{\partial \Phi}{\partial x^i} \right)
  \right]. \\\label{reducedbdmassdipole}
\end{eqnarray}
Taking two time derivatives of this expression
and using the conservation equations (\ref{zerothcomponent}) and
(\ref{ithcomponent}) gives
\begin{eqnarray}\nn
\frac{d^2}{d t^2}\left(\nM_i +
\varepsilon^2\pnM_i\right) &=& - \int_{r<r_-} \left\{\left[
  \,^{\scriptscriptstyle \text{n}}\!{T^{00}} + \varepsilon^2\left(
  \,^{\scriptscriptstyle \text{pn}}\!{T^{00}} +
\,^{\scriptscriptstyle \text{n}}\!{T^{kk}}\right)\right]
\left[\varepsilon^2\frac{\partial}{\partial t} \zeta_i +
  \frac{\partial}{\partial x^i}
\left(\Phi + \varepsilon^2\psi \right) \right] \right.\\\nn
& & +
\left.
\varepsilon^2\,^{\scriptscriptstyle \text{n}}\!{T^{0j}}\left(\frac{\partial}{\partial
  x^{j}}\zeta_{i} - \frac{\partial}{\partial
  x^{i}}\zeta_{j}\right)\right\} d^3x + \varepsilon^2
\frac{d}{d t} \int_{r<r_-} \left(4\Phi
\,^{\scriptscriptstyle \text{n}}\!{T^{0i}} + \,x^i
\,^{\scriptscriptstyle \text{n}}\!{T^{00}}\frac{\partial \Phi}{\partial t}
\right)d^3x \\\label{doubledot}
& & + \varepsilon^2 \frac{d^2}{d t^2} \int_{r<r_-} \left(x^i
x^j - \frac{1}{2} x^k x^k \delta_{ij}
\right)\,^{\scriptscriptstyle \text{n}}\!{T^{00}}\frac{\partial \Phi}{\partial
  x^j}\, d^3x.
\end{eqnarray}
We next substitute in explicit expressions for the gravitational
potentials in which the intrinsic terms are expressed in terms of
integrals over the matter distribution using the field equations
(\ref{WfieldeqA}) -- (\ref{WfieldeqC}):
\begin{eqnarray}\label{newtpot}
\Phi &=& \int_{r < r_-}
\frac{\,^{\scriptscriptstyle \text{n}}\!{T^{00}}(t,\bm{x}^\prime)}{|\bm{x} -
  \bm{x}^\prime|} \, d^3x^\prime  - \sum_{l=0}^\infty
\frac{1}{l!}\nG_L x^{L} , \\\nn
\psi &=& \int_{r<r_-}
\frac{\,^{\scriptscriptstyle \text{pn}}\!{T^{00}}(t,\bm{x}^\prime) +
  \,^{\scriptscriptstyle \text{n}}\!{T^{jj}} (t,\bm{x}^\prime)}{|\bm{x} -
  \bm{x}^\prime|} \, d^3x^\prime
+ \frac{d^2}{d t^2} \int_{r<r_-}
\,^{\scriptscriptstyle \text{n}}\!{T^{00}}(t,\bm{x}^\prime)\frac{|\bm{x} -
  \bm{x}^\prime|}{2} \, d^3x^\prime
 - \sum_{l=0}^\infty \frac{1}{l!}\left[\pnG_L +
  \frac{|\bm{x}|^2}{2(2l+3)}\,^{\scriptscriptstyle
    \text{n}}\!\ddot{G}_L\right]x^{L} ,
\end{eqnarray}
and
\begin{eqnarray}
\zeta_i &=& \int_{r<r_-}
\frac{\,^{\scriptscriptstyle \text{n}}\!{T^{0i}}(t,\bm{x}^\prime)}{|\bm{x} -
  \bm{x}^\prime|} \, d^3x^\prime  - \sum_{l=0}^\infty
\frac{1}{l!}Y_{iL} x^{L}.
\end{eqnarray}
It is a straightforward exercise to show that all the terms involving
double integrals over $\bm{x}$ and $\bm{x}^\prime$ in
Eq.(\ref{doubledot}) cancel out. The laws of motion can thus be
obtained by simply substituting the tidal pieces of the gravitational
potentials into Eq.\ (\ref{doubledot}). The remaining integrals over
$\bm{x}$ can then be expressed in terms of the moments $\nM_L$,
$\pnM_L$ and $Z_{iL}$ via the integral definitions
 (\ref{eq:integraldefine}), (\ref{eq:pnMintegraldefine})
and
\begin{equation}
Z_{iL}(t) = 4 \int_{r<r_-} \,
\,^{\scriptscriptstyle \text{n}}\!{T^{0i}}(t,x^j) x^{<L>}
\,d^3 x.
\label{eq:integraldefineZ}
\end{equation}
This gives
\begin{eqnarray}\nn
\mathcal{F}_i &=& \sum_{l=0}^\infty \frac{1}{l!} \left[\pnM_L\nG_{iL} + \nM_L\left(\pnG_{iL} + \dot{Y}_{iL} - \dot{Y}_{<iL>}\right) - \frac{(l+2)(2l+1)}{(2l+3)}\nM_{iL}\nddotG_L - (2l+1){\,}^{\text{n}}\!\dot{M}_{iL}{\,}^{\text{n}}\!\dot{G}_L - l{\,}^{\text{n}}\!\ddot{M}_{iL}\nG_L \right. \\\label{mainresultII}
& & \left.  + \frac{(2l+1)}{(l+1)(2l+3)}\dot{\mu}_{L}\nG_{iL} + \frac{1}{2}Z_{jL}Y_{[ij]L} - \dot{Z}_{iL}\nG_L - Z_{iL}{\,}^{\text{n}}\!\dot{G}_L\right],
\end{eqnarray}
where we have used Eqs.\ (\ref{newtonianlom}), (\ref{eq:schematic1})
and (\ref{doubledot}).
Using the STF decompositions (\ref{eq:ZiLdecompos}) and
(\ref{eq:YiLdecompos}) of the moments
$Z_{iL}$ and $Y_{iL}$, it is straightforward to check that
Eq.\ (\ref{mainresultII}) is equivalent to Eq.\ (\ref{mainresult}).
\end{widetext}

\section{Formulae for moments in terms of surface integrals}
\label{sec:welldefined}

In this appendix we
show that the various moments are uniquely defined by the expansions
(\ref{basicmodelA}) -- (\ref{reducedzeta}), by
writing down surface integrals from which the moments can be
explicitly computed.
From the definitions (\ref{basicmodelA}), (\ref{basicmodelB}) and
(\ref{basicmodelC}) we obtain
\begin{equation}
\oint_\Sigma n_{<L>}\partial_j\Phi \,d^2\Sigma_j = \frac{1}{(2l+1)R^l}\nM_L -
\frac{l R^{l+1}}{(2l+1)!!}\nG_L,
\label{eq:recover1}
\end{equation}
$\,$
\begin{equation}
\oint_\Sigma n_{<L>}\partial_j\zeta_i \,d^2\Sigma_j =
\frac{1}{(2l+1)R^l}Z_{iL} - \frac{l R^{l+1}}{(2l+1)!!}Y_{iL},
\label{eq:recover2}
\end{equation}
and
\begin{eqnarray}\nn
&& \oint_\Sigma n_{<L>}\partial_j\psi \, d^2\Sigma_j =
\frac{1}{(2l+1)R^l} \pnM_L
\nn \\ \nn
&&+ \frac{1}{(l+1)(2l+3)R^l}\dot{\mu}_L
- \frac{l
R^{l+1}}{(2l+1)!!}\left[\pnG_L - {\dot \nu}_L \right]\\ \nn
& &  -
\frac{l-1}{2 (2l-1)(2l+1)R^{l-2}}
\nddotM_L
-\frac{(l+2)R^{l+3}}{2 (2l+3)!!}
\nddotG_L
. \nn \\
\label{eq:recover3}
\end{eqnarray}
Here $d^2\Sigma_j$ is the natural surface element determined by the
flat metric $(dx^1)^2 + (dx^2)^2 + (dx^3)^2$, and the 2-surface
$\Sigma$ is the coordinate sphere $r = R$.
By evaluating the right hand sides of these equations at several different
values of $R$, and by using the decompositions (\ref{eq:ZiLdecompos})
and (\ref{eq:YiLdecompos}), one can extract explicit expressions for
the moments $\nM_L$, $\nG_L$ [Eq.\ (\ref{eq:recover1})],
$H_L$, $S_L$, $\mu_L$, $\nu_L$ [Eq.\ (\ref{eq:recover2})], and
$\pnM_L$, $\pnG_L$ [Eq.\ (\ref{eq:recover3})] in terms of the surface
integrals and their time derivatives.

\section{Coefficients of final equation of motion after simplification
  using spin evolution equation}
\label{sec:newcoeffs}

In paper II \cite{Racine}, the following spin evolution equation is derived
\begin{eqnarray}
\dot{\bodyS}^A_i &=& \sum_{B\neq A}\sum_{k=0}^\infty\sum_{l=0}^\infty
\frac{(-1)^k}{k!l!}(2k+2l+1)!!\epsilon_{ijm}\bodyM^A_{jL}\bodyM^B_K
\nn \\
&& \times \frac{n^{BA}_{<mKL>}}{r_{BA}^{k+l+2}}.
\label{spinev}
\end{eqnarray}
If we substitute this equation into the full equation of motion (\ref{fulleom}), the following coefficients take the following new values
\begin{widetext}
\begin{eqnarray}
{}^{(3)}{\cal D}_{ijKL}^{AB} &=&
\frac{(-1)^k (2k+2l+1)!!}{k!l! \, \bodyM^A} \bigg\{
\bodyM^A_L\left[- \frac{4}{k+1}v^{BA}_j{\dot\bodyM}^B_{iK}
  -\bodyM^B_K(4v^{BA}_{i} v^{BA}_j + v^A_iv^B_j) \right. \nn \\
\mbox{} && +
 \left. \frac{3}{\bodyM^A}\left(\bodyM^A_{ij}\ddot{\bodyM}^B_K + 2{\dot\bodyM}^A_{ij}{\dot\bodyM}^B_K
  + \ddot{\bodyM}^A_{ij}\bodyM^B_K\right)
+ \epsilon_{ijm}\left(\frac{4(1-\delta_{0k})}{k+2}\dot{\bodyS}^B_{mK} - \bodyS^A_m\frac{{\dot\bodyM}^B_K}{\bodyM^A}\right)
  \right]
\nn\\
& & +2 \bodyM^A_{iL} {\dot\bodyM}^B_K \left[\frac{(l+2)(2l+1)}{(2l+3)} v^{B}_j
  -(l+1) v^A_j \right] \nn\\
& & + {\dot\bodyM}^A_L\left[\epsilon_{ijm}\left(\frac{4}{k+2}\bodyS^B_{mK} -
  \bodyS^A_m\frac{\bodyM^B_K}{\bodyM^A}\right)  +
  \frac{6}{\bodyM^A}\left(\bodyM^A_{ij}{\dot\bodyM}^B_K +
  {\dot\bodyM}^A_{ij}\bodyM^B_K\right)\right]  +
\frac{(2l^2+3l+5)}{(l+1)}v^{BA}_j{\dot\bodyM}^A_{iL}\bodyM^B_K  \nn\\
& &  + \frac{3}{\bodyM^A} \bodyM^A_{ij} \ddot{\bodyM}^A_L\bodyM^B_K +
\frac{4}{l+2}\epsilon_{ijm}\left(\bodyS^A_{mL}{\dot\bodyM}^B_K +
(1 - \delta_{0l})\dot{\bodyS}^A_{mL}\bodyM^B_K\right)
\bigg\}
\end{eqnarray}
and
\begin{eqnarray}\nn
{}^{(5)}{\tilde {\cal D}}_{iKLPQ}^{ABC} &=&
\frac{(-1)^{k+p}(2k+2l+1)!! (2p+2q+1)!!}{k!l!p!q!\, \bodyM^A} \,
\, \Bigg[\frac{(l+2)(2l+1)}{(2l+3)}\bodyM^A_{iL} \bodyM^B_K \frac{\bodyM^B_Q}{\bodyM^B} \bodyM^C_P \\
&& + \frac{4\delta_{0k}}{k+2}\bodyM^A_L\bodyM^B_{iQ}\bodyM^C_P \Bigg].
\end{eqnarray}
\end{widetext}
A new three-body term is generated, which contributes to $\tilde{a}^{ABC}_i$. It can be written as
\begin{equation}
\sum_{l=0}^\infty\sum_{p=0}^\infty\sum_{q=0}^\infty {}^{(7)}{\tilde {\cal D}}_{jLPQ}^{ABC} \frac{n^{CB}_{<iPQ>}}{r_{CB}^{p+q+2}}\frac{n^{BA}_{<jL>}}{r_{BA}^{l+2}}
\end{equation}
where
\begin{eqnarray}
{}^{(7)}{\tilde {\cal D}}_{jLPQ}^{ABC} &=& \frac{(-1)^{p+1}(2l+1)!!(2p+2q+1)!!}{l!p!q!\bodyM^A} 2 \, \bodyM^A_L\bodyM^B_{jQ}\bodyM^C_P.\nonumber \\
\end{eqnarray}
Finally, note that the $\dot{\bodyS}^A_m$ terms in the second and last
lines of Eq.\ (\ref{calDAB3}) cancel each other out. Therefore no
three-body terms are generated from $\dot{\bodyS}^A_m$.

\begin{widetext}

\begin{table}
\caption{
\label{table:symbols}
In this table we list, for ease of reference, some of the symbols
used in the paper in alphabetical order.
We do not list symbols whose meaning is very conventional, or
which are used only in the immediate vicinity of where they are
introduced.  For each item listed, we give a brief description, and
also a reference to the equation or section in the text where the symbol first
appears, or in the vicinity of which the symbol is first introduced.}
\begin{center}
\begin{tabular}{lp{5.5in}r}
\hline
Symbol & \multicolumn{1}{c}{Meaning} & \multicolumn{1}{p{1in}}{First appears in} \\
\hline\hline
${\,}^\text{g}$ & Superscript appended to a symbol denoting that it
is defined with respect to the global coordinate system & (\ref{globalfieldsA})\\
${\,}^\text{n}$ & Superscript prepended to a symbol denoting the
Newtonian piece of a quantity
& \\
${\,}^\text{pn}$ & Superscript prepended to a symbol denoting the
post-1-Newtonian piece of a quantity & \\
$A$ & Index appended to a symbol indicating that is associated with
the $A^{\text{th}}$ body in an $N$-body system ($B$ and $C$ are used
similarly) & Sec.\ \ref{sec:resultseom} \\
$\alpha(t,x^j)$ & Function appearing in gauge transformation which
parameterizes Newtonian-order changes in the time variable
& (\ref{eq:generalcoordtransform})\\
$\alpha_\text{c}(t)$ & Piece of $\alpha(t,x^j)$ that is independent of
spatial coordinates & (\ref{conshgA}) \\
$\beta(t,x^j)$ & Function appearing in gauge transformation which
parameterizes post-1-Newtonian changes in the time variable
& (\ref{eq:generalcoordtransform})\\
$\beta_\text{h}(t,x^j)$ & Piece of $\beta(t,x^j)$ that satisfies
Laplace's equation & (\ref{conshgC})\\
$\chi_{ij}$ & symmetric tensor parameterizing the
post-2-Newtonian spatial metric & (\ref{eq:metricP2N})\\
$\ve$ & Post-Newtonian dimensionless expansion parameter &
(\ref{metric0}) \\
$F_L(t)$ & Tidal moment of order $l$ parameterizing $\psi$ about a
worldline & (\ref{eq:tidalA})\\
$\mathfrak{g}^{\mu\nu}$ & tensor density sometimes called the ``gothic metric'' equal to $\sqrt{-g}g^{\mu\nu}$ & (\ref{gothicmetricdef})\\
$\nG_L(t)$ & Newtonian gravitoelectric tidal moment of order $l$ & (\ref{basicmodelA00}) \\
$\pnG_L(t)$ & Post-1-Newtonian gravitoelectric tidal moment of order
$l$ & (\ref{basicmodelB})\\
$G_L(t)$ & Total gravitoelectric tidal moment of order
$l$ & (\ref{eq:totalGLdef})\\
$h^i(t,x^j)$ & Free function in gauge transformation which parameterizes
post-1-Newtonian translations & (\ref{eq:generalcoordtransform})\\
$h^i_\text{c}(t)$ & Piece of $h^i(t,x^j)$ that is independent of
spatial coordinates & (\ref{conshgB}) \\
${\cal H}^{\mu\alpha\nu\beta}$ & Tensor density appearing in the Landau-Lifshitz formulation of general relativity & (\ref{Einsteinsequations}) \\
$H_L(t)$ & Post-1-Newtonian gravitomagnetic tidal moment of order $l$ & (\ref{reducedzeta})\\
$J_L(t)$ & Tidal moment of order $l$ parameterizing $\psi$ about a
worldline & (\ref{eq:tidalA})\\
$K$ & The multi-index $b_1 b_2 \ldots b_k$ & Sec.\ \ref{sec:notation} \\
$L$ & The multi-index $a_1 a_2 \ldots a_l$ & Sec.\ \ref{sec:notation} \\
$\lambda_L(t)$ & Intrinsic-type multipole moment parameterizing the
harmonic gauge-transformation function $\beta_\text{h}(t,x^j)$ & (\ref{eq:lambdaLdef}) \\
$\Lambda^\Phi_L(t)$ & Inertial moments that appear in the
transformation law of $\nG_L(t)$, nonvanishing for $l=0,1$ only& (\ref{newG})\\
$\Lambda^{\bm{\zeta}}_L(t)$ &
Inertial moments that appear in the
transformation law of $Y_{iL}(t)$, nonvanishing for $l=1,2,3$ only
& (\ref{newY})\\
$\Lambda^{\psi_\text{h}}_L(t)$ &
Inertial moments that appear in the
transformation law of $\pnG_L(t)$, nonvanishing for $l=0,1,2$ only
&(\ref{newpnG})\\
$\nM_L(t)$ & Newtonian mass multipole moment of order $l$ & (\ref{basicmodelA00}) \\
$\pnM_L(t)$ & Post-1-Newtonian mass multipole moment of order $l$ &
(\ref{basicmodelB})\\
$M_L(t)$ & Total mass multipole moment of order $l$ &
(\ref{eq:totalMLdef})\\
$\bodyM_L(t)$ & Mass multipole moment of order $l$ defined in a body-frame non-rotating with respect to distant stars & (\ref{eq:calMdef})\\
$\mu_L(t)$ & Intrinsic gauge moment of order $l$ &
(\ref{basicmodelB})\\
$N$ & The multi-index $a_1 a_2 \ldots a_n$ & Sec.\ \ref{sec:notation} \\
$N_L(t)$ & Intrinsic multipole moment of order $l$ parameterizing $\psi$ about a worldline & (\ref{eq:tidalA})\\
$n^{BA}_j(t)$ & $j^{\text{th}}$ component of a spatial unit vector
pointing from the worldline of body $A$ at time $t$ to the worldline
of body $B$ at time $t$ & Sec.\ \ref{sec:resultseom} \\
$\nu_L(t)$ & Tidal gauge moment of order $l$ & (\ref{basicmodelB})\\
$P$ & The multi-index $c_1 c_2 \ldots c_p$ & Sec.\ \ref{sec:notation} \\
$P^i_\Sigma$ & Momentum enclosed by a surface $\Sigma$ & (\ref{momentum})\\
$P_L(t)$ & Intrinsic multipole moment of order $l$ parameterizing $\psi$ about a worldline & (\ref{eq:tidalA})\\
$\Phi$ & Newtonian potential & (\ref{metric0}) \\
$\psi$ & Post-1-Newtonian correction to the Newtonian potential & (\ref{metric0}) \\
$Q$ & The multi-index $d_1 d_2 \ldots d_q$ & Sec.\ \ref{sec:notation} \\
$R_k(t)$ & Function appearing in gauge transformation which
parameterizes post-1-Newtonian rotations & (\ref{conshgB}) \\
\hline
\end{tabular}
\end{center}

\end{table}

\begin{table}
\begin{center}
\begin{tabular}{lp{5.5in}r}
\hline
Symbol & \multicolumn{1}{c}{Meaning} & \multicolumn{1}{p{1in}}{First appears in} \\
\hline\hline
$r_{BA}$ & coordinate distance between the center of mass worldlines
of bodies $A$ and $B$, defined with respect to the flat metric
$\delta_{ij}$ & Sec.\ \ref{sec:resultseom} \\
$s_A$ & time coordinate of a coordinate system adapted to body $A$ & (\ref{eq:bufferregion}) \\
$S_L(t)$ & Current multipole moment of order $l$ & (\ref{reducedzeta})\\
$\bodyS_L(t)$ & Current multipole moment of order $l$ defined in a
body-frame non-rotating with respect to distant stars & (\ref{eq:calSdef})\\
$t$ & time coordinate of generic, harmonic, conformally Cartesian
coordinate system in Secs.\
\ref{coords} -- \ref{eom} & (\ref{metric0}) \\
$t$ & time coordinate of global coordinate system for an $N$ body system
in Secs.\ \ref{manybody} -- \ref{explicit} & Sec.\ \ref{sec:assumptions} \\
$T^{\mu\nu}$ & Components of the stress-energy tensor & (\ref{eq:st}) \\
${\cal T}^{\mu\nu}$ & Landau-Lifshitz pseudotensor & (\ref{Einsteinsequations}) \\
$\tau_L(t)$ & Tidal-type multipole moment parameterizing the harmonic
gauge-transformation function $\beta_\text{h}(t,x^j)$ & (\ref{eq:lambdaLdef}) \\
${\cal T}^p_N(\bm{z})$ & Taylor coefficients of the function $|\bm{z} - \bm{x}|$ about $\bm{x} = 0$ & (\ref{taylor}) \\
$U^j_i(t)$ & Rotation matrix describing the dragging of asymptotic
rest frames & (\ref{eq:dragging}) \\
$v^A_i$ & Velocity of the $A$th body & (\ref{eq:rbadef}) \\
$v^{AB}_i$ & Relative velocity of bodies $A$ and $B$ &
(\ref{eq:rbadef}) \\
$x^i$ & spatial coordinates of generic, harmonic, conformally Cartesian
coordinate system in Secs.\
\ref{coords} -- \ref{eom} & (\ref{metric0}) \\
$x^i$ & spatial coordinates of global coordinate system for an $N$ body system
in Secs.\ \ref{manybody} -- \ref{explicit} & Sec.\ \ref{sec:assumptions} \\
$\xi_i$ & $i^{\text{th}}$ component of the post-2-Newtonian correction to the gravitomagnetic vector potential & (\ref{eq:metricP2N}) \\
$y_A^j$ & spatial coordinates of a coordinate system adapted to body
$A$ & Sec.\ \ref{sec:resultseom}\\
$Y_{iL}(t)$ &  Tidal moments of order $l$ of the gravitomagnetic potential & (\ref{basicmodelC})\\
$z_i(t)$ & Free function in gauge transformation which parameterizes
Newtonian-order translations & (\ref{eq:generalcoordtransform}) \\
$z_i(t)$ & The center-of-mass worldline of a body, to Newtonian order
& (\ref{coordinatetransformationII})\\
$\cmz_i(t)$ & Center of mass worldline of a body, to
post-1-Newtonian order & Sec.\ \ref{sec:configvars} \\
$Z_{iL}$ & Intrinsic multipole moments of order $l$ of the gravitomagnetic potential & (\ref{basicmodelC})\\
$\zeta_i$ & $i^{\text{th}}$ component of the gravitomagnetic vector potential & (\ref{metric0}) \\
\hline
\end{tabular}
\end{center}

\end{table}
\end{widetext}

\end{document}

\title{Erratum:  Post-1-Newtonian equations of motion for systems of arbitrarily bodies [Phys. Rev. D 71 044010 (2005)]}

\author{\'{E}tienne Racine}
\affiliation{Canadian Space Agency, 6767 route de l'A\'{e}roport, Saint-Hubert, Qu\'{e}bec, Canada J3Y 8Y9}

\author{Justin E. Vines}
\affiliation{Center for Radiophysics and Space Research, Cornell University, Ithaca, New York USA 14853}

\author{\'{E}anna \'{E}. Flanagan}
\affiliation{Center for Radiophysics and Space Research, Cornell University, Ithaca, New York USA 14853}

\begin{abstract}
We correct some errors in
the published paper which were discovered by one of us (JEV), who
joins the two original authors (ER and EEF) as authors of this
erratum.
\end{abstract}

\maketitle

\begin{widetext}

\section{Errors that affected the final results}

\begin{itemize}

\item The expression (6.12c) for the quantity ${\tilde a}^{ABC}_i$ is missing
the following term:
%\begin{widetext}
\be
%\tilde{a}_i^{ABC} \rightarrow \tilde{a}_i^{ABC} +
\frac{1}{2}\sum_{k=0}^\infty\sum_{l=0}^\infty\sum_{p=0}^\infty\sum_{q=0}^\infty
c_{klpq}
\frac{\mathcal{M}^A_L}{\mathcal{M}^A}\mathcal{M}^B_K\frac{\mathcal{M}^B_Q}{\mathcal{M}^B}\mathcal{M}^C_P \, \frac{n^{CB}_{<jPQ>}n^{BA}_{<ijKL>}}{r_{CB}^{p+q+2}r_{BA}^{k+l+1}} \,,
\ee
where
\be
c_{klpq} = \frac{(-1)^{k+p} (2k+2l+1)!! (2p+2q+1)!!}{k!l!p!q!}.
\ee
%\end{widetext}
The original authors simply omitted to transcribe it from their handwritten notes to the source file of the paper, and failed to notice the oversight.

\item Appendix F claimed that no three-body terms are generated from
$\dot{S}^A_m$ when substituting
the spin evolution equation (F1) into Eq.\ (6.13c). This is incorrect, as
it is only the $l=0$ terms in second and last line of (6.13c) that
cancel out.  To correct this error the following term should be added
to the right hand side of Eq. (6.13c):
\be
\frac{(-1)^k(2k+2l+1)!!}{k!l!\mathcal{M}^A}(-2\mathcal{M}^A_L)(1-\delta_{0l})\epsilon_{ijm}\dot{S}^A_m\frac{\mathcal{M}^B_K}{\mathcal{M}^A}\,.
\ee
Upon making use of Eq.(F1), the following new contribution to $a_i^{ABC}$ in Eq.(6.12b) is created
\bea
a_i^{ABC} &\rightarrow& a_i^{ABC} + \nn \\
&& \sum_{k=0}^\infty\sum_{l=0}^\infty\sum_{p=0}^\infty\sum_{q=0}^\infty \!\!{\,}^{(6)}\!\mathcal{D}^{ABC}_{jKLPQ} \frac{n^{CA}_{<iPQ>}}{r_{CA}^{p+q+2}}\frac{n^{BA}_{<jKL>}}{r_{BA}^{k+l+2}}\,, \nn \\
\eea
where
%\begin{widetext}
\be
\!\!{\,}^{(6)}\!\mathcal{D}^{ABC}_{jKLPQ} = \frac{(-1)^{k+p}(2k+2l+1)!!(2p+2q+1)!!}{k!l!p!q!\mathcal{M}^A}(2\mathcal{M}^A_L)(1-\delta_{0l})\mathcal{M}^A_{jQ}\mathcal{M}^C_P\frac{\mathcal{M}^B_K}{\mathcal{M}^A}
\ee
In addition, the coefficient $\!\!{\,}^{(4)}\!\mathcal{D}^{ABC}_{iKLPQ}$, defined in eq.(6.14d), becomes
\be
\!\!{\,}^{(4)}\!\mathcal{D}^{ABC}_{iKLPQ} \rightarrow \!\!{\,}^{(4)}\!\mathcal{D}^{ABC}_{iKLPQ} + \frac{(-1)^{k+p+1}(2k+2l+1)!!(2p+2q+1)!!}{k!l!p!q!\mathcal{M}^A}(2\mathcal{M}^A_L)(1-\delta_{0l})\mathcal{M}^A_{iQ}\mathcal{M}^C_P\frac{\mathcal{M}^B_K}{\mathcal{M}^A}\,.
\ee
%\end{widetext}
The three-body contributions coming from $\dot{S}^B_{m}$ (as opposed
to $\dot{S}^A_m$) given in Appendix F are correct.
\end{itemize}

\section{Errors that did not affect the final results}
\begin{itemize}

\item Equations (3.38c) and (3.38d) for transformed gravitomagnetic
  tidal moments are incorrect, due to a factor of 2 error in Eq.\
  (A19). The second term in the right hand side of Eq.\ (A19) should
  be multiplied by 2.
  Correspondingly, Eqs.\ (3.38c) and (3.38d) should read
%\begin{widetext}
\bea
\bar{H}_L &=& - l! \Lambda^{\bm{\zeta}}_{ij<L-1}\epsilon_{a_l>ij} + \sum_{k=0}^\infty \frac{1}{k!}\Bigg\{H_{LK} + 4\nG_{iK<L-1}\epsilon_{a_l>ij}\dot{z}^j + \frac{4}{k+l+1} \ndotG_{iK<L-1}\epsilon_{a_l>ij}z^j\Bigg\}z^K, \\
\bar{\nu}_{iL} &=& -\tau_{iL} - l!\Lambda^{\bm{\zeta}}_{<iL>} + \sum_{k=0}^\infty\frac{1}{k!}\Bigg\{\nu_{iLK} + \frac{1}{k+l+2}z^j\epsilon_{jm<i}H_{L>mK} + \frac{4}{(k+l+2)(2k+2l+5)}z_jz^j\ndotG_{iLK} \nn \\
&& - \frac{4(2k+1)}{(k+l+1)(2k+2l+3)}z_{<i}\ndotG_{L>K} - 4\dot{z}_{<i}\nG_{L>K}\Bigg\}z^K.
\eea
Note that these expressions were not used in the derivation of the
equations of motion, and were provided instead for completeness.

%\end{widetext}

\item In the paper it was stated that Eq.\ (A18) is valid for tensors
  $T_{iK}$ that are STF on their last $k$ indices.  In fact, the equation
  is only valid for tensors $T_{iK}$ that are STF on all of their indices.

\item Equations (3.39) and (3.40a) should be replaced by
\bea
\barpnM_L &=& -\frac{(2l+1)}{(l+1)(2l+3)}\dot{\bar{\mu}}_L + \dot{\lambda}_L + \sum_{n=0}^l (-1)^{n+l}\frac{l!}{n!(l-n)!}\sigma^{(l)}_{<N}z_{L-N>}\,,
\eea
and
\bea
&& \sigma_N^{(l)} = \pnM_N - \dot{z}^jZ_{jN} + 2\dot{z}_j\dot{z}^j\nM_N + (\alpha_{\rm c} - z_j\dot{z}^j)\ndotM_N - nU_{<a_n}\nM_{N-1>} + \frac{1}{2l+3}\Big[z_jz^j T_N - 2z_jT_{jN}\Big] \nn \\
&& + \left(\frac{2n+1}{2n+3}\right)\left[\frac{1}{n+1}\dot{\mu}_N - \nM_N U_{jj} - n\nM_{j<N-1}U_{a_n>j} + \frac{2n}{2n+1}U_{j<a_n}\nM_{N-1>j} + \dot{z}_j\ndotM_{jN} + (n+2)\ddot{z}_j\nM_{jN}\right]. \nn \\
\eea
%\end{widetext}
Here the tensor $T_N$ is defined as
\bea
T_N &=& \frac{1}{2}\nddotM_N - n \dot{z}_{<a_n}\ndotM_{N-1>} - \frac{n}{2}\ddot{z}_{<a_n}\nM_{N-1>}  + n(n-1)\nM_{<N-2}U_{a_{n-1}a_n>}.
\eea

\item Equation (3.40b) should be replaced by
\be
U_i = h_{\rm c}^i - z_j U_{ij} -z_jz_kV_{ijk}\,,
\ee
i.e. $V_{ij}$ should be replaced by $U_{ij}$.
Again these corrections to Eqs. (3.39), (3.40a) and (3.40b) do not
affect the final results, since these expressions
were not used in the derivation of the
equations of motion.

\item In Eq.\ (6.5), the following replacement must be made
\be
\sum_{l=2}^\infty \mathcal{M}^A_L\mathcal{G}^A_{iL} \rightarrow \sum_{l=2}^\infty \frac{1}{l!}\mathcal{M}^A_L\mathcal{G}^A_{iL} \,.
\ee
This was a typo and our computations were performed using the
correct expression.

\item Equation (6.7) contains two typos. First, in the third term of
  the right-hand side, the quantity $\dot{z}^A_i Y^{{\rm g},A}_{ji}$
  should be replaced by $\dot{z}^A_j Y^{{\rm g},A}_{ji}$. The second
  typo is that the term $\epsilon_{ijk}\bar{f}^A_jR^A_k$ in the third
  line should not be there, as it is absorbed in the definitions
  (5.39) and (6.6) of $\mathcal{M}^A_L$ and $\mathcal{G}^A_L$ when
  going from the laws of motion (4.3) to Eq.\ (6.5).
\end{itemize}

\section{Simplified form of some expressions}

In addition to correcting the above errors,
we take this opportunity to provide simplified forms of some of the
results in the paper:

\begin{itemize}

\item While all the terms in Eq.\ (6.12) are correct as they
currently stand, we provide here simplified
versions for some of them by getting rid of remaining undeveloped
symmetric trace-free (STF) projections. Specifically, the STF
projections appearing in
$\!\!{\,}^{(2)}\!\tilde{\mathcal{D}}^{ABC}_{ijKLPQ}$, defined in
Eq.(6.15b), and in
$\!\!{\,}^{(4)}\!\tilde{\mathcal{D}}^{ABC}_{jKLPQ}$, defined in
Eq.(6.15d), can be rewritten such that the structure of the three-body
acceleration terms in Eq.(6.12c) becomes the following. First, the
term involving $\!\!{\,}^{(2)}\!\tilde{\mathcal{D}}^{ABC}_{ijKLPQ}$
in Eq.\ (6.12c) gets replaced by the following contribution
\be
{\,}^{(2)}\!\tilde{\mathcal{D}}^{ABC}_{KLPQ} \frac{n^{CB}_{<jPQ>}}{r_{CB}^{p+q+2}}\frac{n^{BA}_{<ijKL>}}{r_{BA}^{k+l+3}}\,,
\ee
where
%\begin{widetext}
\bea
{\,}^{(2)}\!\tilde{\mathcal{D}}^{ABC}_{KLPQ} &=& \frac{(-1)^{k+p} (2k+2l+3)!!(2p+2q+1)!!}{k!l!p!q! (2k+2l+7){\mathcal{M}^A}} \mathcal{M}^A_{mL}\mathcal{M}^B_{mK}\frac{\mathcal{M}^B_Q}{\mathcal{M}^B}\mathcal{M}^C_P\,.
\eea
Next, the coefficient $\!\!{\,}^{(3)}\!\tilde{D}^{ABC}_{KLPQ}$ defined in Eq.(6.15c) becomes, after absorbing some terms,
\bea
{\,}^{(3)}\!\tilde{\mathcal{D}}^{ABC}_{KLPQ} &=& \frac{(-1)^{k+p} (8k+8l+11)(2k+2l+1)!!(2p+2q+1)!!}{k!l!p!q! (2k+2l+3)(2k+2l+1){\mathcal{M}^A}}  \mathcal{M}^A_{L}\mathcal{M}^B_{K}\frac{\mathcal{M}^B_Q}{\mathcal{M}^B}\mathcal{M}^C_P\,,
\eea
the coefficient $\!\!{\,}^{(4)}\!\tilde{D}^{ABC}_{jKLPQ}$ defined in (6.15d) becomes
\bea
{\,}^{(4)}\!\tilde{\mathcal{D}}^{ABC}_{jKLPQ} = \frac{(-1)^{k+p+1}(2k+2l+1)!!(2p+2q+1)!!}{k!l!p!q!(2k+3)\mathcal{M}^A}\Big[(k+2)(2k+1)\mathcal{M}^A_L\mathcal{M}^B_{jK} + \mathcal{M}^A_{jL}\mathcal{M}^B_K\Big]\frac{\mathcal{M}^B_Q}{\mathcal{M}^B}\mathcal{M}^C_P\,,
\eea
and the coefficient $\!\!{\,}^{(5)}\!\tilde{\mathcal{D}}^{ABC}_{iKLPQ}$ defined in (6.15e) becomes
\bea
{\,}^{(5)}\!\tilde{\mathcal{D}}^{ABC}_{iKLPQ} = \frac{(-1)^{k+p}(2k+2l+1)!!(2p+2q+1)!!}{k!l!p!q!\mathcal{M}^A}\Bigg[(l+1)\mathcal{M}^A_{iL}\mathcal{M}^B_K + \frac{(\mathcal{M}^A_L\mathcal{M}^B_{iK} - \mathcal{M}^A_{iL}\mathcal{M}^B_{K})}{2k+2l+5}\Bigg]\frac{\mathcal{M}^B_Q}{\mathcal{M}^B}\mathcal{M}^C_P\,.
\eea
%\end{widetext}
The coefficients  $\!\!{\,}^{(1)}\!\tilde{\mathcal{D}}^{ABC}_{KLP}$
and  $\!\!{\,}^{(6)}\!\tilde{\mathcal{D}}^{ABC}_{ijKLPQ}$, defined in
Eqs.\ (6.15a) and (6.15f) respectively, are left unchanged by the reduction of the STF projections.

\item We give an alternative, more transparent version of Eq.\ (3.41), which eliminates awkward interlocking STF projections. The transformed post-1-Newtonian gravitoelectric tidal moment can be written instead as
\be
\barpnG_L = \dot{\bar{\nu}}_L + \dot{\tau}_L - l!\Lambda^\psi_{<L>} + \sum_{k=0}^\infty \frac{1}{k!} \rho_{LK}z^K\,,
\ee
where
%\begin{widetext}
\bea
\rho_{LK} &=& \pnG_{LK} - \dot{\nu}_{LK} + \frac{z^jz_j}{2(2k+2l+3)}\nddotG_{LK} - \dot{z}_j Y_{jLK} + 2\dot{z}_j\dot{z}^j \nG_{LK} + \alpha_{\rm c} \ndotG_{LK} + h_{\rm c}^j \nG_{jLK} \nn \\
&& + \frac{l}{2k+2l+1}z_{<a_l}\nddotG_{L-1>K} + l\,\dot{z}_{<a_l}\ndotG_{L-1>K} + lV_{j<a_l}\nG_{L-1>jk} - l(l-1)\ddot{z}_{<a_l} \nG_{L-1>K}\,.
\eea
%\end{widetext}
Here the quantity $V_{ij}$ is defined in Eq.\ (3.40d) of the paper.

\end{itemize}

\end{widetext}